%% file: Main.tex
\documentclass[a4paper,fleqn,usenatbib]{mnras}

\usepackage{newtxtext,newtxmath}

\usepackage[T1]{fontenc}
\usepackage{ae,aecompl}
\usepackage{listings}

\usepackage{graphicx}	
\usepackage{amsmath}	
\usepackage{amssymb}	
\usepackage{color}      
\usepackage{float}

\usepackage{wrapfig}
\usepackage{lscape}
\usepackage{rotating}
\usepackage{epstopdf}
\graphicspath{{./Graphs/}}

\def\Msol{\mathrel{\rm M_{\odot}}} 

\def\Hal{H$\alpha$ }
\def\Hb{H$\beta$ }

\definecolor{purple}{RGB}{175,0,175}
\definecolor{red}{RGB}{255,0,0}
\definecolor{darkblue}{RGB}{0,0,175}
\definecolor{lime}{RGB}{0,255,0}

\title[KASHz: AGN outflows and star formation]{KASHz: No evidence for ionised outflows instantaneously suppressing star formation in moderate luminosity AGN at $z$$\sim$$1.4$--$2.6$}

\author[J. Scholtz, et al.]{\parbox[h]{\textwidth}{ 
		J.\ Scholtz$,^{\! 1,2}$\thanks{E-mail: honzascholtz@gmail.com}
		C.M. Harrison,$^{\! 2}$\thanks{E-mail: c.m.harrison@mail.com}
		D.J. Rosario,$^{\! 1}$
		D.M. Alexander,$^{\! 1}$
		C-C. Chen,$^{\! 2, 3}$
		D. Kakkad,$^{\! 4}$
		V. Mainieri,$^{\! 2}$ 
		A.L. Tiley,$^{\! 1,5}$
		O. Turner,$^{\! 6}$
		M. Cirasuolo,$^{\! 2}$
		R.M. Sharples,$^{\! 1, 7}$
		S. Stach$^{\! 1}$
	}
	\\
	\\
	$^{1}$ Centre for Extragalactic Astronomy, Department of Physics, Durham University, South Road, Durham,  DH1 3LE, UK\\
	$^{2}$ European Southern Observatory, Karl-Schwarzschild-Str. 2, 85748 Garching b. Munchen, Germany\\
	$^{3}$ Academia Sinica Institute of Astronomy and Astrophysics (ASIAA), No. 1, Section 4, Roosevelt Rd., Taipei 10617, Taiwan \\
	$^{4}$ European Southern Observatory, Alonso de Cordova 3107, Vitacura, Casilla, 19001, Santiago de Chile, Chile\\
	$^{5}$ International Centre for Radio Astronomy Research, University of Western Australia, 35 Stirling Highway, Crawley, WA 6009, Australia \\
	$^{6}$ SUPA, Institute for Astronomy, University of Edinburgh, Royal Observatory, Edinburgh EH9 3HJ \\
	$^{7}$ Centre for Advanced Instrumentation, Department of Physics, Durham University, South Road, Durham DH1 3LE UK
}

\date{XYZ}

\pubyear{2019}

\begin{document}
	\label{firstpage}
	\pagerange{\pageref{firstpage}--\pageref{lastpage}}
	\maketitle
	
	\begin{abstract}
		As part of our KMOS AGN Survey at High-redshift (KASHz), we present spatially-resolved VLT/KMOS and VLT/SINFONI spectroscopic data and ALMA 870$\mu$m continuum imaging of eight $z$=1.4--2.6 moderate AGN ($L_{\rm 2-10 \rm kev}$ = $10^{42} - 10^{45}$ ergs s$^{-1}$). We map [O~{\sc iii}], H$\alpha$ and rest-frame FIR emission to search for any spatial anti-correlation between ionised outflows (traced by the [O~{\sc iii}] line) and star formation (SF; traced by H$\alpha$ and FIR), that has previously been claimed for some high-z AGN and used as evidence for negative and/or positive AGN feedback. Firstly, we conclude that H$\alpha$ is unreliable to map SF inside our AGN host galaxies based on: (i) SF rates inferred from attenuation-corrected H$\alpha$ can lie below those inferred from FIR; (ii) the FIR continuum is more compact than the H$\alpha$ emission by a factor of $\sim 2$ on average; (iii) in half of our sample, we observe significant spatial offsets between the FIR and H$\alpha$ emission, with an average offset of $1.4\pm0.6$ kpc. Secondly, for the five targets with outflows we find no evidence for a spatial anti-correlation between outflows and SF using either H$\alpha$ or FIR as a tracer. This holds for our re-analysis of a famous $z$=1.6 X-ray AGN (`XID\,2028') where positive and negative feedback has been previously claimed. Based on our results, any instantaneous impact on SF by ionised outflows must be subtle, either occurring on scales below our resolution, or on long timescales.

	\end{abstract}
	
	
	
	
	\section{Introduction}\label{Intro}
	
	Supermassive black holes (SMBH) are known to reside at the centre of massive galaxies (\citealt{Kormendy13}). When these SMBHs grow, through accretion events, they become visible as active galactic nuclei \citep[AGN; ][]{Soltan82,Merloni04}. Current theoretical models of galaxy evolution require these AGN to inject significant energy into their host galaxies in order to replicate the basic properties of local galaxies and the intergalactic medium (IGM), such as: the black hole--spheroid relationships, the steep mass function, increased width of sSFR distributions (star formation rate normalized to stellar mass) as a function of stellar mass, galaxy sizes, AGN number densities, galaxy colour bi-modality and enrichment of the IGM by metals \citep[e.g.,][]{Silk98,DiMatteo05,Alexander12, Vogelsberger14,Hirschmann14, Crain15,Segers16,Beckmann17,Harrison17,Choi18,Scholtz18}. The key role of the AGN in these models is to either regulate the cooling of the interstellar medium (ISM) or intracluster medium (ICM), or to eject gas out of the galaxy through outflows. Ultimately this process, usually referred to as ``AGN feedback'', is believed to regulate the rate at which stars can form. However, from an observational perspective, there is still no clear consensus in the literature on the role of AGN in regulating star formation in the overall galaxy population, particularly at high redshift \citep[e.g.,][]{Harrison17,Cresci18}. 
	
	Over the past decade there have been many studies identifying and characterising multiphase outflows \citep[see e.g.,][]{Harrison18,Cicone18}. Indeed, there is now significant evidence that energetic ionised, atomic and molecular outflows are a common property of AGN \citep[e.g.][]{Veilleux05,Morganti05, Ganguly08,Alexander10, Sturm11, Cicone12, Harrison12, Mullaney13,Cicone14,Balmaverde15,Carniani15,Brusa15,Harrison16, Woo16, Leung17, Brusa18, ForsterSch18b,Lansbury18,Fluetsch19,RamosAlmeida19}. 
	
	AGN-driven outflows have been identified on scales between tens of parsecs to tens of kiloparsecs \citep{StorchiBergmann10,Veilleux13,Cresci15,Feruglio15,Kakkad16,McElroy16,Rupke17,Jarvis19}. However, despite observations showing that AGN outflows are common, the impact that they have on star formation is still open to debate. Although, in many of the studies, the most powerful outflows are thought to remove gas at a rate faster than it can be formed into stars, there are still considerable uncertainties in these mass outflow rate calculations due to uncertain spatial scales and the assumptions required to convert emission line luminosities into gas masses \citep[e.g.][]{Karouzos16,VillarMartin16,Husemann16,Rose18,Harrison18}. Measurements are more accurate for the most nearby sources \citep[e.g.][]{Revalski18,Venturi18,Fluetsch19}; however, these samples lack the most powerful AGN which are thought to be the most important for influencing galaxy evolution.  
	
	Another approach to determine the impact of AGN-driven outflows, is to use spatially-resolved observations to map both the outflows and the star formation in or around the outflows. For example, using longslit and integral-field spectroscopy star formation has been detected {\em inside} outflows in local AGN host galaxies, which may be a form of `positive' feedback \citep{Maiolino17, Gallagher19}. On the other hand, \citet{Cresci15} suggest both {\em suppressed} star formation at the location of an ionised outflow (`negative feedback') and {\em enhanced} star formation around the edges of the outflow (`positive feedback') for a $z$=1.6 X-ray identified AGN, commonly referred to as `XID 2028'. Similar findings were presented for three $z$=2.5 extremely powerful (and consequently rare) quasars by \cite{Canodiaz12} and \cite{Carniani16}. These latter works, studying high-redshift AGN, used high-velocity [O~{\sc iii}]$\lambda$5007 emission-line components to map the ionised outflows and H$\alpha$ emission to map the spatial distribution star formation. 
	
	H$\alpha$ emission (as well as ultra-violet continuum) can be used to trace regions of on-going star formation \citep[e.g.][]{Hao11, Murphy11}. However, since this emission is at relatively short wavelengths it is sensitive to dust obscuration. Indeed, significant levels of the star formation in high-redshift galaxies is obscured by dust \citep{Madau96,Casey14,Whitaker14} and sometimes the UV and H$\alpha$ emission can be completely hidden by dust (e.g., \citealt{Hodge16,Chen17}). In these cases the UV and optical light is absorbed by the dust and re-emitted at far-infrared (8--1000\,$\mu$m; FIR) wavelengths. Consequently, the FIR emission is sensitive to on-going {\em obscured} star formation \citep[for reviews see][]{Kennicutt12, Calzetti13}. Importantly for this work, high-redshift AGN and quasar host galaxies have been shown to host significant levels of star-formation obscured by dust \citep[e.g.,][]{Whitaker12, Burgarella13, Stanley15, Stanley18}. In this current study we investigate different possible tracers of star formation, in $z$=1.4--2.6 AGN host galaxies, by combining integral-field spectroscopy, to map the H$\alpha$ emission, with high spatial-resolution observations of the rest-frame FIR from the Atacama Large Millimetre Array (ALMA) 
	
	A limitation of previous work, that investigates the impact of AGN outflows on star formation in distant galaxies, is that they are based on only a few targets and it is consequently unclear how common these effects are in the wider population of more typical high-redshift AGN. Therefore, in this work we make use of our large, representative parent sample of the ``KMOS AGN Survey at High-z'' \citep[KASHz:][Harrison et al. in prep]{Harrison16}. KASHz is a systematic integral field spectroscopy survey designed to spatially-resolve the rest-frame optical emission lines of $\approx$250 X-ray selected AGN, that are representative of the distant ($z=$0.6--2.6) AGN population. KASHz has the benefit of characterising the ionised gas properties of typical distant X-ray AGN and can be used to place into context other studies based on smaller numbers of targets, such as higher spatial-resolution (AO-assisted) integral field unit (IFU) observations \citep[e.g., SUPER survey, see][]{Circosta18}. By combining multi-wavelength photometry from UV--submm we can characterise the star-formation rates, AGN luminosities and stellar masses of the sample, and explore the ionised gas properties, such as the prevalence of outflows, as a function of various AGN and host galaxy properties. KASHz has already demonstrated that: (1) AGN are 5--10 times more likely to host high-velocity outflows ($>$ 600 km s$^{-−1}$) than star-forming, non-active, galaxies; and (2) shown that the most luminous AGN (L$_{\rm X} > 6 \times 10^{43}$ erg s$^{-−1}$) are $\sim 2$ times more likely to host high-velocity outflows than less luminous AGN \citep[]{Harrison16}. Importantly, the sample still contains some relatively extreme sources, both in terms of AGN luminosity and outflow properties \citep[e.g., `XID 2028' presented in ][]{Cresci15}; however, we can place these objects within the context of the overall, more typical, AGN population.
	
	In this pilot study we use sensitive high-resolution ALMA observations and IFU observations of 8 moderate luminosity AGN at redshift z=1.6-2.6. With these data we compare and contrast the FIR continuum and H$\alpha$ as possible star formation tracers in our AGN host galaxies. Combining these possible star formation tracers with the observations of AGN outflows, we then investigate the impact of these outflows on the star formation. In \S2 we describe the sample selection and the data used in our study, in \S 3 we outline the data analyses such as spectral fitting, constructing outflow maps and the analyses of the ALMA data, and in \S4 we present our results and discuss them within the broader context of the impact that AGN outflows have on star formation.
	
	In all of our analyses we adopt the cosmological parameters of
	$\rm H_0 = 67.3 \ km \, s^{-1}$, $\rm \Omega_M = 0.3$, $\rm
	\Omega_\Lambda = 0.7$ \citep{Planck13} and assume a \citet{Chabrier03} initial mass
	function (IMF).
	
	\section{Sample Selection, observations and source properties}
	The primary objectives of our study are to (i) compare the H$\alpha$ and FIR continuum emission as tracers of the star-formation inside AGN host galaxies (at the peak of cosmic star-formation and black-hole accretion; i.e., $z$ = 1--3; \citealt{Madau14}; \citealt{Aird15}) and to (ii) establish if AGN-driven ionised outflows have an instantaneous impact on the star formation within these galaxies. To achieve this, we select a sample of AGN host galaxies with spatially-resolved H$\alpha$ and [O~{\sc iii}] emission from integral field spectroscopy and with ancillary rest-frame FIR data from ALMA. In \S~\ref{sec:Sample} we describe the selection of our sample, in \S~\ref{sec:IFU_obs} and \S~\ref{sec:ALMA_data} we describe the spectroscopic and ALMA observations, respectively, and in \S~\ref{sec:SED} we describe our broad-band SED fitting and investigate how representative our targets are of the parent sample.
	
	\subsection{Sample selection}\label{sec:Sample}
	
	We selected our sample from the KASHz survey, which is an IFU survey of 250, $z$=0.6--2.6 X-ray detected AGN from the fields of CDFS, COSMOS, UDS and SSA22 (\citealt{Harrison16}). The IFU data in KASHz is predominantly from VLT/KMOS, but is also supplemented by archival VLT/SINFONI data. The survey description and the first part of the sample is described in \citet{Harrison16} and the full sample will be described in Harrison et~al. (in prep.). Briefly, the KASHz galaxies were selected based on an X-ray detection and a known archival redshift that places the redshifted H$\alpha$ and/or [O~{\sc iii}]$\lambda$5007 within one of the $YJ$, $H$ or $K$ wavebands; 90\% of the used archival redshifts were spectroscopic. Some targets were observed in only a single grating, whilst other targets were observed in two gratings to obtain data on both emission lines. 
	Relevant for this study are the 53 targets with observations in two gratings, of which 
	39 have detections in both [O~{\sc iii}] and H$\alpha$.
	
	To achieve the objectives of our study, ( i.e, tracing the ionised gas kinematics using the [O~{\sc iii}] line to map ionised outflows; the distribution of H$\alpha$ emission; determining the location of the dusty star formation as traced by the rest FIR emission) we select the KASHz sources with: (1) sufficient quality IFU data to reliably map both the H$\alpha$ and [O~{\sc iii}] emission lines (i.e., both detected with SNR$>$10) and (2) significant detections (SNR$>$4) in archival ALMA images at an observed wavelength $\approx$870 or $\approx$1100 $\mu$m (i.e., ALMA Bands 6 or 7, corresponding to rest-frame wavelengths of $\approx$260--400$\mu$m; see \S~\ref{sec:ALMA_data}). We further required the ALMA data to have a resolution comparable to, or better than, our IFU observations (i.e., typically $\lesssim0.7$\,arcsec; see \S~\ref{sec:IFU_obs}). This final criterion allows us to determine the location of the FIR emission to an accuracy of $\lesssim0.1$\,arcsecond (see \S~\ref{sec:position}). 
	
	Seven KASHz targets met the selection criteria described above (ID\,1--7; Table~\ref{Table:Sample}). For this study we also include ALESS 75.1 (ID\,8), a $z$=2.55 AGN from \citet{Chen19}, which is not part of KASHz, but it has existing IFU and ALMA data, matching the criteria described above. This object was identified as an AGN at mid-infrared wavelengths in previous work (\citealt{Stanley18}), which we confirm here using new SED fitting (\S~\ref{sec:SED}). The 12$\mu$m AGN luminosity of 10$^{46.0}$\,erg\,s$^{-1}$ implies an intrinsic X-ray luminosity of $L_{\rm 2-10\,keV}=$10$^{45.5}$\,erg\,s\,$^{-1}$ for this AGN (following \citealt{Asmus11}).\footnote{We note that this object is covered by, but undetected in, the E-CDFS field with relatively shallow {\em Chandra} X-ray coverage (\citealt{Xue16}). This non-detection implies that this source is a heavily obscured AGN.}
	
	The IDs, sky positions, redshifts, X-ray IDs and X-ray luminosities for our final sample of 8 targets are presented in Table~\ref{Table:Sample}. In the table we also provide other names which have been commonly used in the literature for some of the objects. Indeed, our sample includes well-studied objects, including ID\,5 which has multi-wavelength spatially-resolved observations \cite[see a summary in][]{Loiacono19}. In particular, ID\,6 was presented in \cite{Cresci15} as exhibiting both suppression and enhancement of star formation, traced by H$\alpha$, by an AGN-driven outflow traced by [O~{\sc iii}] \citep[also see][for CO observations]{Brusa18}; we compare our results to the previous work on this source in \S~\ref{sec:xid2028}.  
	
\begin{figure}
		\includegraphics[width=\columnwidth]{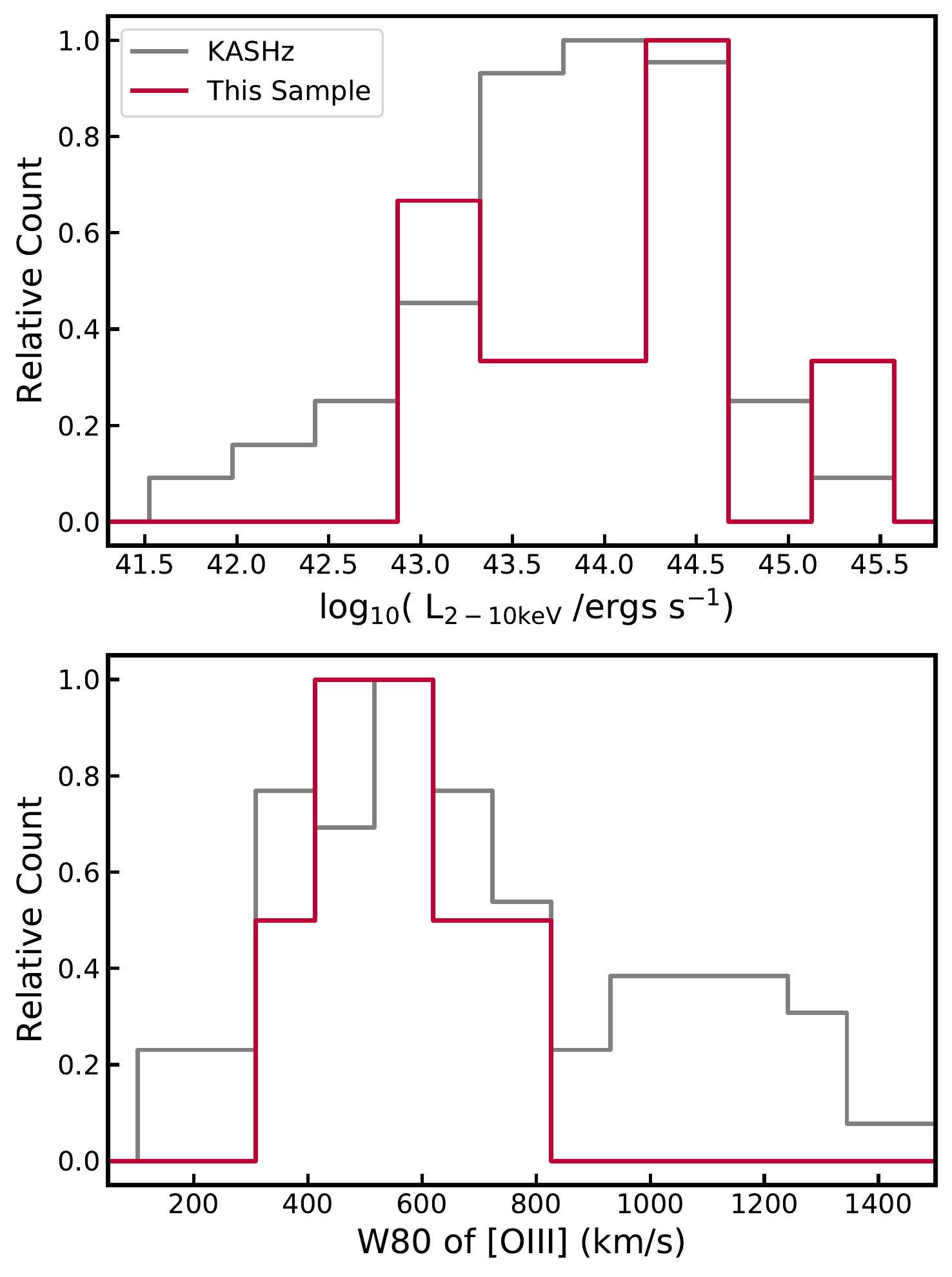}
		\caption[Properties of the IFU sample]{Histograms of the X-ray luminosities ($L_{\rm 2-10keV}$; top panel) and W$_{80}$ velocity width ([O~{\sc iii}] velocity width containing 80 \% of the line flux; bottom panel) for the sample in this work (red) and the parent KASHz sample (grey).}
		\label{fig:Xlum}
	\end{figure}

	Figure \ref{fig:Xlum} places our sources within the context of the overall KASHz sample by showing the relative distributions of X-ray luminosities and [O~{\sc iii}] emission-line widths ($W_{80}$; width of the emission-line containing 80\% of the flux; \citealt{Harrison16}). Our sample covers a similarly wide range of X-ray luminosities, from moderate to luminous AGN, but lacks objects with the most extreme [O~{\sc iii}] line widths ($W_{80}>$800\,km\,s$^{-1}$). However, as we demonstrate in \S~\ref{sec:SF_outflows}, this does not mean that our targets lack outflow signatures in the [O~{\sc iii}] emission-line profiles (\S~\ref{sec:EM_maps} \& \S~\ref{sec:SF_outflows}). We discuss our results in the context of the overall population in \S~\ref{sec:SF_outflows}.
	
	
	\begin{table*}
		\caption[Basic properties of the IFU sample]{Basic properties of our AGN sample. 
			(1) Object ID in this paper; 
			(2) X-ray ID from \citet{Hsu14, Marchesi16,Kocevski18}; 
			(3) other names commonly adopted in the literature; (4,5) Optical coordinates of the objects; 
			(6) Redshift measured from the emission lines (see \S~\ref{sec:em_model}); 
			(7) AGN spectral type (see \S~\ref{sec:em_model}); 
			(8) X-ray luminosity (2-10 keV) from \citet{Hsu14, Marchesi16, Kocevski18}; 
			(9) [O~{\sc iii}] luminosity from our emission line modelling (see \S~\ref{sec:em_model}); 
			(10,11) Stellar mass and FIR luminosity due to star formation from our SED fitting (see \S~\ref{sec:SED}). The stellar mass has a 0.3 dex systematic error. 
			(12) Reference for details of the two reduced IFU cubes for each target's [O~{\sc iii}] and H$\alpha$ emission-line data; 
			(13) Seeing FWHM from the observed PSF stars for the cubes containing [O~{\sc iii}]/H$\alpha$
			(14) Exposure times in ks for cubes containing [O~{\sc iii}]/H$\alpha$.}
		\input{./Tables/Main_Sample_Main.tex}
		\par (a) \citet{Stach19}; (b) \citet{FSchreiber09, Genzel14,Wisnioski17,Popping17,Talia18,Loiacono19}; (c) \citet{Ueda08, Cresci15, Brusa18}; (d) \citet{Hodge13,Simpson15}, Chen et~al 2019; (e) \citet{Harrison16, Stott16}; (f) \citet{Tiley19}; (g) Cirasuolo et~al. in prep.; (h) This paper; (i) \citet{Chen19}; (j) Derived from mid-infrared (\S~\ref{sec:Sample})
		\label{Table:Sample}
	\end{table*}

	\begin{figure}
		\includegraphics[width=1.0\columnwidth]{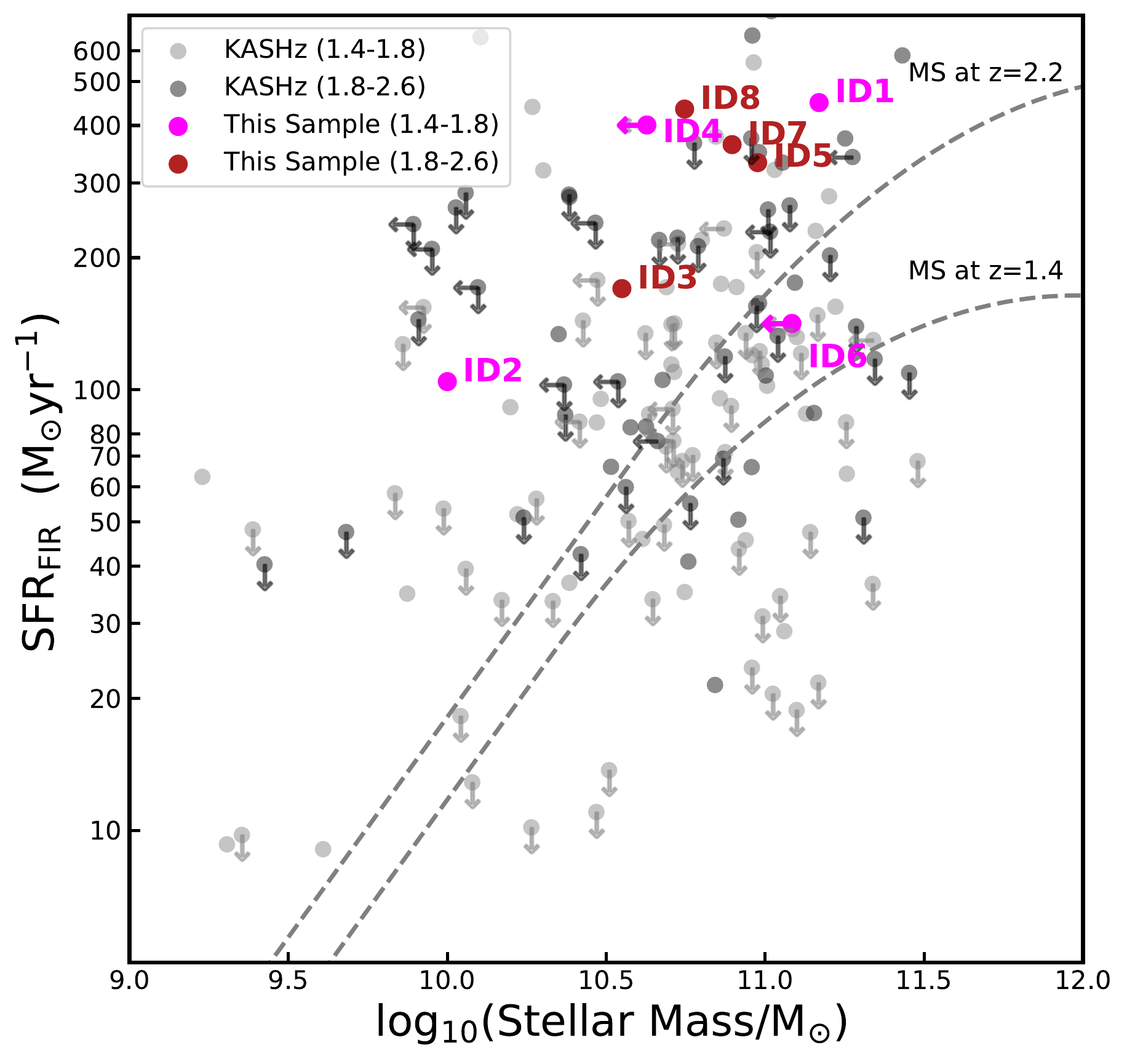}
		\caption[SFR vs stellar mass for the IFU sample]{Star-formation rate (from FIR) versus stellar mass for the AGN used in this work (coloured red for $z$=1.8--2.6 and magenta for $z$=1.4--1.8) compared to the full KASHz sample (black for $z$=1.8--2.6 and grey for $z$=1.4--1.8). The lower and upper dashed curves indicate the `main sequence' of star-forming galaxies \citep{Schreiber15} at $z$=1.4 and $z$=2.2, respectively. Our sample covers a broad range in star-formation rates and stellar masses but is limited to sources on, or above, the main sequence (see \S~\ref{sec:SED}). 
		}
		\label{fig:SFR}
	\end{figure}
	
	\subsection{IFS observations}\label{sec:IFU_obs}
	
	In order to map the H$\alpha$ and [O~{\sc iii}] emission of the targets in our sample, we used data from the near-infrared integral-field spectrograph VLT/KMOS \citep{Sharples04,Sharples13} and VLT/SINFONI \citep{Eisenhauer03, Bonnet04}. A comprehensive overview of the observations and data reduction will be presented in Harrison et~al. (in prep.); however, in Table~\ref{Table:Sample} we give the references to the papers that provide the details of the data and reduction steps for the individual data cubes used in this work. We note that the basic methods used for reducing all of the data were fundamentally the same, and any small differences in the adopted approach in the individual papers are accounted for in our data analysis methods and therefore do no affect our conclusions. That is, when obtaining our measurements and their related uncertainties we take into account the spectral resolution, noise, spatial resolution and imperfect sky subtraction in each data cube (see \S~\ref{sec:ELprops}). Here we provide brief details of the universal approaches taken to obtain and reduce the data.
	
	KMOS has 24 independent IFUs, which can be  centred on targets within a 7.2 arcmin field. Each IFU has a field of view (FoV) of 2.8$\times$2.8 arcseconds with a spatial pixel scale of 0.2 arcseconds. Here we present the results of the $YJ$, $H$ and $K$ gratings with spectral resolutions of R$\approx$3600, 4050 and 4200, respectively. The local spectral resolution (around the emission-lines of interest) were calculated from  sky lines and the instrumental spectral broadening was subtracted off, in quadrature, from the observed emission-line widths during the fitting procedure (\S~\ref{sec:em_model}). Observations were carried out using an ABA observing sequence (where A frames are on-source and B frames are on-sky), with individual exposure times of 600s ($YJ$-band), 300s ($H$-band) and 300s ($K$-band). The total on-source exposure times depend on the individual observing programme during which the observations were taken and are listed in Table \ref{Table:Sample}). The data-reduction process primarily made use of {\sc spark} \citep[Software Package for Astronomical Reduction with KMOS; ][]{Davies13}, implemented using {\sc esorex} \citep[ESO Recipe Execution Tool; ][]{Freudling13}. The {\sc spark} recipes were used to perform dark-frame subtraction, flat-fielding, illumination correction, wavelength calibrations and construct the stacked three-dimensional data cubes. The individual data cubes where then aligned and stacked using the centroids from the dedicated PSF star observations to correct for any offsets. PSF measurements were obtained using observations of stars inside dedicated IFUs that were observed simultaneously to the targets and processed in the same manner as the science observations. Standard star observations were carried out in the same night as the science observations, and processed in an identical manner, in order to flux calibrate the data. 
	
	For two of the targets the IFU data were obtained using the SINFONI integral field spectrograph (ID6 and ID8). The observations presented here were all observed using the 8$\times$8\,arcsec field of view which is divided into 32 slices of width 0.25\, arcsec with a pixel scale of 0.125 arcsec along the slices. SINFONI has a comparable spectral resolution to KMOS, ranging from $\approx$2000--4000; again, the spectral resolution was taken into account during the analyses. Our analyses of the $J$-band data for ID\,6 were first presented in \citet{Harrison16} (also see \citealt{Cresci15}) and the $HK$-band data for ID8 were presented in \cite{Chen19}. Here we present, for the first time, $H$-band data of ID6 which was taken under ESO Programme ID 094.B-0286(A), with 5.4\,ks of on-source exposure time. For a more direct comparison to the analyses presented in \cite{Cresci15} for this source (see \S~\ref{sec:SF_outflows}) we also re-reduced the archival $HK$-band data for ID6 that was first presented in that publication. 
	
	Following \citet{Harrison16} and \citet{Chen19}, all SINFONI data reduction was carried out using the standard procedures within {\sc esorex}. Centroids of individual exposures were found by creating white-light images from the datacubes and then individual datacubes were stacked using these centroids. Solutions for flux calibration were derived using the {\sc iraf} routines {\sc standard}, {\sc sensfunc} and {\sc calibrate} on the standard stars, which were observed on the same night as the science observations. These standard star observations were also used to estimate the PSF of the observations. Whilst this is not as reliable as the simultaneous PSF measurements we made for KMOS (see above), we note that we used the broad-line region H$\alpha$ for the final constraint of the PSF for ID\,6 (see \S~\ref{sec:IFU_size}). Although ID\,8 is type-2 source observed with SINFONI, it is very clearly extended in H$\alpha$ emission (\S~\ref{sec:SF_res}). The final pixel scale of the reduced SINFONI cubes is 0.125''/pixel. Overall the PSF of the IFU observations range from 0.6-1.0'' and are tabulated in Table \ref{Table:Sample}. 
	
	The latest versions of the SINFONI and KMOS pipelines were used to reduce the data at the time of the various references provided in Table 1. We verified that the measured continuum flux is calibrated correctly, by measuring continuum flux of the PSF stars against the catalogue values, which were well-matched to their true magnitudes.

	\subsection{ALMA observations and imaging}\label{sec:ALMA_data}
	To map the rest-frame FIR emission for our AGN host galaxies, we make use of observations from ALMA. We queried the ALMA archive for all observations of our targets performed with Band 6 or 7 and at a resolution of $\le$0.7\,arcsec (see \S~\ref{sec:Sample}). Here we describe the observations used in this work and how we produced the images.
	
	\subsubsection{ALMA observations and data reduction}
	The ALMA observations used in this work come from our own Cycle 1\&2 programmes \citep[ID\,3,4]{Mullaney15, Scholtz18, Stanley18}, the AS2UDS survey \citep[ID\,1]{Stach19}, follow-up observations of the ALESS survey \citep[ID\,8]{ Hodge13, Simpson15, Chen19} and other observational campaigns:  \citep[ID 2]{Jin18, Santini19}, \citep[ID 5]{Talia18}, \citep[ID 6]{Brusa18} and \citep[ID 7]{Barro17}. Due to the archival nature of this study, the on-source exposure times are wide ranging (between 40 and 14\,000\,s; where the longest observations were designed to detect CO emission lines). The individual programme IDs and central wavelengths of the observations are provided in Table~\ref{Table:Sample_ALMA}.
	
	We reduced the data by creating the calibrated measurement sets using the standard ALMA pipeline provided in the archive and the corresponding version of Common Astronomy Software Application (CASA) used during the generation of these scripts. Before creating images, we performed manual checks in CASA on the calibrated measurement sets to verify that all calibrations (such as phase calibrations) and flagging of bad antennae pairs had worked correctly during the reduction process.
	
	\begin{table*}
		\caption[Properties of the ALMA observations]{Table summarising the ALMA observations. The (IFM) and (HR) indicate IFU matched and high resolution maps (see \S~\ref{sec:ALMA_img}). (1) Object ID in this paper; (2) ALMA programme ID; (3) Central wavelength and ALMA Band of the observations; (4) Synthesised beam size of the IFU matched resolution maps; (5) RMS of the IFU matched resolution maps ($1\sigma$ map depth); (6) Beam size of the high resolution maps; (7) RMS of the high resolution map ($1\sigma$ map depth); (8) Signal-to-noise of the peak continuum measured from the IFU matched maps; (9) Flux density of the continuum measured in the uv plane (see \S~\ref{sec:ALMA_analysis}).}
		\input{./Tables/Main_Sample_ALMA.tex}
		\label{Table:Sample_ALMA}
	\end{table*}
	
	\subsubsection{Imaging the ALMA data}\label{sec:ALMA_img}
	
	The calibrated ALMA measuring sets were imaged using CASA version 5.1.2. The uv-visibilities in the measuring set were Fourier transformed to create dirty images and these dirty images were subsequently cleaned using a similar technique to that described by \citet{Hodge13}, using the \texttt{tclean} command in CASA.\footnote{Cleaning is a common technique applied to interferometric data to reduce the strength of the side lobes from bright sources to allow for the detection of faint sources.} We measured the RMS in off-source regions of the dirty maps and then cleaned the maps down to a $3 \sigma$ depth around the sources from the IFU data or any bright sources identified in the FOV. We verified that the spectral windows used to create the continuum images did not contain any visible emission lines ([C~{\sc ii}], CO, etc). 
	
	We created two sets of clean images using natural weighting whenever possible and a summary of the resulting resolution and RMS noise of all of the maps is provided in Table~\ref{Table:Sample_ALMA}. The first set of images was created to, as closely as possible, match the resolution of the IFU data containing the H$\alpha$ emission line (see Table~\ref{Table:Sample}). This was done by applying a Gaussian taper\footnote{Tapering is a process during the imaging which reduces the weight of the longest baselines. This results in a reduction of the spatial resolution of the images; however, at the cost of not including all the data and consequently increasing the RMS noise in the maps.} of an appropriate width to match the size of the resulting ALMA synthesised beam to the width of the PSF during the H$\alpha$ IFU observations (i.e., $\approx$0.6--1.0\,arcseconds;  see Table~\ref{Table:Sample_ALMA}). The ALMA maps created in this process are labelled as "IFU matched ALMA maps" (IFM). If the object was observed by ALMA at a resolution higher than the resolution of IFU data, we also created ALMA maps without any tapering called "High Resolution ALMA maps" (HR), which have a final resolution of 0.15--0.35\,arcseconds (see Table~\ref{Table:Sample_ALMA}). We used these two sets of maps to assess the impact of differing spatial resolutions upon our measurements of the location of the peak emission (see \S~\ref{sec:position}).
	
	The final IFU matched ALMA maps have an RMS between 0.02\,mJy  and 0.69\,mJy and a median RMS of 0.24 mJy. The quoted signal-to-noise measurements in Table~\ref{Table:Sample_ALMA} are derived from these maps by dividing the peak flux density by the RMS of the map. The final high resolution ALMA maps have the RMS between 0.10\,mJy and 0.37\,mJy (median value of 0.11\,mJy). By selection (see \S~\ref{sec:Sample}), we detected all our of our sources in the IFU-matched ALMA maps with a SNR$>$4 (see Table~\ref{Table:Sample_ALMA}). Six of the eight targets have SNR$>$8, and are the most reliable for measuring the sizes of the rest-frame FIR emission (\S~\ref{sec:ALMA_analysis}). 
	
	\subsection{SED fitting and sample properties}\label{sec:SED}
	We compiled multi-wavelength photometry from UV to FIR wavelengths and performed SED (spectral energy distribution) template fitting to measure the SFRs, stellar masses and the dust attenuation of our targets. 
	
	All three extragalactic survey fields that contain our sources (CDFS, COSMOS and UDS) are covered by {\it Herschel} and {\it Spitzer} imaging in the infrared waveband. We make use of public catalogues from the PEP and HERMES {\it Herschel} surveys for FIR fluxes over 100--500 $\mu$m \citep{lutz11, Oliver12}, and catalogues from the FIDEL and SCOSMOS {\it Spitzer} programmes for MIPS 24 $\mu$m fluxes (available from NASA IPAC). The UV, optical, and NIR photometry are taken from public versions of the multi-wavelength catalogues available from the CANDELS, MUSYC (CDF-S), COSMOS and UKIDSS/UDS survey consortia \citep[][ and O. Almaini, priv. comm.]{Guo13,Cardamone10, laigle16}. Details of each dataset and the processing of the public photometry will be described in Rosario et al (in prep.). We additionally used the band 6/7 continuum fluxes from the ALMA data in the SED fitting. We describe the measurement of these fluxes in \S \ref{sec:ALMA_analysis}.
	
	The multi-wavelength SEDs of the targets were modelled using the Bayesian SED code \texttt{FortesFit} \citep{Fortes}. Four SED components were used in the modelling:
	\begin{itemize}
		
		\item a stellar component of fixed solar metallicity from the \citet{bruzual03} library, with a star-formation history modelled as a delayed exponential with a range of ages (0.001-1 Gyr) and exponential timescales (0.01-2 Gyr). A variable screen extinction following a Milky Way law was applied (A$_{\rm V}<$ 10 mag).
		
		\item an AGN accretion disc with a range of spectral slopes (-1.1 to 0.75) as prescribed by the models of \citet{Slone12} with a variable extinction following a Milky Way law (B-V reddening up to 1 mag).
		
		\item an AGN dust emission component with a range of shapes as prescribed by the empirical templates from \citet{Mullaney11} (short wavelength slope: -3 to 0.8, long wavelength slope: -1 to 0.5, turnover wavelength: 20 to 60 microns). 
		
		\item dust emission heated by star-formation following the one-parameter template sequence (0--4) from \citet{dale14}.
	\end{itemize}
	
	Probabilistic priors were used to constrain the luminosity of the accretion disc and AGN dust emission components based on the X-ray luminosity. \texttt{FortesFit} generates full marginalised posterior distributions of stellar mass ($M_{\star}$), FIR luminosity from star formation ($L_{\rm IR,SF}$; over 8--1000$\mu$m) and stellar dust attenuation ($A_{V}$), as well as other parameters that are not used in this work. We present the individual SEDs and the resulting fits in the Supplementary Material. The $L_{\rm IR,SF}$ and $M_{\star}$ values are provided in Table \ref{Table:Sample}, along with their uncertainties. From the FIR luminosities, we estimate star formation rates (SFR(FIR)) using the calibration from \citet{Kennicutt12}, and these are discussed in \S~\ref{sec:Gal_prop}. The dust attenuation, and the impact that this has on the observed H$\alpha$ fluxes from the IFU data, is discussed in \S~\ref{sec:Dust_cor}.  
	
	Six objects in our sample are detected in the radio at 1.4\,GHz \citep[][]{Simpson06,Schinnerer10,Miller13}. The corresponding rest-frame 1.4\,GHz radio luminosities for all but one of the sample are $L_{\rm 1.4GHz}\lesssim$2$\times$10$^{24}$\,W\,Hz$^{-1}$ (assuming a spectral index of $\alpha=-0.7$; defined as f$_{\nu} \sim \nu^{\alpha}$). This is consistent with these seven targets being `radio quiet', and following \cite{Kennicutt12}, their radio luminosities imply reasonable star formation rates of a few hundred -- 1500\,M$_{\odot}$\,yr$^{-1}$ , although we can not rule out low-level radio jets e.g \citep[e.g.,][]{Jarvis19}. The one exception is ID\,2 which has a luminosity of $L_{\rm 1.4GHz}$=4$\times$10$^{25}$\,W\,Hz$^{-1}$ which we discuss further below. 
	
	For this work we require that the 870$\mu$m--1100$\mu$m emission is uncontaminated by processes other than star formation (e.g., synchrotron emission from radio jets). For the radio-quiet sources, this is supported by the fact that the sub-mm fluxes (Table~\ref{Table:Sample_ALMA}) would imply unrealistic (highly inverted) spectral indices of $\alpha>0.4$ if they arose from synchrotron emission. Additionally, for ID\,6 the 870um/1.1\,mm flux ratio is fully consistent with star-formation heated dust; see Appendix of \citet{Brusa18}. In the case of ID\,2, which has a higher radio luminosity, we also consider the 2.3\,GHz and 5.5\,GHz data from \citet{Zinn12} and \citet{Huynh12}, which together imply a spectral index of -0.3. Extrapolating this radio slope to ALMA band 7 suggests the ALMA emission could have a roughly equal contribution from star formation and synchrotron emission; however, we note that some contamination to the ALMA flux for this single source does not influence our main conclusions in this work. Based on these assessments and our SED results where we decomposed AGN and star formation components, we argue that the ALMA 870 and 1100 $\mu$m emission provides a good tracer of the dust dust obscured star formation in these sources. 
	
	In Figure \ref{fig:SFR} we show the SFR vs stellar mass plane for the parent KASHz sample and highlight the targets used in this work. We also show the star-forming galaxy main sequence, at two representative redshifts, as turquoise and orange dashed lines \citep{Schreiber15}. We find that our target galaxies have SFRs which are either on, or above, the main-sequence of star-forming galaxies. The distribution to relatively high SFRs for the targets in our sample, compared to the parent sample, is due to our requirement for a strong detection in both H$\alpha$ and rest-frame FIR (\S~\ref{sec:Sample}). We discuss the implications for this on our results in \S~\ref{sec:SF_outflows}. 
	
	\section{Analyses}
	To achieve the goals of our study, we perform the following analyses: (1) compare galaxy-wide star-formation measurements inferred from the rest-frame FIR with those inferred from H$\alpha$; (2) map the star formation within the galaxies as inferred from maps of both H$\alpha$ emission and rest-frame FIR emission and (3) compare the location of AGN-driven ionised outflows with the distribution of star formation. In this section we describe how we achieved this by extracting galaxy-wide (unresolved) and spatially-resolved emission-line measurements from the IFU data (\S~\ref{sec:ELprops}), analysing the maps of the rest-frame FIR emission that were created using the ALMA data (\S~\ref{sec:ALMA_analysis}) and by measuring the offsets between the FIR and H$\alpha$ emission (\S~\ref{sec:position}). 
	
	\subsection{Emission-line properties}\label{sec:ELprops}
	
	Each of our targets have two sets of IFU observations (see \S~\ref{sec:IFU_obs}), one covering the [O~{\sc iii}] emission line (also H$\beta$ in some cases) and one covering the H$\alpha$ and [N~{\sc ii}]6548,6583 emission lines. The emission-line profiles for each of our targets are shown in Figure~\ref{fig:HST_spec}. Here we describe how we used the IFU data to: (1) extract galaxy-integrated spectra from each data cube to obtain global properties (\S~\ref{sec:em_model}); (2) obtain constraints on the star-formation rates using H$\alpha$ emission (\S~\ref{sec:Dust_cor}); (3) map the distribution of the H$\alpha$ emission and [O~{\sc iii}] outflows (\S~\ref{sec:EM_maps}) and; (4) measure the sizes of the H$\alpha$ emission (\S~\ref{sec:IFU_size}). 
	
	\subsubsection{Extracting spectra and emission-line fitting procedure}\label{sec:em_model}
	
	We extracted galaxy-integrated spectra with the primary goals of identifying [O~{\sc iii}] emission-line outflows (e.g., following \citealt{Mullaney13}) and calculating total narrow H$\alpha$ fluxes (to infer star-formation rates). To do this, we first found the peak of the continuum emission in the data cube by creating median wavelength collapsed images of our targets, excluding any spectral channels contaminated by sky-lines or the emission lines. We then fitted a single 2D Gaussian model to the wavelength collapsed continuum image to find the peak of the continuum emission. The 2D Gaussian is a sufficient model of the continuum since our seeing-limited continuum images are dominated by the point source from the central AGN (for the Type~1 AGN) or the stellar light from the galaxy which is the strongest towards the nucleus (for the Type~2 AGN). 
	
\begin{figure*}
		\includegraphics[width=0.8\paperwidth]{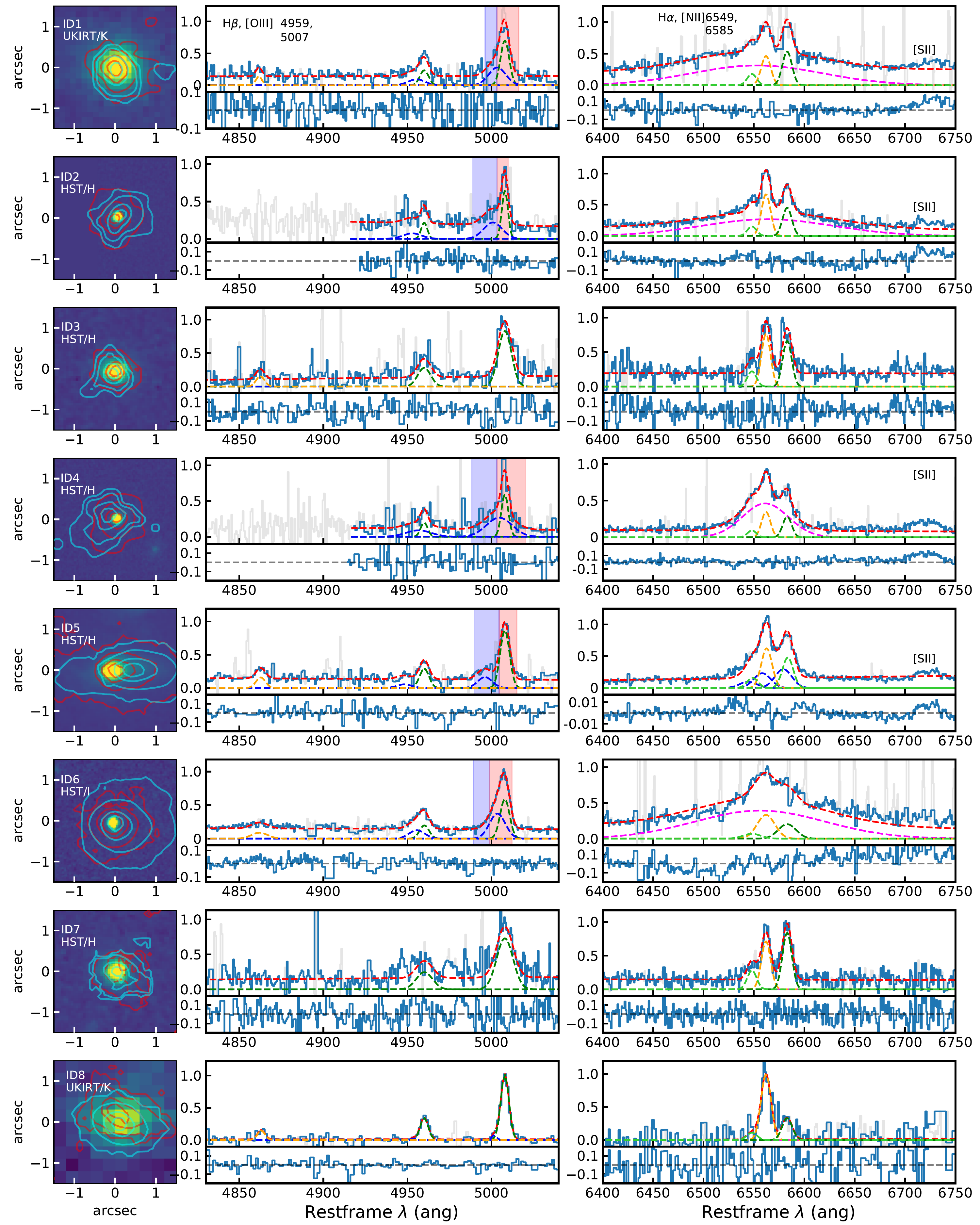}
		\caption[Rest frame optical images and optical spectra]{Rest-frame UV or optical images (left panel), and [O~{\sc iii}] (centre panel) and H$\alpha$ (right panel) emission-line profiles extracted from the inner 5\,kpc nuclear spectra for the eight objects in our sample. As labelled, the images are from {\rm HST} WFC3 $H$-band or $I$-band, when available, or UKIRT K-band images. North is up and East is left. The red and cyan contours show the narrow H$\alpha$ and narrow [O~{\sc iii}] flux maps, respectively, with levels of 90, 68, 32 and 10 \% of the peak flux in the map. For the emission-line profiles the light blue curves show the data and the grey curves show the masked sky-line residuals. For the [O~{\sc iii}] spectra the green, dark blue and red dashed curves show the narrow line, outflow components and total fits, respectively. The red and blue shaded regions indicate the wavelength slices of the non-outflowing and outflowing ionised gas, respectively, as defined in \S~\ref{sec:EM_maps}. Overlaid on the H$\alpha$ profiles the yellow, magenta, dark green, light green, blue and red curves show the narrow H$\alpha$, BLR H$\alpha$, [N~{\sc ii}]  doublet (6583 \AA and 6548 \AA), an outflow in H$\alpha$ and [N~{\sc ii}] doublet,  and the total fit, respectively.}
		\label{fig:HST_spec}
	\end{figure*}

	From each data cube we extracted spectra from two different circular apertures centred on the continuum peak: (a) a nuclear aperture within 5\,kpc diameter (i.e., approximately within one PSF 
	\footnote{The PSF for the ID 1 is larger than 5kpc (9 kpc). However, for consistency we used the same aperture for this object as for the others.}) to characterise the emission-line profile shapes and to search for outflows (see Figure~\ref{fig:HST_spec}) and (b) a `maximum' aperture to obtain total fluxes with a diameter of 1.2--2.4"; see Table \ref{Table:Sample_SF}, for which the sizes were determined by extracting spectra from increasingly large apertures until maximum emission-line fluxes were obtained (see column 5 in Figure~\ref{fig:Sizes_calc}). In Table~\ref{Table:Sample_IFU} we provide the key measured parameters from the former spectra (i.e., the emission-line flux ratios and velocity widths). The total H$\alpha$ luminosities, extracted from the latter spectra, are provided in Table~\ref{Table:Sample_SF}. We present all of the fitting results of the 5 kpc aperture in Table 1 of the Supplemantary data.
	
	To model the H$\beta$, [O~{\sc iii}]$4959,5007\AA$, H$\alpha$ and [N~{\sc ii}]$6548,6583\AA$ emission-line profiles, each line was fitted with one or two Gaussian components, with the centroids, FWHM and fluxes (normalisation) as free parameters. In each case the continuum was well characterised by fitting a straight line with a normalisation and slope as a free parameter.\footnote{We note that we see no significant Fe complexes in our spectra. This is likely due to the lack of very luminous Type~1 sources in our sample.} 
	
	Best-fit solutions, and the uncertainties, for the free parameters were obtained using the Python \texttt{lmfit} least-square library. The noise of the spectra was estimated as the RMS of the emission-line free region in the spectrum.
	During the fitting procedures we masked wavelengths which were effected by strong sky-line residuals. To construct the skyline residual masks we extracted a sky spectrum by summing all of the object-free (sky only) spatial pixels in the cube and identifying the strongest skyline residuals by picking any spectral pixels outside $1\sigma$. Visual inspection showed this method to be effective (see grey regions in Figure~\ref{fig:HST_spec}). We estimated the errors using the Monte Carlo approach. With this method we added random noise (with the same RMS as the noise in the spectra) to the best fit solution from the initial fit and then we redid the fit. We repeated this 500 times to build a distribution of all free parameters. The final values and errors on the parameters (median and standard deviation of the distribution) are consistent with the errors estimated by the \texttt{lmfit}.
	
	For the [O~{\sc iii}]4959,5007\AA \space emission-line doublet we simultaneously fit [O~{\sc iii}]$4959\AA$ and [O~{\sc iii}]$5007\AA$, using the respective rest-frame wavelengths of  4960.3\AA \space and 5008.24\AA. We tied the line widths and central velocities of the two lines and fixed the [O~{\sc iii}]$\lambda$5007/[O~{\sc iii}]$\lambda$4959 \space flux ratio to be 2.99 \citep{Dimitrijevic07}. We initially fit a single Gaussian component per emission line, then, we refit with a second Gaussian component. We use the BIC
	\footnote{The Bayesian Information Criterion \citep{Schwarz78}, which uses $\Delta \chi^{2}$ but also takes into the account the number of free parameters, by penalising the fit for more free parameters. BIC is defined as BIC=$\Delta \chi^{2} + k \log(N)$, where N is the number of data points and k is the number of free parameters.}
	to choose whether the fit needs a second broad component (using $\Delta$BIC$>2$ as boundary for choosing a more complex model); for verification we also performed a visual inspection of the residual spectra after subtracting the narrow component. A `broad' [O~{\sc iii}] component was required to fit nuclear spectra for five of the targets (see Figure~\ref{fig:HST_spec}), which are consequently the targets with the strongest evidence for ionised outflows (e.g., \citealt{Mullaney13}; see \S~\ref{sec:SF_outflows}).

	The H$\beta$ emission line is covered by the datacubes that also contain the [O~{\sc iii}] emission line. However, in 2 of the 8 targets the H$\beta$ emission falls within very strong atmospheric telluric features and we can not obtain any meaningful constraints (see Figure~\ref{fig:HST_spec}). Overall, we detected H$\beta$ at $>$3$\sigma$ in the nuclear spectra of 5 targets. Due to the limited signal-to-noise ratio of the H$\beta$ detections, we fitted the \Hb emission line using only a single Gaussian component and were not able to disentangle the broad-line region from the narrow-line region components.\footnote{We note that we do not tie the H$\beta$ and [O~{\sc iii}] kinematics as we often see that the H$\alpha$ line (which will follow the same kinematic structure as H$\beta$) does not follow the kinematics of the [O~{\sc iii}] line (see \S~\ref{sec:sampleSummary}).}
	Due to this limitation we only have meaningful H$\beta$ measurements of the narrow-line regions for three targets (ID\,3, ID\,5 and ID\,8), which we use for emission-line ratio diagnostics (\S~\ref{sec:Gal_prop}) and calculating a Balmer Decrement (\S~\ref{sec:Dust_cor}). 
	
	For characterising the H$\alpha$ emission-line profile we first identified the Type~1 sources as those with an H$\alpha$ broad-line region (BLR) component in the nuclear spectrum (i.e., a broad component of FWHM$>$2000\,km\,s$^{-1}$ that is not seen in the [O~{\sc iii}] or [N~{\sc ii}] emission lines; see Figure~\ref{fig:HST_spec}). Reassuringly, the Type~1/Type~2 classification is consistent with the presence of a UV--optical accretion disk component identified in our broad-band SED fitting (see \S~\ref{sec:SED}). For both Type~1 and Type~2 AGN we treat the narrow-line emission the same. That is, we simultaneously fitted the H$\alpha$ and neighbouring [N~{\sc ii}]$6548,6583\AA$ emission-line doublet, adopting the same approach as for the [O~{\sc iii}] emission line doublet. The central velocity and line width for all three emission-line profiles of [N~{\sc ii}]$6548\AA$, H$\alpha$, [N~{\sc ii}]$6583\AA$ were tied, with rest-frame wavelengths of 6549.86\AA, 6564.61\AA \space and 6585.27\AA, respectively. This approach, which assumes that the H$\alpha$ and [N~{\sc ii}] emission comes from the same gas, is commonly used in high-redshift observations to limit the number of free parameters \citep{FSchreiber09,Genzel14, Harrison16,ForsterSch18a}. During the fitting the H$\alpha$ and [N~{\sc ii}]$6583\AA$ fluxes were free to vary but the [N~{\sc ii}]$6548\AA$/[N~{\sc ii}]$6583\AA$ flux ratio was fixed to be 3.06 (based on the atomic transition probability; \citealt{Osterbrock06}). For the Type~1 sources an additional broad H$\alpha$ component was included with a free central velocity, line width and flux, associated with the BLR emission.
	
	We attempted to also characterise the outflow visible in [O~{\sc iii}] emission line in  H$\alpha$ and [N~{\sc ii}]. For this case we fitted an additional Gaussian component to the H$\alpha$ and [N~{\sc ii}] doublet (with parameters coupled as above) to characterise the outflowing component. However, except for the ID 5, we did not detect any outflow component in these emission lines. In the unique case of the Type~2 source ID 5, an additional broad component can be identified in both the H$\alpha$ and [N~{\sc ii}] emission-line doublet (also see \citealt{Genzel14}, who previously identified this as an outflow).
	Only the narrow component was considered to be tracing the total `narrow' H$\alpha$ emission (i.e., these are not part of the broad-line region or outflow) when exploring the total H$\alpha$ luminosities in \S~\ref{sec:Gal_prop}.

	\begin{table*}
		\caption[Properties of the optical emission lines]{Table of the key emission-line properties for our sample. The spectra have been extracted from the inner 5\,kpc (nuclear) region (see \S~\ref{sec:ELprops}).
			(1) Object ID in this paper; 
			(2) FWHM of narrow [O~{\sc iii}]; 
			(3) FWHM of broad [O~{\sc iii}]; 
			(4) velocity offset between the narrow and broad components of the [O~{\sc iii}];
			(5) FWHM of narrow H$\alpha$;
			(6) FWHM of BLR H$\alpha$;
			(7) log$_{10}$ flux ratio of narrow H$\alpha$ and H$\beta$ used for the Balmer decrement; 
			(8) log$_{10}$ flux ratio of narrow [N~{\sc ii}] and H$\alpha$ ; 
			(9) log$_{10}$ flux ratio of total [O~{\sc iii}] and narrow H$\beta$. }
		\input{./Tables/Main_Sample_spec.tex}
		\par $^{\rm x}$ velocity FWHM of the H$\alpha$ outflow. $^{\star}$ For the Type~1 AGN we do not use the Balmer Decrement to correct the H$\alpha$ emission for dust obscuration (\S~\ref{sec:Gal_prop}). 
		\label{Table:Sample_IFU}
	\end{table*}
	
	\subsubsection{Dust-corrections to H$\alpha$ emission and the derived star-formation rates}\label{sec:Dust_cor}
	
	In \S~\ref{sec:Gal_prop} we compare the star-formation rates inferred from the measured H$\alpha$ luminosity (excluding the BLR; SFR(H$\alpha$)) with those inferred from the FIR (SFR(FIR)). To estimate SFR(H$\alpha$) we converted from the measured L$_{\rm H\alpha}$ by using the calibration from \citet{Kennicutt12}. However, it is important to also consider the dust-correction to the H$\alpha$ luminosities. The preferred approach to constrain this is to measure the nebular dust attenuation (A$_{\rm V,HII}$) using the Balmer decrement \citep[the H$\alpha$/H$\beta$ flux ratio;][]{Reddy15}. For three objects (ID 3, ID 5, \& ID 8), for which we have reliable narrow \Hb detections in the nuclear spectra (see \S~\ref{sec:em_model}), we can measure the A$_{\rm V,HII}$ directly. We assume the \citet{Calzetti00} extinction curve and an intrinsic H$\alpha$/H$\beta$ ratio of 2.86, and consequently correct the total H$\alpha$ luminosities for dust attenuation. These correction factors are 1.5 -- 18 (see Table~\ref{Table:Sample_SF}). Unfortunately, we are unable to reliably correct for dust-obscuration for the other five sources. Although we can obtain some handle on the obscuration of stellar light from the SED fitting (A$_{\rm V,stellar}$; \S~\ref{sec:SED}) we choose not to use these to correct the H$\alpha$ luminosities because: (1) they are poorly constrained due to the challenges with fitting the uv--optical SEDs of AGN host galaxies \citep[e.g. ][]{Alexander12,Hickox18}; and (2) the stellar light and emission lines are often found to be obscured by different amounts, requiring a further uncertain correction factor to obtain A$_{\rm V,HII}$ \citep{Wild11, Kashino13, Price14, Reddy15, Puglisi16}. In \S~\ref{sec:Gal_prop} \& \ref{sec:SF_res} we discuss the various challenges in using H$\alpha$ has a star formation tracer in AGN host galaxies, considering both the dust correction and the contribution of the AGN itself (in addition to the star formation) to illuminating the gas.

	\subsubsection{Emission-line maps}\label{sec:EM_maps}
	To map the H$\alpha$ and [O~{\sc iii}] emission in our AGN host galaxies, we performed spaxel-by-spaxel fitting of the emission lines. We binned the spectra by averaging the nearby spaxels within radius of 0.2 arcsec. This significantly increases the SNR of the spaxels' spectra, while maintaining the seeing limited spatial resolution of $\sim$0.6--1.0\,arcsec. We fitted the [O~{\sc iii}] and H$\alpha$ emission lines in the binned spatial spaxels using the same overall procedure as described in \S~\ref{sec:em_model}. The final maps were re-binned to 0.1\,arcsec.
	
\begin{figure*}
		\centering
		\includegraphics[width=0.85\paperwidth]{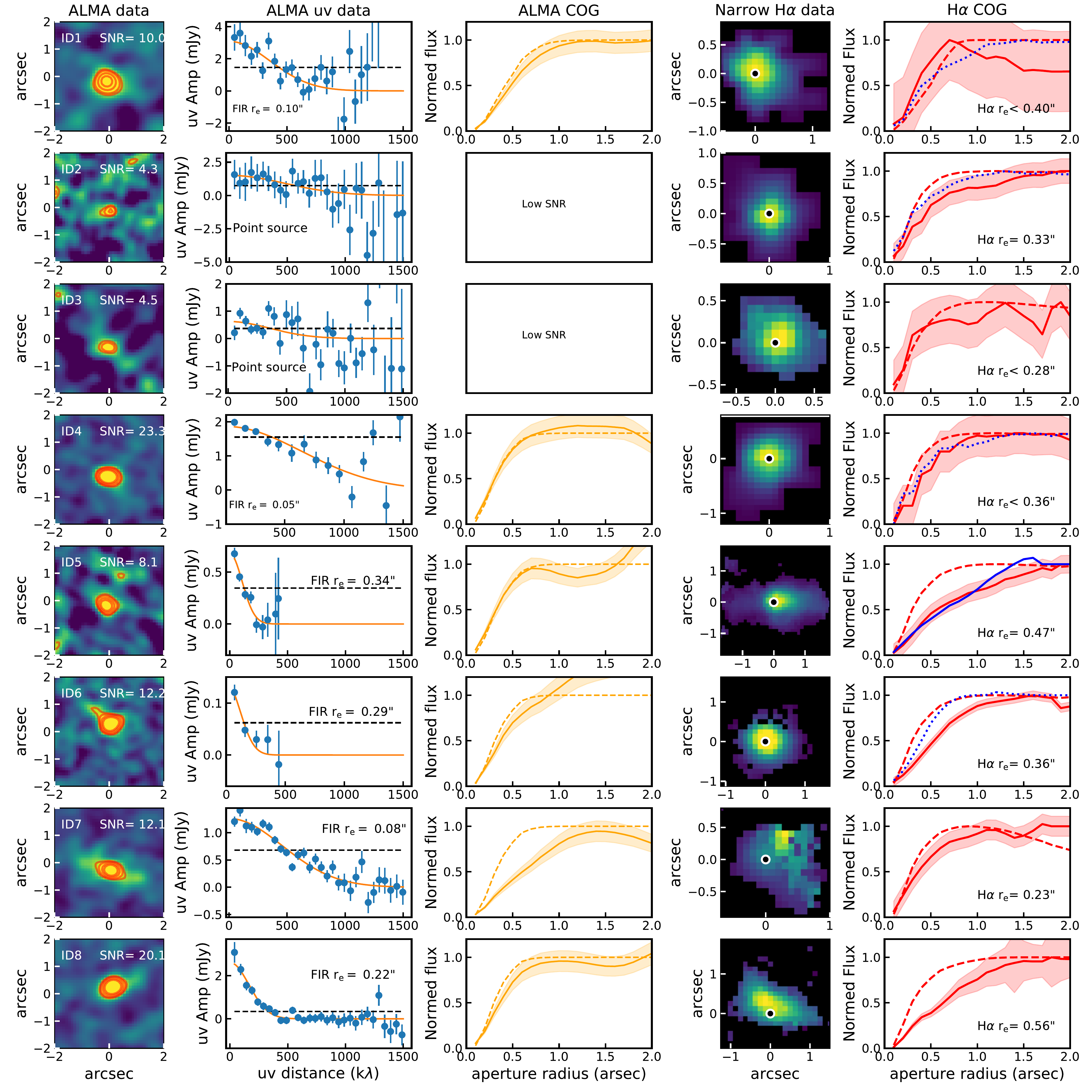}
		\caption[Measuring the FIR and H$\alpha$ sizes]{Size analyses of the ALMA and narrow H$\alpha$ images. Columns from left to right: 
			Column 1: ALMA data imaged at the resolution of the IFU data (IFU matched). The red contours indicate 2.5, 3, 4, 5 $\sigma$ levels of the data. North is up and East is left.
			Column 2: The uv amplitude data vs the uv distance binned per 50k$\lambda$. The orange solid curves, and black dashed curves show the resolved and unresolved model fits. We show the half-light radii when measured or indicate whether it is a point source in the panel.
			Column 3: The curve of growth (COG) for the IFU-matched FIR continuum (from ALMA; yellow solid curve) normalized to the flux estimated from the uv-plane. The yellow shaded region shows the 1 $\sigma$ uncertainty on the flux density. The yellow dashed curves show the COG for the ALMA beam. We do not present the COG for objects which are detected by ALMA at SNR<$8$. 
			Column 4: The narrow H$\alpha$ maps. The white circle show the centre of the continuum from HST or ground-based images. North is up and East is left.
			Column 5: The COG for the H$\alpha$ emission normalized to the peak of the COG, where the solid red curve shows the COG for the narrow H$\alpha$ with the shaded region indicating the 1 $\sigma$ uncertainty and the red dashed curves show the COG for the PSF. The dotted blue curves indicate the COG of the broad line region (BLR) and the solid blue curve shows the COG of the H$\alpha$ outflow for ID 5.}
		\label{fig:Sizes_calc}
	\end{figure*}

	For the spaxel-by-spaxel fitting of the H$\alpha$ emission line we have taken into account the emission coming from the BLR in Type 1 AGN that will contaminate multiple pixels (due to the PSF spreading out the emission). For these targets, we fixed the central velocity and line-width of the BLR component to be the same as that obtained from the nuclear spectrum (Figure~\ref{fig:HST_spec}), leaving only the flux of the BLR as a free parameter. The resulting flux map of the BLR also serves as a measurement of the PSF inside these data cubes, as it is intrinsically a point source. We found reasonable agreement between the spatial profile of the BLR and the PSF stars (see \S~\ref{sec:IFU_size}; see column 4 in Figure~\ref{fig:Sizes_calc}), with a median ratio of the resulting sizes of $1.1 \pm 0.2$ (see \S~\ref{sec:IFU_size}). In each case, we fit the same models as in \S \ref{sec:em_model} to H$\alpha$ spaxel spectra ( i.e., continuum model, narrow H$\alpha$ and [N{\sc ii}] doublet). In case of ID5 we also fitted a model for the outflow visible in the H$\alpha$ and [N{\sc ii}] doublet in the nuclear spectrum.  The maps of the narrow H$\alpha$ emission component (i.e., after the broad-line region and continuum emission have been subtracted) are shown in Figure~\ref{fig:Sizes_calc}, fourth column and Figure~\ref{fig:SF_img}. 
	
\begin{figure}
		\includegraphics[width=0.99\columnwidth]{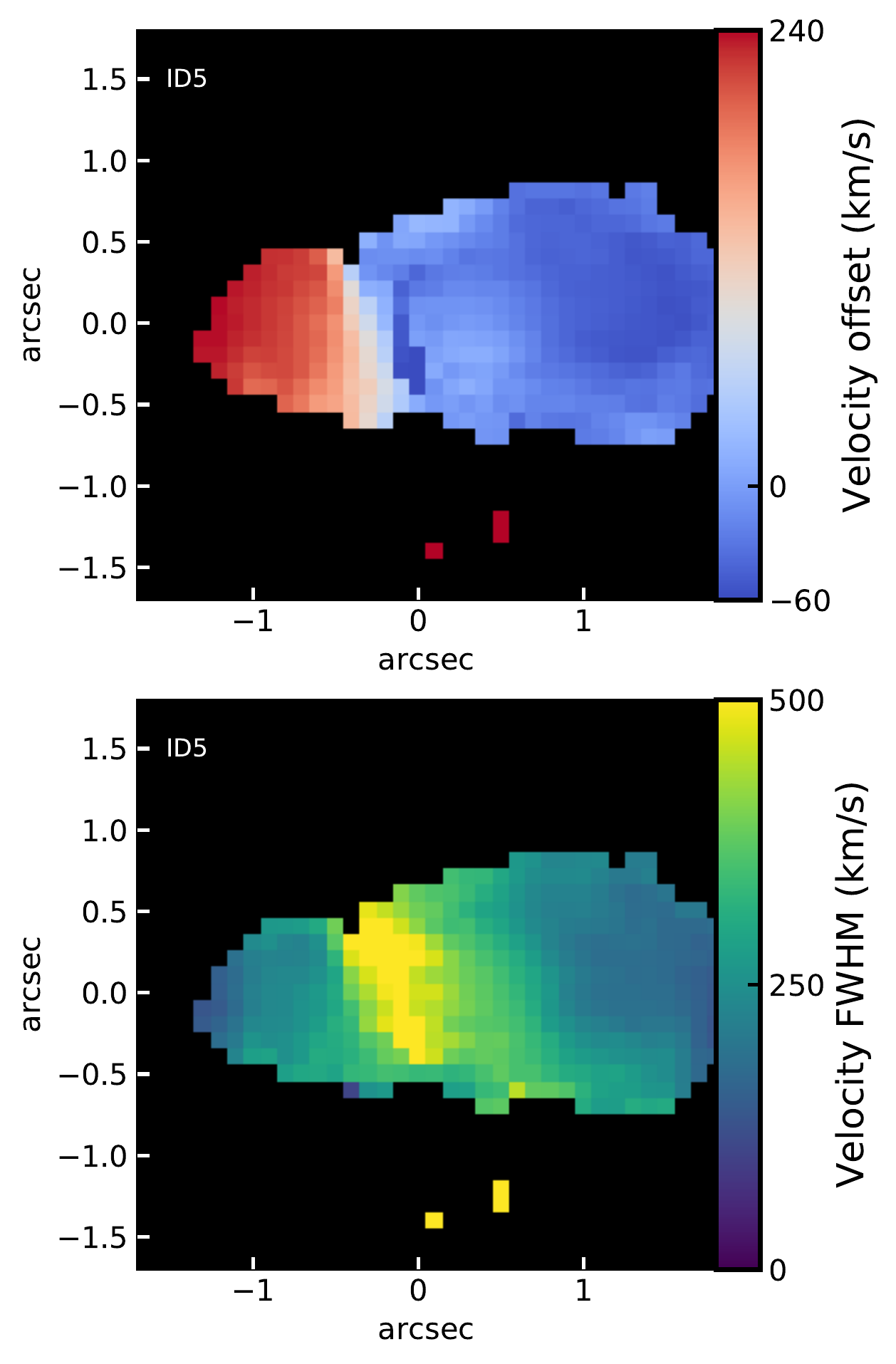}
		\caption{Example of the velocity field used to find the region of spectra dominated by host galaxy 
			dynamics versus that by the outflow as traced by the [O~{\sc iii}] emission line (see \S \ref{sec:EM_maps}). 
			Top panel: Velocity offset map. Bottom panel: Velocity dispersion. North is up and East is left.}
		\label{fig:Vel_field}
	\end{figure}

	In case of the [O~{\sc iii}] we were only able to fit a single component to the spaxel-by-spaxel spectra due to the low signal-to-noise ratios. This was even true for the 5 targets where we identified a second `outflow' component in the nuclear spectra (see Figure~\ref{fig:HST_spec}). Therefore, we employed a different method to map the outflow for these 5 targets by creating a narrow-band image in the spectral region of the outflow. We first subtracted the continuum from the [O~{\sc iii}] cubes by fitting a continuum model only to the emission-line free spectral regions. To define the velocity band to create this outflow narrow-band image, we first considered the underlying [O~{\sc iii}] velocity map (which is dominated by the narrow component, and likely galaxy dynamics; example shown in Figure \ref{fig:Vel_field}). We define the velocity range of the underling velocity structure as the maximum and minimum velocity in the map $\pm$0.5$\times$FWHM of the narrow component. These velocity ranges are shown as the red shaded region on the [O~{\sc iii}] profiles in Figure~\ref{fig:HST_spec}). We then define the outflow velocity slice as any [O~{\sc iii}] emission blue-ward of this (see blue shaded regions on the [O~{\sc iii}] profile in Figure \ref{fig:HST_spec}). The blueward limit of the outflow velocity slice was defined from the galaxy integrated spectra, as 2$\times$ FWHM from the centre of the outflow component. Visual inspection reveals that this definition of the outflow is dominated by the broad blue-shifted components. Furthermore, we confirmed that our results on the relative location of the outflow region compared to the H$\alpha$ and FIR emission (presented in \S~\ref{sec:SF_outflows}) are not sensitive to the exact definition of the velocity slice for the outflow. The final outflow maps are presented in \S~\ref{sec:SF_outflows}.

	\begin{table*}
		\caption[Star formation properties of the IFU sample]{Star formation properties and size measurements of our sample. 
			(1) Object ID in this paper; 
			(2) SFR derived from FIR luminosities; 
			(3) Dust extinction in the V-band of the nebular lines (derived from narrow Balmer decrement, see \S~\ref{sec:Dust_cor}). ID 2 \& 4 do not have reliable upper limits on the H$\beta$ and therefore we were unable to determine even an upper limit on Av$_{\rm HII}$; 
			(4) H$\alpha$ luminosity (narrow component) derived from the total flux; 
			(5) SFR derived from the observed H$\alpha$ luminosity; 
			(6) SFR derived from the dust corrected H$\alpha$ luminosity; 
			(7) Half-light radii of the H$\alpha$ emission (see \S~\ref{sec:IFU_size}); 
			(8) SNR of the FIR continuum; 
			(9) The half-light radii of the FIR continuum derived using the curves of growth method (see \S~\ref{sec:IFU_size}); 
			(10) The half-light radii of the FIR continuum derived in the uv-plane (see \S~\ref{sec:ALM_flx_size}). The (P) indicates if the FIR emission is unresolved; 
			(11) Projected physical offset between the H$\alpha$ and FIR emission regions;
			(12) Diameter of the aperture to obtain the total H$\alpha$ flux.}
		\input{./Tables/Main_Sample_SF.tex}
		\par $^*$ The FIR in this object may be contaminated by AGN radio emission (see \S~\ref{sec:SED}).
		\label{Table:Sample_SF}
	\end{table*}
	
	\subsubsection{H$\alpha$ sizes}\label{sec:IFU_size}
	
	To measure the extent of the narrow H$\alpha$ emission we used a curves-of-growth (COG) method \citep[e.g.,][]{Chen17,ForsterSch18a}. Due to using large apertures out to the edge of the field-of-view, the COG method allows us to search for flux in the outer regions that can not be detected in individual spatial pixels in the emission-line maps described above. We measured the total flux enclosed in a series of increasingly large circular apertures, where the apertures were centred on narrow H$\alpha$ peak. The narrow H$\alpha$ peak is defined as the brightest pixel on the narrow H$\alpha$ maps. We further discuss the definition of the location of the narrow H$\alpha$ in \S \ref{sec:offsets}. For each aperture, we extracted spectra and fit the emission-line profiles following \S~\ref{sec:em_model}. To reduce the degeneracies during the fitting procedures, for Type~1 AGN, we locked the FWHM and central velocity of the H$\alpha$ BLR Gaussian component in each aperture. This is a reasonable approach for such point source emission because only the {\em flux} in these BLR components will vary with distance, following the PSF. The errors on the COG of the objects are estimated the same way as for the galaxy integrated spectra using the errors from the Python's \texttt{lmfit}.
	
	We repeated the COG process on both the science observations and the observations of the corresponding PSF stars; however, for the PSF stars we measured the {\em continuum} in each aperture (as opposed to the emission line flux). The errors on the PSF are negligible, since all of the PSF stars are well detected and their COG profile can be accurately established. Figure \ref{fig:Sizes_calc} shows the comparison of the COG for the narrow H$\alpha$ emission (solid red lines), PSF star (dashed red line), BLR H$\alpha$ emission (blue dotted line, for the Type~1 AGN) and H$\alpha$ outflow component (blue solid line, only applicable for ID\,5). We used linear splines to interpolate between the data points and we measured the half light radii (radius containing 50 \% of the total flux). We derived the objects {\em intrinsic} sizes ($r_{e}$) by subtracting off, in quadrature, the size of the associated PSF \citep[see e.g.,][]{Chen17,ForsterSch18a}. For the Type~1 AGN we used the BLR as the PSF measurement (because it comes from the exact same datacube) and for the Type~2 AGN we use the corresponding PSF star. We note that for the KMOS observations of the Type~1 AGN (3 objects), when we have both measurements of the PSF star, we found that the BLR sizes are 10 \% larger than the PSF stars. We define whether a target is resolved in the COG by comparing the half-light radii of the object and the PSF. If a measured narrow Ha half-light radius is bigger than the half-light radius of the corresponding PSF by 1$\sigma$, we consider it resolved. By this criterion, 5 of our 8 targets are resolved in H$\alpha$. Uncertainties on the final H$\alpha$ sizes are calculated by considering the full range of possible radii for the 1$\sigma$ range of fluxes at each radii (see shaded curves in Figure~\ref{fig:Sizes_calc}). We note that we obtain consistent results for the $r_{e}$ of the narrow H$\alpha$ emission compared to \cite{Chen19} for ID\,8, despite their use of slightly different approaches (e.g., the use of non-ciruclar apertures).  The original intrinsic sizes of the narrow H$\alpha$ emission, and their corresponding uncertainties, are provided in Table \ref{Table:Sample_SF}.

	\subsection{Flux density and size measurements from ALMA data}\label{sec:ALMA_analysis}\label{sec:ALM_flx_size}
	
	In this section we describe how we measured the total flux densities and sizes of the FIR emission from the ALMA data. To obtain reliable fluxes and sizes of the FIR emission, we made measurements from the data in the image plane (the images are described in \S~\ref{sec:ALMA_img}) as well as directly from the calibrated visibilities in the $uv$ plane. As described in detail below, in Figure~\ref{fig:Sizes_calc} we show the ALMA maps (see \S~\ref{sec:ALMA_img}), the COG on these images and the spatially-binned visibilities in the amplitude--$uv$ distance plane (see below).
	
	Our preferred method to obtain total flux density measurements and sizes from the ALMA data is to use the visibilities directly, as it does not rely on the choices made during the imaging process. We first phase centred our data to the objects' central coordinates using CASA's \texttt{fixvis}.\footnote{The objects' central coordinates were determined from the peak of the High Resolution images described in \S~\ref{sec:offsets}}
	We then extracted the visibility amplitudes, binning across the $uv$ distance in steps of 50 k$\lambda$ (see Figure \ref{fig:Sizes_calc}; second column). We modelled these binned visibility amplitudes either as a constant over $uv$-distance (describing a point source) or as Gaussian centred at 0\,k$\lambda$ (describing a resolved 2D Gaussian source). \footnote{In the Fourier space the large $uv$ distance corresponds to a small spatial scale in the image plane. As a result, a point source has constant amplitude across all $uv$ distances, while for any resolved emission the amplitude is decreases with $uv$ distance \citep[see e.g.][]{Rohlfs96}.} We fitted these models using the Scipy's \texttt{curvefit} and we estimated the 1$\sigma$ uncertainty on the position using the covariance matrix from this fitting routine. We used the Bayesian Information Criterion (BIC) to choose the best-fit model, only accepting the Gaussian extended model if $\Delta$BIC $\geq 2$ (see Figure \ref{fig:Sizes_calc}; for more discussion about BIC see \S \ref{sec:em_model}). With this method we found that 6 of the 8 targets are extended in the ALMA data. We note, however, that the two sources that are consistent with being point sources, are also the two sources with the lowest signal-to-noise ratios with SNRs$\approx$4.5, for which it has been shown that sizes can not be reliably determined (see \citealt{Simpson15} for more details). For these objects, we used the size of the beam as a conservative upper limit on the size. Reassuringly we obtain consistent result on which of the sources are extended by using CASA's \texttt{uvmodelfit} routine which directly fits to the calibrated $uv$ visibilities. The intrinsic source sizes and their uncertainties, as determined from fitting the Gaussian models (shown in Figure~\ref{fig:Sizes_calc}, second column), are provided in Table~\ref{Table:Sample_SF}. 
	
	
	As a further verification of our results, we measured flux densities and sizes from the ALMA data in the image plane. Because we are interested in comparing directly these sizes to the H$\alpha$ sizes (see \S~\ref{sec:IFU_size}) we make use of the resolution-matched (``IFU matched'') ALMA maps described in \S~\ref{sec:ALMA_img} (Figure~\ref{fig:Sizes_calc}, first column). To obtain the total flux density measurements we used CASA's \texttt{IMFIT} routine to fit a single elliptical Gaussian model convolved with the synthesised beam. These fits reproduced consistent flux densities (within the 1$\sigma$ errors) that were obtained directly from the visibilities described above. 
	
	We then proceeded to measure the rest frame FIR sizes using a curve-of-growth method on the ``IFU matched'' ALMA maps, in order to be consistent with the method used to obtain H$\alpha$ emission sizes (see Figure \ref{fig:Sizes_calc}, third column). However, we do not perform the curve-of-growth analyses on the two objects which are classified as unresolved in the analysis of the visibilities above, which have low SNRs of $<8$. For the other six sources, as with the H$\alpha$ maps, we calculated the total flux in the ALMA maps using apertures with increasing size where the apertures were centred on the location of peak emission. The COG are normalised to the total flux densities obtained from the \texttt{IMFIT} fitting results. We note that the upturn seen in the curve-of-growth for ID5 beyond 1.5\,arcseconds is caused by a faint companion seen to the North of the main sources and in ID6 there is a faint tail of emission extending to the North East \citep[also see][]{Brusa18}.
	
	Following the analysis on the IFU data cubes (\S~\ref{sec:IFU_size}), we also performed the COG analysis on the synthesised beam (see Figure \ref{fig:Sizes_calc}; third column; dashed curves) and used this measurement to de-convolve the observed size measurements to obtain intrinsic sizes. The errors on the COG were estimated as the RMS of the maps.
	
	The rest frame FIR sizes from both methods (amplitude--$uv$-distance fitting method and COG to the image place) are provided in Table \ref{Table:Sample_SF}. We note that we obtain consistent size measurements for both H$\alpha$ and the rest-frame FIR as presented in \cite{Chen19} for ID8. Furthermore, there are only two targets where the the two different size measurements are not consistent within their 1\,$\sigma$ uncertainties: ID1 and ID7. For the reminder of this work we favour the sizes from the amplitude--$uv$-distance fitting method, but highlight results from both methods in the relevant figures. The different sizes observed in H$\alpha$ and rest-frame FIR for our targets are discussed in \S~\ref{sec:SF_res}. 
	
	\subsection{Alignment of the astrometric frames and measuring spatial offsets}\label{sec:position}
	We aim to measure the physical offsets between the FIR emission, the H$\alpha$ emission and the AGN outflows in our targets. These offsets have two main sources of uncertainty: (a) the relative astrometric calibrations of the IFU data cubes and the ALMA maps and; (b) the data quality in the images (i.e., both their resolution and sensitivity). In the following subsections we discuss how we addressed these issues by aligning the astrometric frames (\S~\ref{sec:astrometry}) before carefully measuring the final spatial offsets and their corresponding uncertainties (\S~\ref{sec:offsets}). 
	
	\subsubsection{Astrometric Alignment of the IFU and ALMA maps}\label{sec:astrometry}
	The absolute astrometric accuracy of ALMA depends on the frequency, baseline and calibration; however, in the case of our observations it is negligible at $\approx$20--30\,mas (ALMA Cycle 7 Technical Handbook\footnote{https://almascience.eso.org/documents-and-tools/cycle7/alma-technical-handbook}
	). However, the astrometric calibration of the IFU data is less accurate and requires additional calibration. Due to the limited field of view of the KMOS and SINFONI instruments (see \S~\ref{sec:IFU_obs}), it is not possible to calibrate the absolute astrometry by identifying known stars in the field of view with known, accurate positions. Instead, we aligned the IFU astrometry on the object itself by using supplementary high-resolution images from {\em HST} or UKIRT of the targets (e.g., see Figure \ref{fig:HST_spec}). To determine the central position of the AGN in the IFU date cubes we created white-light images by collapsing the data over the same wavelength range as the corresponding broad-band images. We then identified the central position of the source in the IFU data cube by fitting a 2D Gaussian model using the Scipy's \texttt{curvefit} and we estimated the 1$\sigma$ uncertainty on the position using the covariance matrix from this fitting routine. The RA and Dec of this central position in then determined by the position of the source in the corresponding broad-band images ({\em HST} or UKIRT).

	As with many previous studies \citep{Miller08,Hsu14,Dunlop17,Elbaz18, Scholtz18} we noticed a systematic offset between the optical astrometric frame (e.g., in {\em HST}) and the radio astrometric frame (e.g., from VLA or ALMA) in the CDFS field. This affects six of our eight targets in our sample which lie in this field. Previous studies typically corrected for this difference by applying a global shift to the astrometry in the optical frame. However, it has been found that this offset is not constant across the field (\citealt{Elbaz18}) and for the purpose of this study we require the most precise correction possible. To accurately align the ALMA and IFU cubes we used the spatially varying second order corrections adopted in \cite{Elbaz18} (M. Dickinson; private communication). Briefly, these corrections were obtained by using Pan-STARRS DR1 data to search for offsets in the different regions of GOODS-South field. The distortions from the different regions were then applied to the HST catalogue of GOODS-South sources.

	For our six targets in this field the average correction of the optical astrometry frame is $+$0.19 and $-$0.23\,arcseconds in RA and Dec, respectively. To calculate the final positional uncertainties we propagated the errors of the 2D Gaussian fitting, used to locate the source in the IFU data cubes, and the astrometric uncertainties on the broad-band images. Overall, we are able to constrain the astrometric positions in the IFU datacubes with 0.1\,arcsec accuracy (i.e., 0.8 kpc at z$\sim 2$). We note that the alignment of the individual exposures of the IFU observations does not introduce a large uncertainty, since these mis-alignement primarily influence the size of the PSF, rather than the location of the center of the emission.

	\subsubsection{Measuring the projected offsets}\label{sec:offsets}
\begin{figure*}
		\includegraphics[width=0.85\paperwidth]{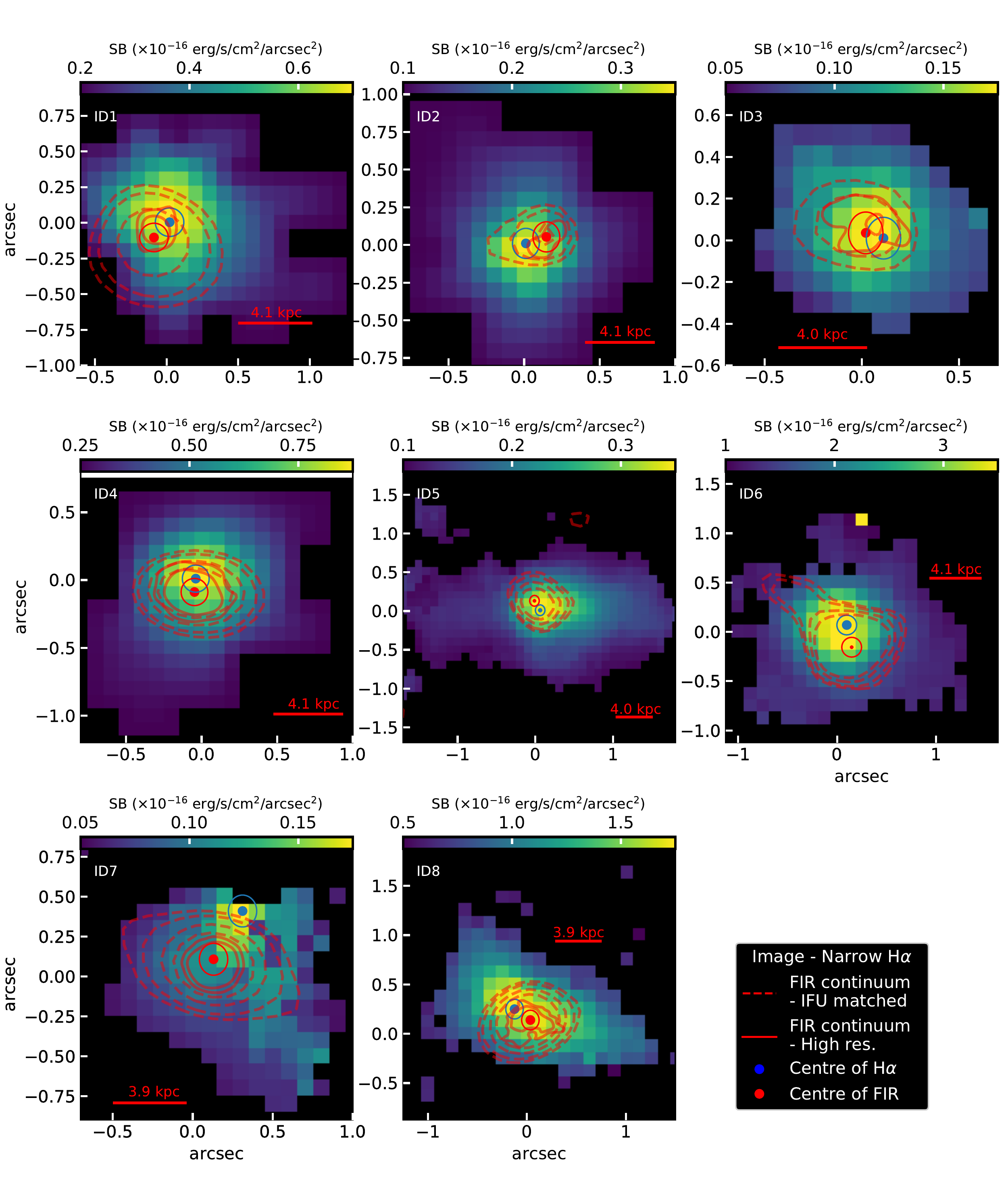}
		\caption[Maps of FIR and narrow H$\alpha$ emission]{A comparison of the spatial distribution of the FIR emission and the narrow H$\alpha$ emission for our AGN host galaxies. The images show the narrow H$\alpha$ emission (see \S~\ref{sec:EM_maps}). The red solid line represents the major-axis size of the PSF of the IFU observations, labelled with the corresponding physical scale in kiloparsec. Red contours show the FIR continuum  (see \S~\ref{sec:ALMA_img}), where the dashed and solid contours are from the IFU-matched (comparable spatial resolution) and high-resolution ALMA maps (where applicable), respectively, with levels of  2.5, 3, 4, 5 $\sigma$. The colourbar indicates the surface brightness of the narrow H$\alpha$ map.  The blue and red solid circles show the centres of H$\alpha$ and FIR emission, respectively. We discuss the alignment between the two sets of data in \S~\ref{sec:astrometry}.  There is a range of H$\alpha$ and FIR morphologies, with four targets showing significant spatial offsets between the two sources of emission (ID 5, 6, 7 and 8). North is up and East is left.}
		\label{fig:SF_img}
	\end{figure*}

	To determine the offsets between the H$\alpha$ and FIR emission we first needed to find the location of the peak emission in the H$\alpha$ and ALMA maps. Since we have cases where the H$\alpha$ emission is extended in one direction or has a complex morphology (for example ID 7, see Figure \ref{fig:SF_img}), we cannot apply a simple 2D model to determine the peak position accurately. Instead, we determined the centre of the H$\alpha$ emission by finding the brightest pixel. We note that we masked the skylines during the emission-line fitting to produce these maps (see \S \ref{sec:EM_maps}) therefore these peak positions are not affected by strong skyline residuals. To find the centre of the FIR emission we used the same technique, identifying the peak pixel in the ALMA maps. In Figure~\ref{fig:SF_img} we show H$\alpha$ maps with contours from the ALMA overlaid (``IFU matched'' as dashed contours and ``High Resolution'' as solid contours); the peak positions, with 
	1$\sigma$ error circles are shown in red and blue for H$\alpha$ and rest-frame FIR, respectively. The positional uncertainties for the peak position in the ALMA maps were determined by relating the signal-to-noise ratio of the emission and the size of the PSF (or beam), following $\delta$pos = PSF/(2 $\times$ SNR) \citep{Condon97}. The positions of the H$\alpha$ are dominated by the 0.1\,arsec systematic (see above). 
	
\begin{figure}
		\includegraphics[width=1.05\columnwidth]{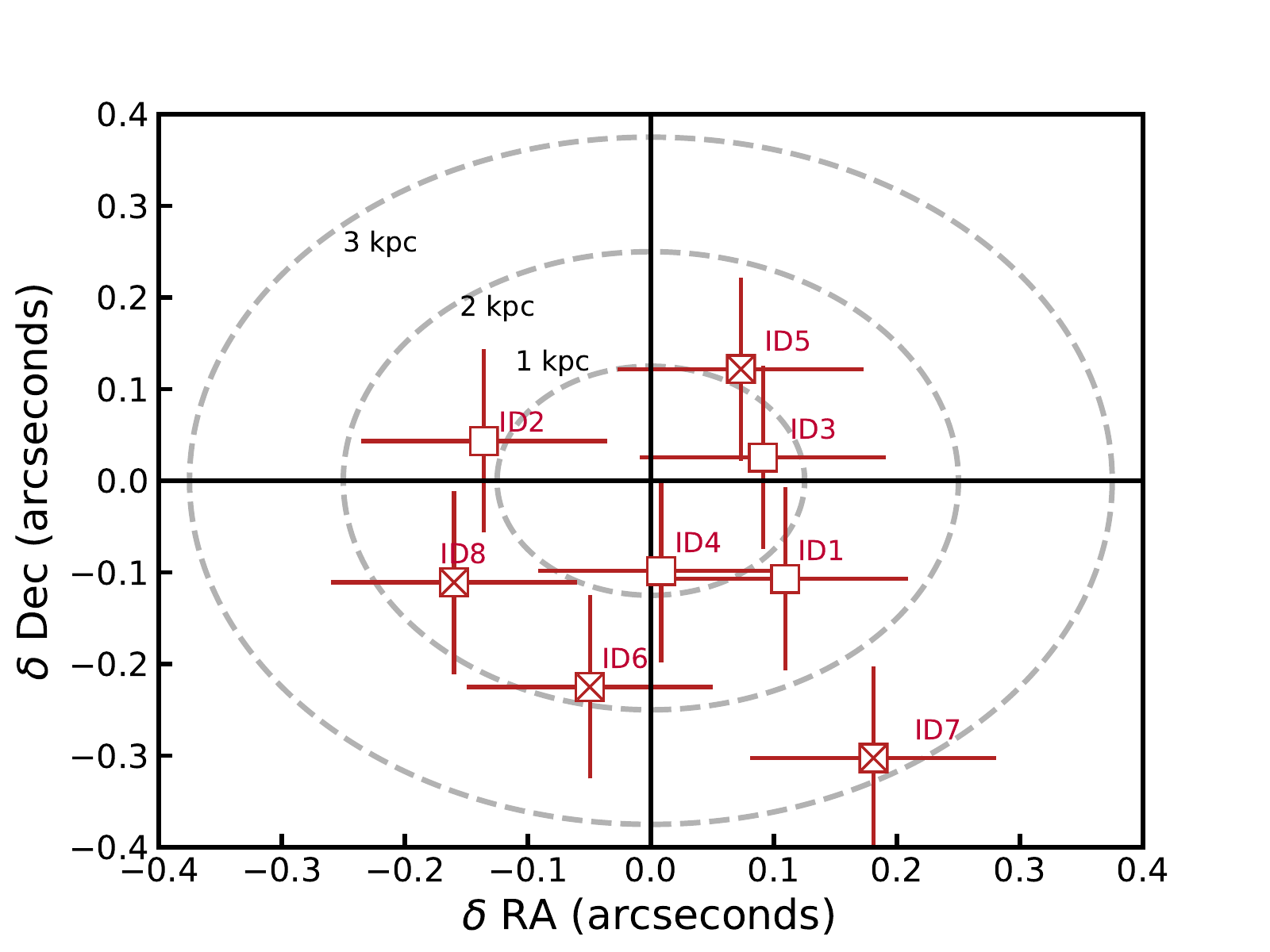}
		\caption[Measured offset between FIR and narrow H$\alpha$]{Offset between the FIR and narrow H$\alpha$ emission of our AGN after correcting the astrometry (see \S~\ref{sec:astrometry}). The grey circles indicate the projected physical offset of 1, 2, and 3 kpc. The squares filled with crosses indicate the four objects with significant projected {\em radial} offsets between FIR and H$\alpha$ emission (see \S~\ref{sec:SF_res} and Figure ~\ref{fig:SF_img}).} 
		\label{fig:Hal_ALM_off}
	\end{figure}

	We present the measured offsets in RA and Dec between the narrow H$\alpha$ and FIR emission in Figure~\ref{fig:Hal_ALM_off} where the final uncertainties on the offsets between the the peaks of the two emission are determined by combining the individual uncertainties on the two positions using a bootstrap method. We draw 1000 random positions from a 2D Gaussian distributions, centred on the individual H$\alpha$ or FIR positions and with width of the positional uncertainty. We calculated the offsets for all 1000 random positions. The final values in Figure \ref{fig:Hal_ALM_off} are the median value of the offsets. The errors are calculated as 1 $\sigma$ of the offset distributions. In this figure we highlight the four sources with crosses which have significant offsets between the H$\alpha$ and FIR emission (i.e., those where the positional error circles do not overlap in Figure~\ref{fig:SF_img}). 
	For ID 5, due to the large size of the host galaxy does not provide a clear look at the positional error circles.
	The final {\em projected} offsets range from 0.8--2.8\,kpc and are provided, with their uncertainties in Table~\ref{Table:Sample_SF}.\footnote{We note that using either the ``IFU-matched'' and ``High Resolution'' ALMA maps, results in consistent results for the final projected offsets.}
	The spatial offsets between the two sources of emission are discussed in \S~\ref{sec:SF_res}.

	\section{Results and Discussion}\label{sec:Results}
	In this section we present the results of our analyses of the IFU and ALMA observations for the eight $z$=1.4--2.6 AGN in our sample. Our study is motivated by previous work that has used IFU observations to map star formation, using H$\alpha$, and AGN outflows, using high-velocity components of [O~{\sc iii}]
	\citep[e.g., ][]{Canodiaz12,Cresci15,Carniani16}. Here, in addition to H$\alpha$ and [O~{\sc iii}] constraints we also include maps of the rest-frame FIR emission of our targets to trace the obscured star formation. After giving an overview of the emission-line properties of our sample (\S~\ref{sec:sampleSummary}), we present results that address our two main objectives: (1) to test H$\alpha$ as a star-formation tracer (both galaxy-integrated and spatially resolved) in our high-$z$ AGN host galaxies (see \S~\ref{sec:Gal_prop} \& \ref{sec:SF_res}), and (2) to search for evidence that AGN outflows suppress and/or enhance star formation in their host galaxies (\S~\ref{sec:SF_outflows}). In \S~\ref{sec:implications} we discuss the wider implications of our results for understanding the relationship between AGN outflows and star formation. 
	
	\subsection{Overview of the emission-line properties}\label{sec:sampleSummary}
	
	In Figure \ref{fig:HST_spec} we present the the H$\beta$, [O~{\sc iii}], H$\alpha$ and [N~{\sc ii}] emission-line profiles for our sample (extracted from a 5\,kpc diameter aperture; see \S~\ref{sec:em_model}). Our targets have representative emission-line properties of the parent sample from which they were selected \citep[][ see Figure \ref{fig:Xlum} and \S~\ref{sec:Sample}]{Harrison16}. For example, they have total [O~{\sc iii}] luminosities of $\log[L_{\rm [O III]}/$erg\,s$^{-1}]$= 42.2--43.4, which is expected for their X-ray luminosities (see Table~\ref{Table:Sample}) based on the L$_{\rm [O~{III}]}$- L$_{\rm 2-10 \rm kev}$ relation of $z\approx1$ X-ray AGN \citep{Harrison16}. Furthermore, they have typical [O~{\sc iii}] emission line widths ($W_{80}$; Figure \ref{fig:Xlum}, bottom panel). In this respect they represent typical X-ray AGN at this redshift range; however, see \S~\ref{sec:implications} for more discussion on the sample in terms of their star-formation rates. Here we describe the emission-line profiles in more detail. The key emission-line properties are summarised in the Table~\ref{Table:Sample_IFU}. 
	
	As can be seen in Figure~\ref{fig:HST_spec}, three of the targets have [O~{\sc iii}] emission-line profiles characterised with single Gaussian components (ID 3, 7, 8), and five of the targets require two components (ID 1,2,4,5 and 6; see \S~\ref{sec:em_model}). These latter five targets have second, broad components with FWHM$=$ 500--950\,km\,s$^{-1}$ (with blue-shifted velocity offsets of 230--600 \,km\,s$^{-1}$ with respect to the narrow line) and are those targets which we define here as clearly having AGN-driven ionised outflows. The `broad' outflow component in ID1 has a number of degenerate solutions. Regardless of the exact velocity widths of the each components, the fit requires an additional component to account for the blue wing in the emission line. For these targets we are able to define velocity slices in the wings which are most-likely not due to gravitational motions (\S~\ref{sec:EM_maps}; see blue shaded regions in Figure~\ref{fig:HST_spec}). However, we note that although ID\,7 is adequately described with a single component fit, the high-velocity width of FWHM$=$720\,km\,s$^{-1}$ would strongly suggest contributions from gas motions which are non gravitational \citep[e.g.,][]{Liu13b,Harrison16}. 
	
	During the fitting of the emission lines we did not tie the redshifts of the narrow components of [O~{\sc iii}], H$\alpha$ and H$\beta$ emission lines. However, the median velocity offset across the sample between [O~{\sc iii}]-H$\alpha$ and [O~{\sc iii}]-H$\beta$ is -36\,km\,s$^{-1}$ and - 44\,km\,s$^{-1}$, respectively.
	
	
	The total narrow H$\alpha$ luminosities of the sample are in the range $\log(L_{\rm H\alpha}$/erg\,s$^{-1}$)= 42.1-43.2 and are discussed in  \S~\ref{sec:Gal_prop}. The nuclear H$\alpha$ kinematics from the narrow-line region (i.e., after removing broad-line region components in Type 1 AGN) are typically more modest than those seen in [O~{\sc iii}], with FWHM= 350--640\,km\,s$^{-1}$ (see Figure \ref{fig:HST_spec}). Only in one source do we see strong evidence for an outflowing component in H$\alpha$ and [N~{\sc ii}] (ID 5; FWHM=900\,km\,s$^{-1}$; see Figure \ref{fig:Sizes_calc}; also see \citealt{Genzel14}). Narrower H$\alpha$ compared to [O~{\sc iii}] has been noted before for both high-$z$ and low-$z$
	AGN \citep{Harrison16,Kang17}. Outflow components can be stronger in [O~{\sc iii}] when compared to H$\alpha$ if the outflows are co-located with the AGN ionisation cones (perpendicular to the disk) whilst the H$\alpha$ is strongly dominated by star-forming disks (as has been seen in local AGN host galaxies; e.g., \citealt{Venturi18}). However, we also note that the complexities and degeneracies of simultaneously fitting the [N~{\sc ii}] doublet and H$\alpha$ with broad and narrow components makes it very difficult to isolate, potentially weak, outflow components in these lines. We compare the spatial distribution of the [O~{\sc iii}] and H$\alpha$ emission for our targets in \S~\ref{sec:SF_outflows}. 
	
	\subsection{Comparison of star-formation rates from FIR and H$\alpha$}\label{sec:Gal_prop}
	
\begin{figure}
		\includegraphics[width=1.05\columnwidth]{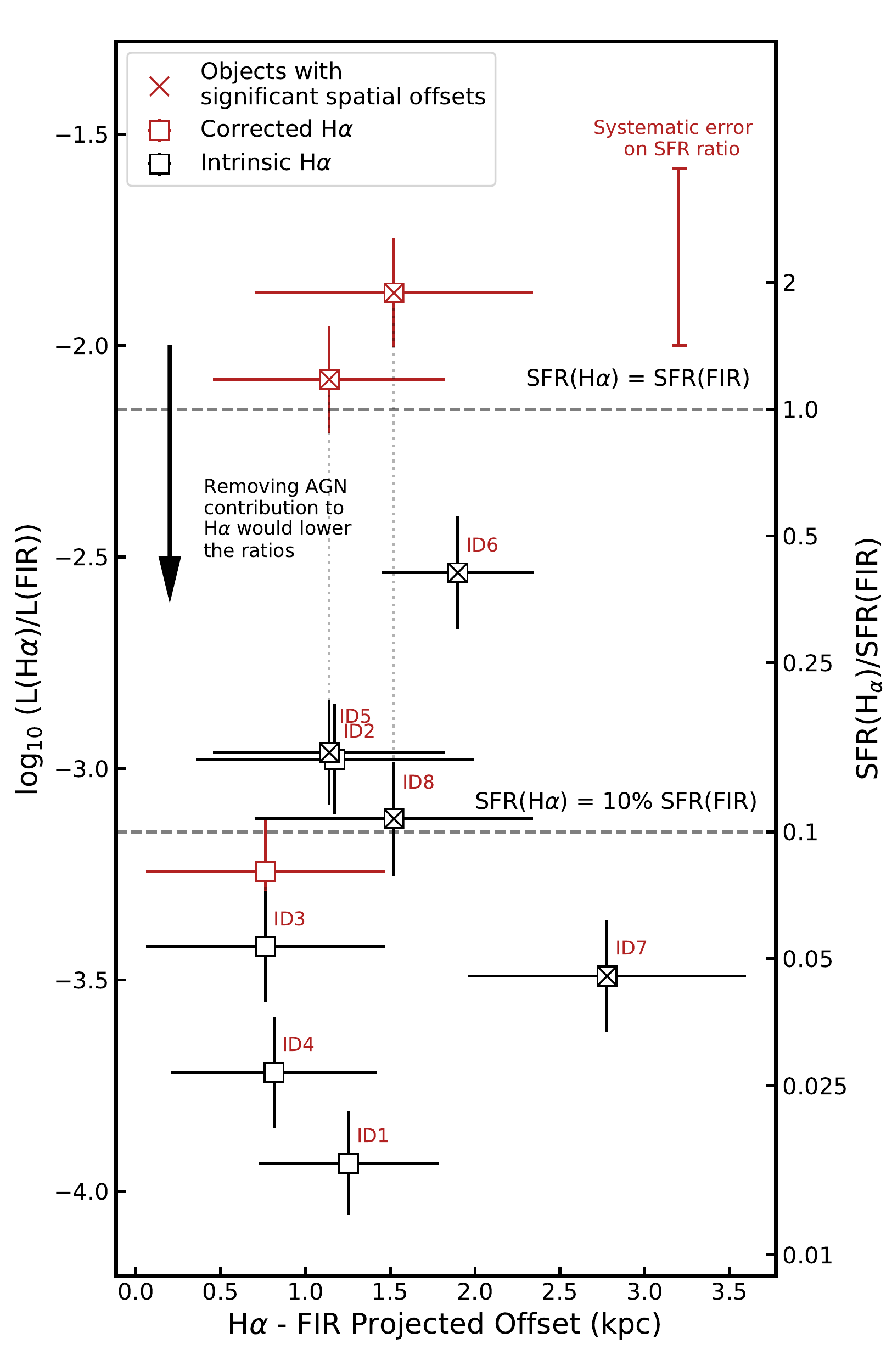}
		\caption[SFR(FIR)/SFR(H$\alpha$) vs emission offset]{ The ratio of L$_{\rm H\alpha}$ and L$_{\rm FIR}$ (left axes) and, equivalently, the ratio of SFR$_{\rm H\alpha}$ and SFR$_{\rm FIR}$ (right axis)  as a function of projected physical offset between the H$\alpha$ and FIR emission (see Figure \ref{fig:Hal_ALM_off}). The black and red symbols indicate dust attenuation uncorrected and corrected H$\alpha$ data, respectively (see \S~\ref{sec:Dust_cor}). The crosses show the objects with significant offsets between the FIR and H$\alpha$ emission from Figure \ref{fig:SF_img}. The red error bar indicates the systematic error on SFR ratios due to calibrations (0.42 dex). H$\alpha$ luminosities uncorrected for dust dramatically underestimate the SFRs, and in one case even after a dust-correction (ID 3). Applying a correction for a contribution from the AGN to the H$\alpha$ emission would introduce a further discrepancy between the two tracers (see black arrow and \S~\ref{sec:Gal_prop}).}
		\label{fig:Hal_LIR}
	\end{figure}

	In Figure~\ref{fig:Hal_LIR} we compare the star-formation rates inferred from the H$\alpha$ luminosity, SFR(H$\alpha$), to those inferred from the FIR luminosity, SFR(FIR) as calculated in \S~\ref{sec:SED} and \ref{sec:Dust_cor}. By performing SED fitting on multi-wavelength photometry (UV-submm), the FIR emission used here has had the AGN contribution removed (\S~\ref{sec:SED}). If we convert the observed H$\alpha$ luminosities directly to star-formations rates the median ratio of the two SFR tracers is SFR(FIR)/SFR(H$_{\alpha}$)=14.5, with a range of $\approx$2.5--65 (black squares in Figure~\ref{fig:Hal_LIR}). However, these ratios suffer from two important effects: (1) obscuring dust which will {\em lower} the observed H$\alpha$ fluxes; (2) the contribution from AGN photoionisation, which will {\em increase} the H$\alpha$ fluxes above that produced by star-formation alone. Although H$\alpha$ and FIR can trace star formation on different timescales, we assume that the global star formation has not changed significantly in the past 100 Myr ($\sim$ 1 dynamical time) which is a reasonable assumption for massive galaxies like our host galaxies.

	To investigate the effect of dust attenuation, we make use of the observed Balmer decrement (i.e., the H$\alpha$/H$\beta$ flux ratios), where possible, to calculate a single (flux-weighted average) Av$_{\rm HII}$ value per galaxy (\S~\ref{sec:Dust_cor}). Although we detect H$\beta$ in 5 out of 8 objects (ID 1, 3, 5, 6, and 8; Figure~\ref{fig:HST_spec}), it was not possible decompose the broad and narrow components in the Type 1 AGN (ID 1 \& 6), therefore we only have direct Balmer decrement constraints for 3 targets (see \S~\ref{sec:Dust_cor}). For these targets the correction factors to the fluxes are $\approx$1.5--18 (see red squares in Figure~\ref{fig:Hal_LIR}). 
	After correcting for dust obscuration, the SFR(H$\alpha$) values of ID 5\&8 are a factor 1.1--1.8 higher than the SFR(FIR).\footnote{We note that, using independent analyses, \cite{Loiacono19} also find SFR(H$\alpha$) is higher than SFR(FIR) for ID\,5, possibly due to AGN contamination.} Although this discrepancy is within the systematic error on the SFR calibrations, the SFR(H$\alpha$) should be considered an upper limit on the SFR, due to possible photo-ionisation from the AGN (see discussion below). Even after the dust correction, the SFR(H$\alpha$) of ID3 is a factor of 12 lower than SFR(FIR). For this source the total SFR, as inferred from FIR emission, can not be recovered from the H$\alpha$ emission. Similar results have been seen for sub-mm galaxies and may be due to a different spatial distribution of obscured and unobscured star-forming regions and/or star-forming regions being completely undetected in the optical/near-infrared data due to the obscuring dust (\citealt{Hodge16,Chen17,Chen19}). 
	
	Using H$\alpha$ as a star-formation rate indicator in AGN host galaxies is a well known challenge, and high-redshift data typically lack the diagnostic power to carefully decompose the relative constributions to the H$\alpha$ luminosity from AGN photoionisation, star formation photoionisation and shocks \citep{Davies14a,Davies14b,DAgostino19}. Previous work using IFU data on AGN host galaxies presented low [N~{\sc ii}]/H$\alpha$ emission-line ratios as evidence that the H$\alpha$ emission is star-formation dominated in off-nuclear regions for those specific targets \citep[e.g.,][]{Canodiaz12,Cresci15,Carniani16}; however, we re-assess this for one of these literature sources (our ID\,6) in \S~\ref{sec:xid2028}. Based on the spectra shown in Figure~\ref{fig:HST_spec} the emission-line flux ratios of log$_{10}$ ([N~{\sc ii}]/H$\alpha$) range between -0.64 and 0.20 (median of 0.05) for our sample. For the 5 objects with detected H$\beta$, the emission-line ratio of log$_{10}$ ([O~{\sc iii}]/H$\beta$) ranges between 0.97 and 1.47 (median value of 1.02) We do not detect H$\beta$ in the outer parts of the host galaxies, however using the upper limits, these region still lie in the AGN dominated parts of the BPT diagram.
	
	In summary, whilst our sample may be biased to those with particularly high levels of dust (due to the pre-selection of a detection in the ALMA data; \S~\ref{sec:Sample}), we have shown the H$\alpha$ luminosities uncorrected for dust could dramatically under predict the true values. Furthermore, in one target the SFR inferred from H$\alpha$ is still an order of magnitude lower than that inferred from the FIR after a dust correction. On the other hand, we have shown that H$\alpha$ emission is likely to have a significant ionisation contribution from the AGN which would result in the SFRs inferred from H$\alpha$ being {\em higher} than the true values (also see black arrow in Figure~\ref{fig:Hal_LIR}). In conclusion, we find that the narrow H$\alpha$ emission does not provide a reliable census of the total SFRs within our AGN host galaxies. We shown the importance of having FIR measurements and/or emission-line ratio diagnostics to assess the true SFRs in AGN host galaxies. In the following sub-section we explore the differences between H$\alpha$ and FIR further by utilising the spatially-resolved information in our data.
	
	\subsection{Spatially-resolved comparison of H$\alpha$ and FIR emission}\label{sec:SF_res}
	
	In Figure \ref{fig:SF_img}, we compare the spatial distribution of H$\alpha$ emission (background maps) and FIR continuum (contours). The red and blue points with their respective error circles around, show the locations of peak narrow H$\alpha$ and FIR emission, respectively (see \S~\ref{sec:position}). Except for ID\,7 we find that the H$\alpha$ emission is centrally concentrated. However, we see a variety of {\em sizes} of the H$\alpha$ emission, with ID\,5 showing a particularly impressive 20\,kpc wide H$\alpha$ emitting region elongated in a East-West direction.\footnote{ID 5 has been considered a compact star-forming galaxy, progenitor of compact quiescent galaxies \citep{Popping17,Talia18}. Despite this, we measure the r$_{\rm e,}\rm H\alpha$ to be 4 kpc and both H$\alpha$ and [O~{\sc iii}] are detected on scales up to 20 kpc. We note that these are extraordinary sizes; however, they may be due the additional photoionisation by the AGN.}
	The FIR emission is also mostly centrally concentrated; however, for ID\,6 we see a tail of rest-frame FIR emission to the North East which (see Figure \ref{fig:Sizes_calc}), as shown by \citet{Brusa18}, is extended towards a companion galaxy that is detected in the K-band LUCI+ARGOS data.
	
	For four out of the eight targets we find a significant projected spatial offset between the peak in H$\alpha$ emission and the peak in the FIR emission. That is, the positional error circles do not overlap for the two sources of emission in ID 5, 6, 7 and 8 (Figure~\ref{fig:SF_img}). These conclusions are consistent if we use either the ``High resolution'' or ``IFU matched'' ALMA maps (see \S~\ref{sec:ALMA_img}). In Figure \ref{fig:Hal_ALM_off} we show the positional offsets in Right Ascension and Declination between the two sources of emission. Across the full sample the projected offsets range from 1.3\,kpc to 2.8\,kpc, where the median offset is $1.4 \pm 0.6$ kpc (see Table~\ref{Table:Sample_SF}).  We could not find previous work which clearly quantifies the spatial offsets between H$\alpha$ emission and FIR continuum for high-$z$ galaxies to compare to. However, offsets between optical continuum and dust continuum have previously been reported in a qualitative way in several works \citep[e.g.,][]{Hodge16,Chen17,Elbaz18}

\begin{figure}
		\includegraphics[width=0.9\columnwidth]{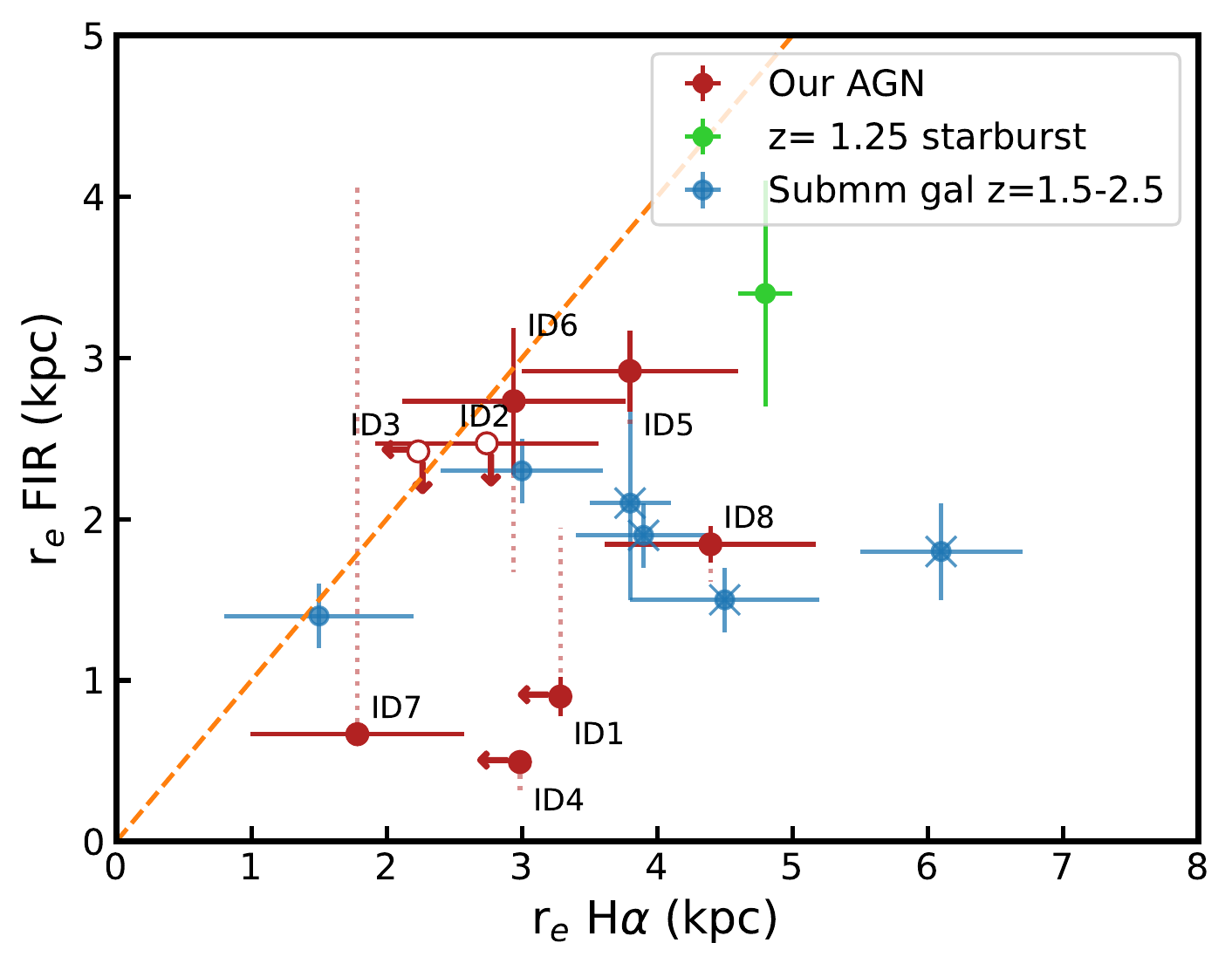}
		\caption[Comparison of the SF emission sizes]{Comparison of the H$\alpha$ and FIR emission sizes. The red circles represent our AGN sample (filled - FIR SNR>8, empty - FIR SNR<8). The red dotted lines indicate the range of FIR sizes between the $uv$ and COG methods (\S~\ref{sec:ALMA_analysis}). In each case we took into account of the smearing by the beam/PSF.  The blue and green points show sub-mm galaxies \citep{Chen19} and a starburst galaxy at z=1.5 \citep{Nelson19}, respectively, where the blue crosses indicate sub-mm galaxies confirmed to host an AGN (X-ray or MIR).  The orange dashed line indicates the one-to-one ratio between the H$\alpha$ and FIR sizes. On average, the FIR emission is more compact than the H$\alpha$ emission, similar to that observed in submm and starburst galaxies.}
		\label{fig:Sizes}
	\end{figure}

	In Figure~\ref{fig:Sizes} we compare the half-light radii of H$\alpha$ and FIR emission. These are calculated as described in \S~\ref{sec:IFU_size} and \ref{sec:ALM_flx_size} and the values are provided in Table~\ref{Table:Sample_SF}. For the five targets for which we were able to make a direct measurement we obtained H$\alpha$ sizes of 1.8--4.4\,kpc with an average value of 3.1 kpc. These H$\alpha$ sizes for our targets are consistent with those measured by \citet{ForsterSch18a}, who targeted massive optically/NIR selected galaxies at z$\sim2$ using VLT/SINFONI and  KMOS, finding H$\alpha$ sizes between 1--8 kpc with a median value of $2.9 \pm 1.5$ kpc. For six of our targets we have a direct size measurement from the ALMA data (i.e., those with SNRs$>$8), and find FIR sizes of 0.5--2.9\,kpc, using our preferred method of obtaining the sizes from the visibility data (see \S~\ref{sec:ALMA_analysis}), with an average value of 1.6\,kpc. These FIR sizes agree well with the $\approx$0.6--2.5 kpc sizes previously found for X-ray AGN host galaxies \citep{Harrison16Alm} and sub--mm and star-forming galaxies \citep[e.g.][]{Ikarashi15,Simpson15,Hodge16, Spilker16, Tadaki17,Fujimoto18, Lang19, Chen19}. In summary, the H$\alpha$ and FIR sizes that we observe for our AGN host galaxies do not appear to be exceptional compared to other redshift-matched, mostly FIR bright, galaxy samples in the literature.

	We find that the H$\alpha$ sizes are factor of $\approx$2\,times larger than the FIR sizes and in the four targets that we can make this comparison directly, the H$\alpha$ sizes are 1.1--2.6$\times$ larger than the FIR sizes. In Figure~\ref{fig:Sizes} we compare these different size measurements of our sample to the $z$=1.5--2.5 sub-mm galaxies from \cite{Chen19} (blue points) and a z$=$1.25 starburst galaxy from \citet{Nelson19}. These samples also exhibit H$\alpha$ sizes which are $\approx$2$\times$ larger than the FIR sizes. Also consistent with this are other studies of high-$z$ galaxies which have found that FIR continuum sizes to be $2-3\times$ smaller than the rest-frame {\em optical} sizes \citep{Hodge16,Tadaki17,Elbaz18,Fujimoto18,Lang19}, which also implies FIR sizes which are $2-3\times$ smaller than H$\alpha$, because broad-band optical and H$\alpha$ sizes typically agree within $\approx$30\% \citep[]{Nelson12,ForsterSch18a}. 
	
	Overall, based on the above comparison to the literature, H$\alpha$ sizes that are a factor of 2--3 bigger than the FIR continuum are somewhat expected. However, what is particularly striking in Figure~\ref{fig:Sizes} is that the sub-mm galaxies which host an AGN (see crossed blue points) are those with the largest H$\alpha$ sizes. Although in \citealt{Chen19} they find that the [N~{\sc ii}]/H$\alpha$ ratios are generally low, potentially indicating a low AGN contribution to ionising the gas. In our targets we are not able to rule out that AGN have a strong contribution to producing the most extended H$\alpha$ emission. In the outer regions ($>$0.6\,arsec) of the galaxies the $\log$([N~{\sc ii}]/H$\alpha$) ratios remain high, ranging from -0.4--0.4, which indicates AGN dominating the ionisation in the extended regions at least for some of the targets (unfortunately H$\beta$ is too weak in the outer part of the galaxy to be realibly detected). Future work which is able to de-couple the contribution of the AGN and the star formation components on larger samples is needed to fully understand the contribution of the overall AGN to producing the observed H$\alpha$ emission sizes. 
	
	Based on (1) the discrepancy between star-formation rates inferred from H$\alpha$ compared to those from FIR; (2) the different sizes and distributions of the FIR (tracing obscured star formation) compared to the H$\alpha$ distribution and; (3) the challenges in decoupling the contribution of star-formation from the AGN contribution to producing the H$\alpha$ emission, we conclude that H$\alpha$ emission alone is not a reliable tracer of the star-formation in the AGN host galaxies in our sample. These challenges can be overcome, at least to some degree, in IFU observations of local AGN since the high spatial resolution observations can result in maps of multiple emission-line ratio diagnostics \citep[e.g.,][]{Venturi18,DAgostino19}. However, with the current observational facilities this is rarely possible for high-$z$ systems and caution, and a careful case-by-case assessment is required when using H$\alpha$ emission to trace star-formation in high-$z$ AGN host galaxies.

	\subsection{Star formation and AGN driven outflows} \label{sec:SF_outflows}
	
	Despite the need for AGN feedback in cosmological simulations, we still lack a consensus on what impact AGN outflows have on star formation from observations. This is despite a lot of work in the literature that has searched for such an impact by comparing AGN-driven outflow properties with the star-formation rates and molecular gas measurements within the host galaxies. This is attempted both from a statistical point of view using large samples  \citep[e.g.,][]{Woo16, Wylezalek16, Lanzuisi17,Harrison17,Perna18,Scholtz18,Kirkpatrick19} and from detailed, spatially-resolved observations of individual objects \cite[e.g.,][]{Alatalo15,Cresci15b,Husemann19,Shin19}. Of particular relevance for this work is the reported spatial anti-correlation between the AGN driven outflows (traced through [O~{\sc iii}]) and the star formation (traced through H$\alpha$) in $z$=1.5--2.5 AGN \citep{Canodiaz12, Cresci15, Carniani16}. Unlike in the previous studies, we use multiple potential star formation tracers (FIR emission and H$\alpha$) to search for the impact of AGN ionised outflows on the star formation within their host galaxies of our sample, which also has representative luminosities and ionised gas kinematics of the parent AGN population (see Figure~\ref{fig:Xlum}). 
	
	\begin{figure*}
		\includegraphics[width=0.85\paperwidth]{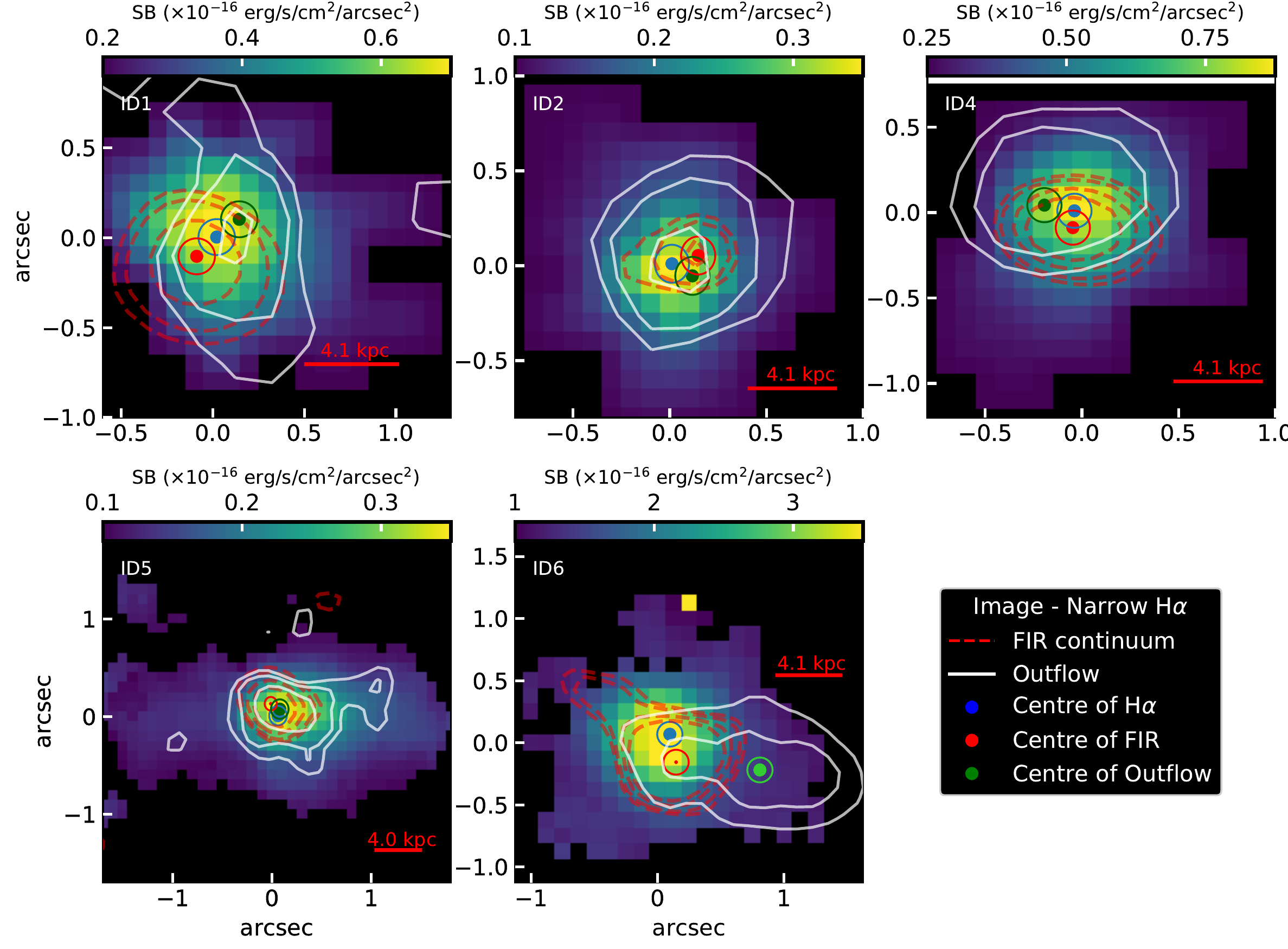}
		\caption[Comparison of SF regions and AGN driven outflows]{Maps to show surface brightness distribution of the narrow line H$\alpha$ components, with the red-dashed contours showing the distribution of FIR emission (as described in Figure \ref{fig:SF_img}) for the five targets where we identified outflows. The red solid line represents the major axis the PSF of the IFU observations, labelled with the corresponding size in kiloparsec. The white contours show the distribution of the ionised outflow (3,4,5 $\sigma$ levels), as defined by the high-velocity wings of the [O~{\sc iii}] emission line (Figure \ref{fig:HST_spec}). The blue, red and green points show the peak of the H$\alpha$, FIR and the outflow, respectively. We do not see significant anti spatial correlation between the H$\alpha$ and outflows as found for a few high-$z$ AGN (\citealt{Canodiaz12,Cresci15,Carniani16}). In Figure~\ref{fig:2028_UV} we provide further insight into ID\,6 which also appears in \citealt{Cresci15}. Overall, we do not see any strong evidence for the outflows instantaneously suppressing star formation in our sample. North is up and East is left.}
		\label{fig:SF_outflow}
	\end{figure*}
	
	We detected ionised gas outflows in five out of the eight objects in our sample (63 \%, see Figure \ref{fig:HST_spec}; \S~\ref{sec:sampleSummary}). In Figure~\ref{fig:SF_outflow} we present maps of the [O~{\sc iii}] outflows as white contours (produced as described in \S~\ref{sec:EM_maps}).\footnote{We note that, given the deep observations of ID 5, we detected outflows in both H$\alpha$ and [O~{\sc iii}] \citep[also see ][]{Genzel14,Loiacono19} Comparing these two outflows, we found that they differ in both outflow kinematics (Figure~\ref{fig:HST_spec}) and spatial extent, with the H$\alpha$ being more extended (up 4\,kpc scales see Fig~\ref{fig:Sizes_calc}). However, it is not the focus of this work to characterise, in detail, the differences or origin of these two outflow components.}
	Three of the targets (ID1, ID5 and ID6) show significant [O~{\sc iii}] outflows elongated beyond the central regions. In this figure, we also show maps of the H$\alpha$ emission (background map) and rest-frame infrared (dashed contours). Similarly to the narrow H$\alpha$ and FIR emission, we found the center of the outflow as a location of the brightest pixel in the outflow map. We represent the peak locations of the H$\alpha$, FIR and outflow emission as red, blue and green points, respectively. We do not see any strong evidence that the outflows suppress the star formation; i.e., either through cavities in the H$\alpha$ emission at the location of the ionised outflows (cf. \citealt{Canodiaz12,Cresci15,Carniani16}) or cavities in the rest-frame FIR emission. Similarly to the offsets between H$\alpha$ and FIR emission, we also measured the position of the peak of the outflow emission. Based on the positional uncertainties (see circles in Figure~\ref{fig:SF_outflow}), in three sources (ID1, ID4 and ID6) we see a significant offsets of 1.7--6.4\,kpc between the outflows and the FIR emission (with a median value of 2.3$^{+2.6}_{-1.3}$\,kpc across the full sample). However, this could just be due to differential obscuration by the dust (i.e., [O~{\sc iii}] is more obscured where the dust is located); unfortunately, we do not have the required signal-to-noise in the H$\beta$ emission lines to map the Balmer decrement. Alternatively outflows may preferentially escape away from the dusty regions \citep[e.g.,][]{Gabor14}. Only in ID\,6 do we see a significant offset between the peak of the H$\alpha$ emission and the [O~{\sc iii}] outflow, but this is just because the outflow is so extended beyond the centrally concentrated H$\alpha$ emission. This source was originally presented with IFU observations in \citet{Cresci15} as showing evidence for positive and negative feedback. We do not conclude the same here, and discuss this source in detail in \S~\ref{sec:xid2028}. For three objects without any detected outflow, we do not see any systematic different star formation morphologies compared to those with detected AGN-outflow. This further indicates that the presence (or lack-there-of) of ionised outflows does not impact upon the distribution of star formation within the host galaxies in our sample. 
	
	Overall, we do not see any strong evidence that ionised outflows are suppressing star formation (or enhancing it) in the host galaxies of our AGN host galaxies. This is in contrast to the results on three luminous $z$=2.5 quasars \citep[][]{Canodiaz12,Carniani16}. These observations are quite similar to ours, in terms of using seeing-limited ground-based IFU observations to map both the H$\alpha$ and [O~{\sc iii}] emission. Although, in these works the H$\alpha$ may be a more reliable tracer of star formation than for our targets (see \S~\ref{sec:SF_res}), it is worth noting that they do not include an analysis of the rest-frame FIR emission which may yet reveal dusty ``obscured'' star formation at the location of the observed deficit in H$\alpha$ emission. It is also worth noting that these quasars represent some of the most powerful AGN in the Universe (L$_{\rm bol}\sim 10^{47.5}$ ergs s$^{-1}$), which are a factor of $\sim$ 100--1000 higher than our targets. Furthermore, the  [O~{\sc iii}] FWHM of the quasars are 700--1500  kms$^{-1}$, representing the most extreme outflow systems (Figure~\ref{fig:Xlum}). Therefore, it possible that the AGN in our sample lack the required power to rapidly impact upon the host galaxy properties, and it is only the most extreme systems where this effect can be observed.  Clearly, similar observations on a much larger sample are now warranted to establish if galactic outflows driven by powerful quasars are uniquely responsible instantaneously suppressing star formation inside their host galaxies.

	\subsubsection{No clear evidence of feedback in ID 6 - XID 2028}\label{sec:xid2028}
	
	IFU data for target ID\,6 was previously presented by \citet{Cresci15}, where they identified a cavity in the H$\alpha$ emission at the location of the AGN driven [O~{\sc iii}] outflow, and enhanced H$\alpha$ emission around the outflow edges. We do not observe similar features, instead finding that the H$\alpha$ emission is spatially extended, but centrally concentrated (Figure~\ref{fig:SF_outflow}). However, we note that in this work we present the H$\alpha$ observations using the SINFONI $H$-band grating (ID 094.B-0286(A); not previously published), while the \citet{Cresci15} work used the earlier lower spectral resolution and shallower $HK$-grating observations (ID 383.A-0573(A)). Therefore, we repeated our analyses on the $HK$ grating data, obtaining consistent conclusions to those seen in Figure~\ref{fig:SF_outflow} (discussed in more detail below). Regardless of  the exact H$\alpha$ morphology, there is still sufficient FIR continuum to imply significant star formation spatially-coincident with the outflow.
	
\begin{figure}
		\includegraphics[width=1.0\columnwidth]{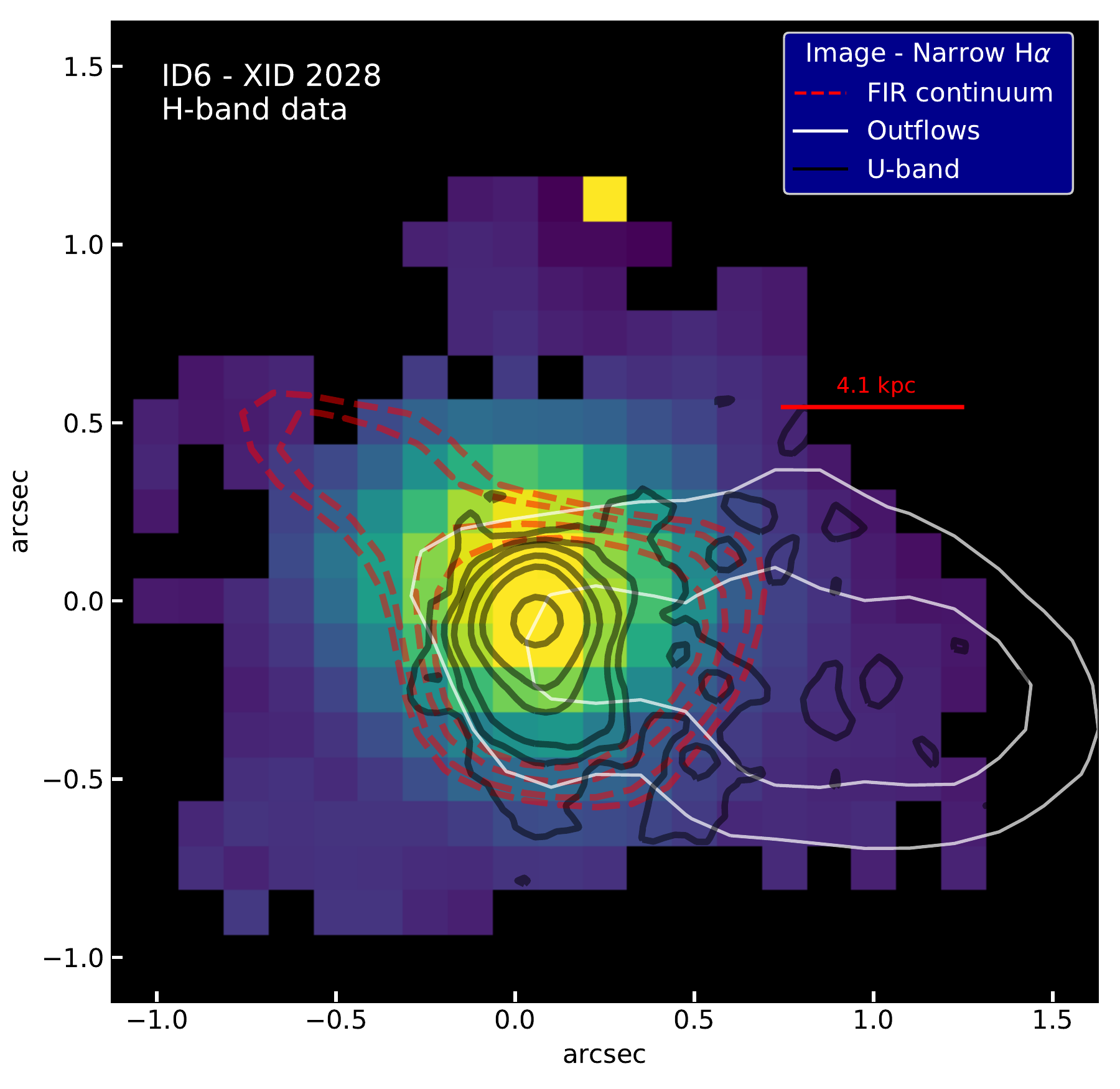}
		\includegraphics[width=1.0\columnwidth]{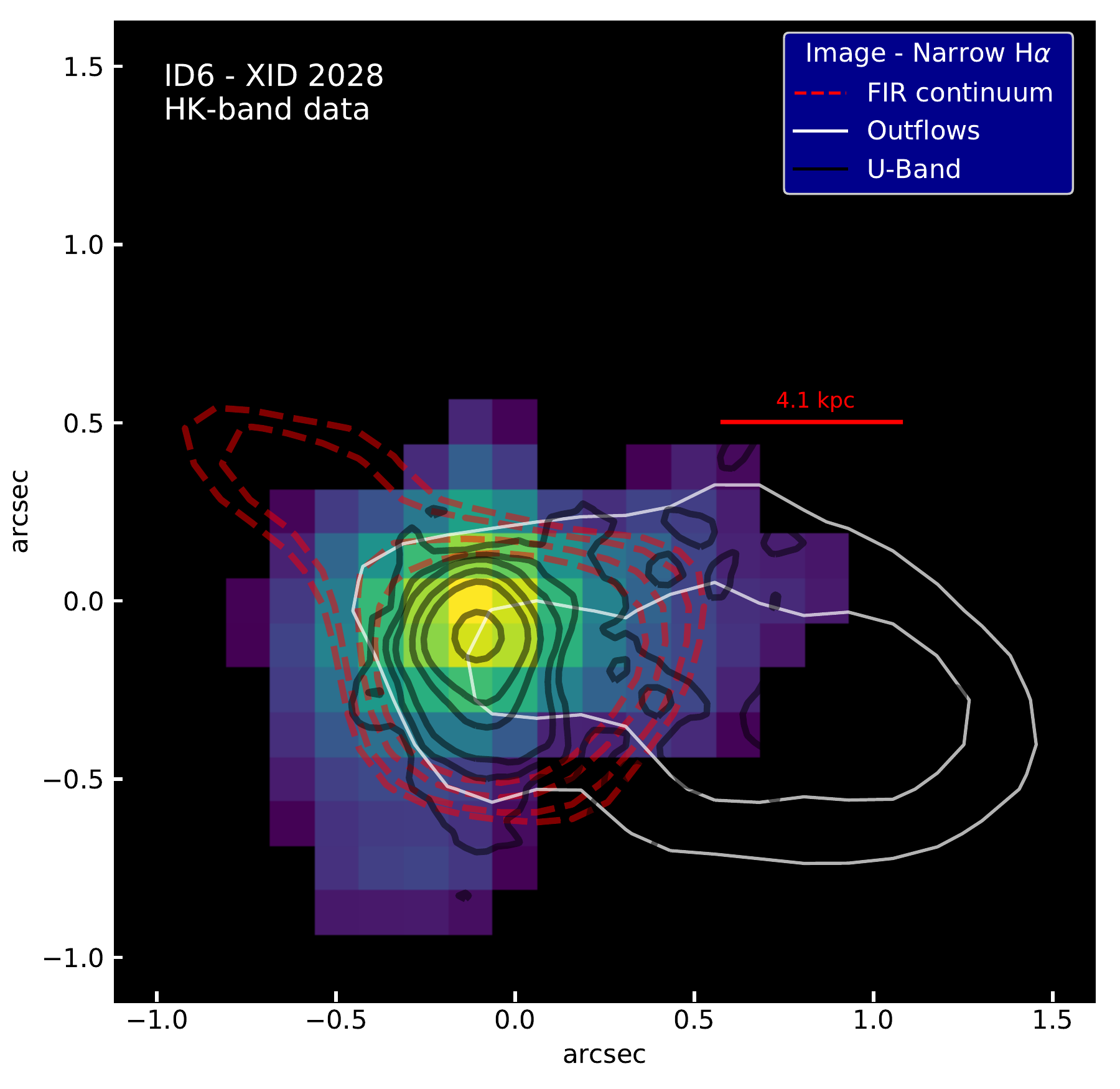}
		\caption[HK and H-band observations of XID 2028]{A comparison between the various emission discussed in this work for ID6 (also known as ``XID\,2028'') from various observations. Top panel: H-band high-spectral resolution observations. Bottom panel: HK-band low-spectral observations. In both panels the maps show the narrow line H$\alpha$ emission, the red-dashed contours show the distribution of FIR emission (2.5, 3, 5 $\sigma$ levels) and the red solid line shows the size of the PSF (all as described in Figure \ref{fig:SF_img}). The white contours show the distribution of the ionised outflow (3,4,5 $\sigma$ levels), as defined by high-velocity wings of the [O~{\sc iii}] emission line (Figure \ref{fig:HST_spec}). The black contours show \textit{HST} I-band image (rest frame U-band; contour levels of 0.008, 0.015, 0.022, 0.05, 0.1, 0.5 relative to the peak). We observe marginally extended U-band emission, FIR continuum and narrow H$\alpha$ emission all in the Western direction of the giant ionised outflow. We do not see evidence for suppressed star formation, instead, our results could indicate star-formation in the direction of the outflow and/or indicate the preferential direction of the ionising radiation from the AGN (\S~\ref{sec:xid2028}). North is up and East is left.}
		\label{fig:2028_UV}
	\end{figure}

	The difference in the results of the H$\alpha$ emission between our results and those of \citet{Cresci15} could be the result of different analysis methods; for example, the adopted approach to account for the broad H$\alpha$ emission. Therefore, we also performed similar analyses to those presented in \citet{Cresci15} by first fitting and subtracting the continuum and H$\alpha$ broad-line region pixel-by-pixel from the cube before making a narrow-band image of the residual narrow-line component. We note that we applied this additional method to create narrow H$\alpha$ maps to H$\alpha$ data of the rest of the sample and we did not observe any differences in both flux and morphology of the narrow H$\alpha$ maps.
	To be fully consistent, we performed this on the $HK$-band data and the results are presented in the bottom panel of Figure~\ref{fig:2028_UV}. Although we do not detect H$\alpha$ over the large scales measured using the deeper $H$-band data, this analysis still reveals a possible extension of the H$\alpha$ emission to the West. Even-so this extension is {\em within} the [O~{\sc iii}] outflow, in contrast to that presented by \citet{Cresci15}, where the extended H$\alpha$ emission is outside of the region covered by the [O~{\sc iii}] outflow (Figure~\ref{fig:SF_img}). The difference between the maps in Figure \ref{fig:2028_UV} is caused by the quality of the $H$-band and $HK$-band data.
	We find that it is only when we use the same fully-reduced $HK$ data cube as that used by \citet{Cresci15} that we are able to observe a cavity in H$\alpha$ at the location of the outflow (G.~Cresci, priv. communication). This implies that the differences found in this work to those in \citet{Cresci15} are not dominated by the adopted analyses methods but, instead, in the intermediate data reduction steps (e.g., sky subtraction or frame stacking). However, this source is scheduled for observations with {\em JWST}/NIRSpec, through an Early Release Science programme \citep{Wylezalek17JWST},\footnote{http://www.stsci.edu/jwst/observing-programs/approved-ers-programs/program-1335} which will provide sensitive and improved spatial resolution IFU observations of this source, resulting in the most definitive description of this source's H$\alpha$ morphology. 
	
	We further investigate our results for ID\,6 by comparing to the morphology of the rest-frame U-band emission as determined from the {\em HST}, $I$-band image (F814W filter) using the available 1 orbit of observations (\citealt{Koekemoer07}; Figure~\ref{fig:2028_UV}, top panel). It can be seen that the U-band emission is slightly extended in the Western direction, within the region of the [O~{\sc iii}] outflow. Furthermore, there is a possible slight extension of the FIR emission in this direction \citep[in addition to the ``tail'' to the North East; also see][]{Brusa18}, possibly implying star formation is located in the general direction of the outflow. Narrow H$\alpha$ emission is also detected over the extent of the outflow; however, we find that the [N~{\sc ii}]/H$\alpha$ and [O~{\sc iii}]/H$\beta$ emission-line flux ratios are consistent photo-ionisation dominated by an AGN in this region. In summary, we find FIR emission, rest-frame U-band emission, and H$\alpha$ emission all co-spatial with the [O~{\sc iii}] outflow and consequently find no evidence for suppressed, or enhanced star formation due to the outflow in this source.

	\subsection{Implications of our results}\label{sec:implications}
	
	Our work has shown that H$\alpha$ emission must be used with caution as a star-formation tracer for AGN host galaxies, even when a global Balmer decrement is available to correct for dust obscuration (which is often not the case for high-$z$ studies). Future, sensitive and high spatial-resolution IFU observations, e.g., with VLT/ERIS, {\em JWST}/NIRSpec or ELT/HARMONI, will make it possible to map the ionisation conditions and Balmer decrements, and separate the contribution from AGN and star-formation in high-$z$ AGN host galaxies. Except in exceptional cases of adaptive optics assisted IFU observations of lensed galaxies \citep{Fischer19}, this is currently only possible for local galaxies \citep[e.g., ][]{DAgostino19}. Furthermore, for a complete census of the star formation we suggest it is necessary to also use spatially-resolved FIR observations to map the dust-obscured star formation. 
	
	Our sample is representative of typical AGN luminosities and outflow properties for $z\approx$1--2 AGN; however, it is limited to sources with existing detections in FIR and H$\alpha$ emission, resulting in all of the sources lying on, or above, the `main sequence' of star formation (Figure \ref{fig:SFR}).  We should also caution that, consequently, the systems where the star formation has rapidly shutdown may not be in our sample; however, ID 6 is a strong star-forming galaxy where suppressed star formation was previously suggested. 
	
	A key development of our study over previous work is that we focus on more common moderate luminosity AGN. However, it possible that our moderate luminosity AGN do not have sufficient power to rapidly change the star formation in their host galaxies, compared to their more powerful quasar counterparts \citep[][]{Canodiaz12,Carniani16}. A more complete survey covering the full AGN luminosity -- star-formation rate -- stellar mass parameter space is now required to place more comprehensive constraints. 
	
	Useful insight to interpret our results can come from observations of nearby AGN host galaxies. Recently, \citet{Shin19} observed both positive and negative feedback in NGC 5728, a nearby Seyfert like galaxy. The IFU and ALMA observations, showed enhanced star formation on the edges of the outflow in the very core of the galaxy as well as a lack of molecular gas in the outflow in the outskirts. However, both effects were observed on scales of $<1$ kpc scale. Indeed, although the samples lack the most powerful AGN, observations of local systems find that any impact by outflows and/or jets on the star formation, or molecular gas, is localised to small scales and is only affecting a small fraction of the the total star formation or gas content in the host galaxy \citep[e.g.,][]{Alatalo15,Cresci15b,Rosario19}. 
	
	Based on our work, we therefore do not find any evidence that outflows from moderate luminosity AGN instantaneously influence the in-situ star formation inside their host galaxies at least on $\approx$4\,kpc scales. However, impact from these outflows could be occurring on spatial scales below those to which we are sensitive (i.e., $<$a few kiloparsec) and maybe subtle, only influencing a small region of the galaxy 
	\citep[e.g., ][]{Croft06,Alatalo15,Cresci15b,Querejeta16,Rosario19,Shin19,Husemann19}. Alternatively, the AGN outflows may have an impact over longer timescales, without an instantaneous influence on the star formation, for example, by removing low entropy gas which is later prevented from re-accreting onto the host galaxy \citep[][]{McCarthy11,Gabor14, Harrison17, Scholtz18}.
	
	\section{Conclusions}
	In this work we present integral field spectroscopy (VLT/KMOS and VLT/SINFONI) and rest-frame FIR observations (ALMA) for eight z=1.4-2.6 moderate luminosity AGN (L$_{\rm 2-10 \rm kev}$ $\approx$ $10^{42} - 10^{45}$ ergs s$^{-1}$). Our study is designed to build upon previous work that has claimed evidence for suppression and/or enhancement of star formation by high-$z$ AGN by using integral field spectroscopy to spatially-resolve ionised outflows (using the [O~{\sc iii}] line) and to map star formation (using the H$\alpha$ line; \citealt{Canodiaz12}; \citealt{Cresci15}; \citealt{Carniani16}). In this work, we also used rest-frame FIR observations to map the dust-obscured star formation. We are able to assess how representative our targets are of the overall AGN population (see \S~\ref{sec:SED}) by utilising KASHz, an IFU survey of $\approx$250 AGN, as our parent sample. 
	
	We performed SED fitting on the compiled multi-wavelength photometry (UV-sub-mm) to measure the star-formation rates as traced by the FIR emission (SFR(FIR)) and confirm that the ALMA continuum traces dust-obscured star formation. We extracted galaxy-integrated H$\alpha$ emission-line profiles to infer star-formation rates from H$\alpha$ (SFR(H$\alpha$)). Where possible, the level of dust attenuation ($A_{V}$) was measured using H$\alpha$/H$\beta$ ratios. Furthermore, we produced maps of the: (1)  narrow component H$\alpha$ emission; (2) rest-frame FIR emission and; (3) [O~{\sc iii}]-identified ionised outflows. On the basis of our analyses we obtained the following results:
	
	\begin{enumerate}
		\item For all of our targets, the total SFR inferred from the observed H$\alpha$ luminosities is lower than that inferred from the FIR, by a factor of 2.5--65, with a median factor of 14.5. After applying a correction to the H$\alpha$ luminosities for dust attenuation (possible for three targets), the SFR(H$\alpha$ corr) is still a factor of 12 lower than SFR(FIR) for one target. Furthermore, accounting for the AGN photo-ionisation contribution to the narrow H$\alpha$ emission causes further uncertainty in using this as a reliable star-formation tracer in our targets (see \S~\ref{sec:Gal_prop}; Figure \ref{fig:Hal_LIR}). 
		
		\item We found that the projected spatial extent of the H$\alpha$ emission is typically larger than that of the FIR continuum, by an average factor of $\approx$2. This is similar to that observed in sub-mm galaxies, particularly those hosting AGN, and is possibly due to dust-obscured star formation generally being more compact than unobscured star formation and/or additional photo-ionisation by the AGN to the H$\alpha$ emission (\S~\ref{sec:SF_res}; Figure~\ref{fig:Sizes}). Additionally, in half of our sample we observe significant, $\approx$1--3\,kpc, projected offsets between the peak of the FIR emission and the peak of the narrow H$\alpha$ emission. The average projected offsets across the full sample of eight targets is $1.4 \pm 0.6$ kpc (see \S~\ref{sec:SF_res}; Figure~\ref{fig:SF_img}; Figure \ref{fig:Hal_ALM_off}). 
		
		\item We detected ionised outflows in five out of the eight AGN in our sample, traced by broad [O~{\sc iii}] emission-line components (FWHM$=$610--950\,km\,s$^{-1}$; Figure~\ref{fig:HST_spec}). Based on the spatial distribution of star formation and ionised outflows we see no strong evidence that the AGN outflows are rapidly suppressing or enhancing in-situ star formation in the host galaxies. The same conclusion for a lack of impact on star formation is found whether considering either the FIR or H$\alpha$ emission as possible star-formation tracers; i.e., we see no ``cavities'' in the star formation at the location of the outflows. In three targets the [O~{\sc iii}] outflows are offset from the peak of the FIR emission; however, this could be due to differential dust obscuration or the outflows preferentially escaping away from the dusty regions (see \S~\ref{sec:SF_outflows}; Figure \ref{fig:SF_outflow}). 
		
		\item One of AGN in our sample, ID\,6, is a well studied $z$=1.6 X-ray AGN where a spatial anti-correlation of H$\alpha$ emission and the [O~{\sc iii}] outflow has previously been claimed as evidence for positive and negative feedback \citep[`XID 2028' from ][]{Cresci15}. We are able to reproduce the observations of a spectacular $\approx$10\,kpc outflow in this source; however, based on a re-analyses of the H$\alpha$ data, including new high spectral resolution IFU observations, we do not observe any spatial anti-correlation between the outflow and H$\alpha$. We find significant star-formation (traced through FIR continuum) coincident with the outflow. Furthermore, the H$\alpha$ emission, rest-frame U-band, and AGN outflow are all roughly co-spatial in the Western regions, consistent with an ionisation cone, or star formation located within the outflow (see \S~\ref{sec:xid2028}; Figure~\ref{fig:2028_UV})
	\end{enumerate}
	
	Overall, we have highlighted the challenges in using H$\alpha$ to map the star formation in typical $z$=1.4--2.6 AGN host galaxies. We advocate using multiple possible tracers of star formation for a complete consensus such as FIR continuum. Within our sample we see no evidence that ionised outflows from moderate luminosity AGN are instantaneously having an impact upon the star formation inside their host galaxies. However, impact from these outflows could be occurring on spatial scales below those to which we are sensitive ($<$a few kiloparsec). Alternatively, the outflows may have an impact over longer timescales, for example by removing low entropy gas, without an instantaneous impact on the current rate of star formation.
	
	\section*{Acknowledgements}
	
	We thank Marcella Brusa, Giovanni Cresci and Alessandro Marconi for helpful discussions and sharing their datacube. We gratefully acknowledge support from the Science and Technology Facilities Council (JS through ST/N50404X/1; DJR and DMA through grant ST/L00075X/1) and from a European Southern Observatory studentship (JS). ALT acknowldges support from STFC (ST/L00075X/1 and ST/P000541/1), the ERC advanced Grant DUSTYGAL (321334) and a Forrest Research Foundation Fellowship. This paper makes use of ALMA data: 2015.1.01528.S, 2015.1.01074.S, 2013.1.00884.S, 2012.1.00869.S, 2015.1.01379.S, 2015.1.00299.S, 2015.1.00907.S, 2015.1.00664.S and 2016.1.00735.S. 
	ALMA is a partnership of ESO (representing its member states), NSF (USA) and NINS (Japan), together with NRC (Canada) and NSC and ASIAA (Taiwan), in cooperation with the Republic of Chile. The Joint ALMA Observatory is operated by ESO, AUI/NRAO and NAOJ.
	
	
	
	\bibliographystyle{mnras}
	\bibliography{refs} 
	

	
	
	\appendix
	
	\begin{table*}
		\caption{Properties of the emission lines in the inner 5 kpc aperture.
			(1) Object ID in this paper; 
			(2) narrow H$\alpha$ flux;
			(3) Broad line region H$\alpha$ flux; 
			(4) narrow [N~{\sc ii}] flux; 
			(5) narrow H$\alpha$ FWHM;
			(6) Broad line region H$\alpha$ FWHM;
			(7) narrow H$\alpha$ centre;
			(8) Broad line region H$\alpha$ centre;
			(9)narrow [O~{\sc iii}] flux; 
			(10) broad [O~{\sc iii}] flux; 
			(11) narrow [O~{\sc iii}] FWHM;
			(12) broad [O~{\sc iii}] FWHM;
			(13) narrow [O~{\sc iii}] centre;
			(14) broad [O~{\sc iii}] centre;
			(15) narrow H$\beta$ flux;
			(16) narrow H$\beta$ FWHM;
			(17) narrow H$\beta$ centre;
		}
		\input{./Tables/Table_appendix_fnl.tex}
		\label{Table:Sample_all}
	\end{table*}
	
	\begin{figure*}
		\centering
		\includegraphics[width=0.54\paperwidth]{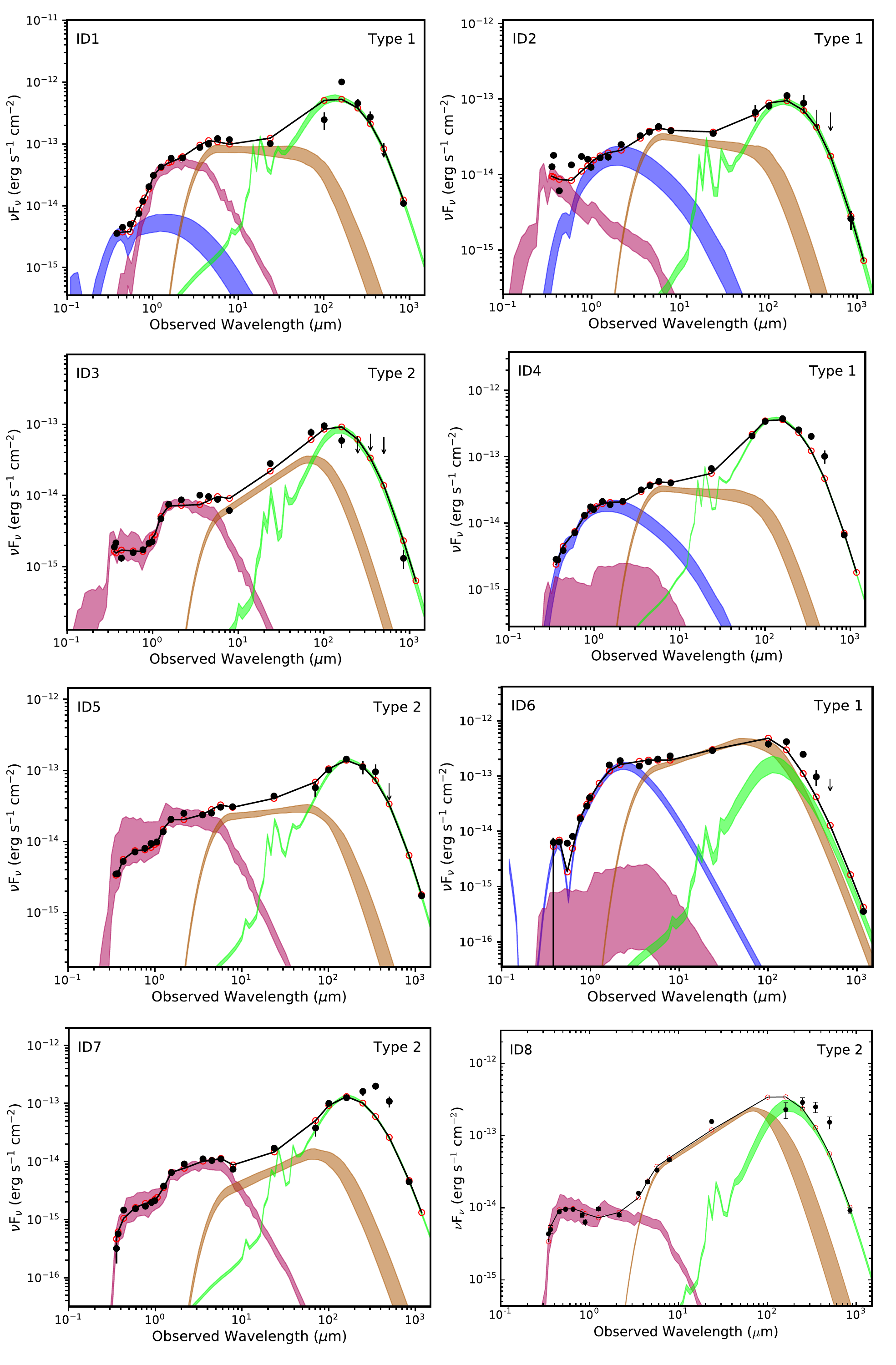}
		\caption{SEDs for all our objects from UV to FIR. The black and red points represent the measured and modelled photometry, respectively. We note that \texttt{FortesFit} also models photometry for which we do not have a measured photometry. The black line shows the total SED. The shaded region represents 1$\sigma$ uncertainty on the fitted components. We fit the following component: AGN accretion disk (blue region), stellar component (red region), AGN torus addition (brown region) and cold dust emission from star formation (green region). The details of the SED fitting can be found in \S~2.4.}
		\label{fig:SEDs}
	\end{figure*}

	
	\bsp	
	\label{lastpage}
\end{document}

%% file: Tables/Main_Sample_Main.tex
\resizebox{0.8\paperwidth}{!}{\begin{tabular}{@{}lccccccccccccccc@{}} 
\hline 
\hline 
(1) & (2 & (3) & (4) & (5) & (6) & (7)   & (8) & (9) & (10) & (11) & (12) & (13) & (14)\\
ID & X-ray ID & Other names & RA        & DEC      & z & AGN  &log$_{10}$   & log$_{10}$     & log$_{10}$                         &log$_{10}$ & IFU data & H$\alpha$/[OIII] & H$\alpha$/[OIII]\\
   &          &             &  (optical & (optical)&   & Type & (L$_{\rm X}$ /ergs s$^{-1}$)& (L$_{\rm [OIII]}$ /ergs s$^{-1}$) &(M$_{*}$/$\Msol$ )  & (L$_{\rm FIR,SF}$/ergs s$^{-1}$) & reduction & Seeing (") & exp time (ks)\\
\hline 
ID1 & xuds 481  & UDS 354.0 $^{\rm (a)}$&34.657385 & -4.98064 & 1.41 & 1 & 44.3 & 42.4$\pm 0.1$ & 11.2 &  $ 46.07^{+0.04}_{-0.04} $  & (e,f) & 0.98/0.85 & 8.4/13.2 \\
ID2 & XID 208  & - &53.045467 & -27.73748 & 1.61 & 1 & 44.3 & 42.5$\pm 0.1$ & 9.0 & $ 45.44^{+0.04}_{-0.04} $  & (e,f) & 0.55/0.52 & 9.0/14.1 \\
ID3 & XID 419  & - &53.097649 & -27.71527 & 2.14 & 2 & 43.0 & 42.5$\pm 0.1$ & 10.5 &  $ 45.65^{+0.04}_{-0.04} $  & (f,g) & 0.61/0.45 & 14.1/82.5\\
ID4 & XID 449  & - &53.104855 & -27.70521 & 1.61 & 1 & 43.8 & 42.6$\pm 0.1$ &  < 10.6 &  $ 46.02^{+0.01}_{-0.01} $  & (f,g) & 0.55/0.85 & 7.8/14.1  \\
ID5 & XID 587  & GS3-19791, KD20-ID5 $^{\rm (b)}$&53.131081 & -27.77309 & 2.22 & 2 & 42.8 & 42.7$\pm 0.1$ & 11.0 &  $ 45.94^{+0.02}_{-0.02} $  & (g,g) & 0.59/0.58 & 20.4/65.7  \\
ID6 & lid 1565  & XID 2028 $^{\rm (c)}$&150.546996 & 1.61846 & 1.59 & 1 & 45.1 & 43.2$\pm 0.1$ &  < 11.1 &  $ 45.57^{+0.21}_{-0.13} $   & (e,h) & 0.68/0.55 & 25.2/5.4\\
ID7 & XID 614  & - &53.137566 & -27.70009 & 2.45 & 2 & 43.3 & 42.2$\pm 0.1$ & 10.9 &  $ 45.98^{+0.02}_{-0.02} $  & (g,g) & 0.56/0.57 & 20.4/88.2 \\
ID8 & -  & ALESS 75.1 $^{\rm (d)}$&52.863303 & -27.93093 & 2.55 & 2 & 45.5$^{\rm j}$ & 43.4$\pm 0.1$ & 10.4 &  $ 46.32^{+0.07}_{-0.06} $  & (i,i) & 0.59/0.58 & 16.8/16.8\\
\hline 
\end{tabular}}

%% file: Tables/Main_Sample_ALMA.tex
\resizebox{0.8\paperwidth}{!}{\begin{tabular}{@{}lcccccccc@{}} 
\hline 
\hline 
(1) & (2 & (3) & (4) & (5) & (6) & (7)   & (8) & (9) \\
ID & Prog &$\lambda$/Band &  Beam (IFM)  & RMS (IFM)& Beam (HR) & RMS (HR)&ALMA &Flux density\\
   &      & $\mu$m/-      &  arcsecond & (mJy)  & arcsecond& (mJy)  & SNR &(mJy)\\
\hline 
ID1 & 2015.1.01528.S & 870/7&0.83x0.75 &0.685 & 0.21x0.20 &0.367&10.0 & $ 3.09\pm 0.41$\\
ID2 & 2015.1.01074.S & 870/7&0.61x0.50 &0.344 & 0.19x0.16 &0.268&4.3 & $ 0.74\pm 0.21$\\
ID3 & 2013.1.00884.S & 870/7&0.76x0.62 &0.244 & 0.28x0.23 &0.143&4.5 & $ 0.37\pm 0.11$\\
ID4 & 2012.1.00869.S & 870/7&0.68x0.47 &0.224 & 0.28x0.24 &0.215&23.3 & $ 1.87\pm 0.10$\\
ID5 & 2015.1.01379.S & 1100/6&0.66x0.56 &0.031 &  - & - &8.1 & $ 0.68\pm 0.05$\\
ID6 & 2015.1.00299.S & 1100/6&0.63x0.53 &0.018 &  - & - &12.2 & $ 0.14\pm 0.02$\\
ID7 & 2015.1.00907.S & 870/7&0.68x0.51 &0.190 & 0.16x0.16 &0.114&12.1 & $ 1.26\pm 0.05$\\
ID8 & 2016.1.00735.S & 870/7&0.57x0.54 &0.323 & 0.17x0.12 &0.110&20.1 & $ 2.61\pm 0.21$\\
\hline 
\end{tabular}}

%% file: Tables/Main_Sample_spec.tex
\resizebox{0.8\paperwidth}{!}{\begin{tabular}{@{}lcccccccc@{}} 
\hline 
\hline 
(1) & (2) & (3) & (4) & (5) & (6) & (7)   & (8) & (9) \\
ID & [O~{\sc iii}] Narrow & [O~{\sc iii}] broad  &[O~{\sc iii}] $\Delta$v &  \Hal Narrow & \Hal BLR   & log$_{10}$(H$\alpha$/H$\beta$) & log$_{10}$([N~{\sc ii}]/H$\alpha$) & log$_{10}$([O~{\sc iii}]/H$\beta$) \\
   & FWHM (km/s)   & FWHM (km/s)   &       km/s      &  FWHM (km/s) &  FWHM (km/s) & ratio                    & ratio                           & ratio                    \\
\hline 
ID1 &$339 \pm 50$ & $614 \pm 70$ & -234 $\pm $ 30 & $478 \pm 72$ & $5947 \pm 70$  & 1.16 $\pm$ 0.1$^{\star}$ &  0.20 $\pm$ 0.1& 1.47 $\pm$ 0.1 \\
ID2 &$226 \pm 60$ & $792 \pm 90$ & -403 $\pm $ 40 & $363 \pm 48$ & $4622 \pm 80$  &  -  &  -0.06 $\pm$ 0.1&  -  \\
ID3 &$329 \pm 50$ & $-$ & - & $483 \pm 63$ & $-$  & 0.53 $\pm$ 0.1 &  0.06 $\pm$ 0.1& 0.88 $\pm$ 0.1 \\
ID4 &$429 \pm 70$ & $747 \pm 70$ & -613 $\pm $ 60 & $399 \pm 48$ & $2291 \pm 90$  &  -  &  0.09 $\pm$ 0.1&  -  \\
ID5 &$316 \pm 50$ & $951 \pm 90$ & -294 $\pm $ 40 & $437 \pm 67$ & 897 $\pm$ 70$^{x}$  & 0.79 $\pm$ 0.1 &  0.04 $\pm$ 0.1& 0.97 $\pm$ 0.1 \\
ID6 &$383 \pm 50$ & $647 \pm 60$ & -262 $\pm $ 50 & $640 \pm 138$ & $5945 \pm 90$  & 0.94 $\pm$ 0.1$^{\star}$ &  -0.16 $\pm$ 0.1& 1.06 $\pm$ 0.1 \\
ID7 &$724 \pm 100$ & $-$ & $-$ & $468 \pm 80$ & $-$  & > 0.91 &  0.20 $\pm$ 0.1& > 1.15 \\
ID8 & $374 \pm 60$ & $-$ & $-$ & $529 \pm 80$ & $-$  & 0.94 $\pm$ 0.1 &  -0.64 $\pm$ 0.1& 1.02 $\pm$ 0.1 \\
\hline 
\end{tabular}}

%% file: Tables/Main_Sample_SF.tex
\resizebox{0.8\paperwidth}{!}{\begin{tabular}{@{}lccccccccccc@{}} 
\hline 
\hline 
(1) & (2 & (3) & (4) & (5) & (6) & (7)   & (8) & (9)  & (10) & (11) & (12)\\
ID & SFR(FIR)            & Av$_{\rm HII}$ & log$_{10}$ &SFR(H$\alpha$ uncor)     &    SFR(\Hal   cor)  & \Hal r$_{\rm e}$ & SNR  & FIR r$_{\rm e}$ (COG) & FIR r$_{\rm e}$ (uv)& H$\alpha$ - FIR offset & Total H$\alpha$\\
   & ($\Msol$ yr$^{-1}$) &                 & (L$_{\rm H\alpha}$/ergs s$^{-1}$)&($\Msol$ yr$^{-1}$) & ($\Msol$ yr$^{-1}$) & (kpc)             & (FIR)& (kpc)       &   (kpc)      & (kpc) & aperture (")\\
\hline 
ID1 & $ 459^{+477}_{-239} $  & > 5.72 &  42.1$\pm 0.1$ &  $ 7^{+8}_{-4} $  &  $ - $  & $< 3.3$&10.0 & $ 1.9\pm 0.8$&$ 0.9\pm 0.1$&$1.3 \pm 0.5$ & 1.2\\
ID2 & $ 107^{+111}_{-55} $$^*$  & - &  42.5$\pm 0.1$ &  $ 15^{+16}_{-8} $  &  $ - $  & $ 2.7\pm 0.8$ &4.3 & $< 2.4$&(P)&$1.2 \pm 0.8$ & 2.0\\
ID3 & $ 174^{+181}_{-90} $  & 0.57 &  42.2$\pm 0.1$ &  $ 9^{+10}_{-5} $  &  $ 14^{+16}_{-8} $  & $< 2.2$&4.5 & $< 2.9$&(P)&$0.8 \pm 0.7$ & 1.4\\
ID4 & $ 409^{+426}_{-212} $  & - &  42.3$\pm 0.1$ &  $ 11^{+12}_{-5} $  &  $ - $  & $< 3.0$&23.3 & $ 0.3\pm 0.8$&$ 0.5\pm 0.1$&$0.8 \pm 0.6$ & 1.4\\
ID5 & $ 336^{+349}_{-174} $  & 2.71 &  43.0$\pm 0.1$ &  $ 51^{+53}_{-25} $  &  $ 379^{+428}_{-213} $  & $ 3.8\pm 0.8$&8.1 & $< 2.6$&$ 2.9\pm 0.3$&$1.1 \pm 0.7$ & 2.4\\
ID6 & $ 145^{+150}_{-75} $  & > 3.93 &  43.0$\pm 0.1$ &  $ 58^{+60}_{-29} $  &  $ - $  & $ 2.9\pm 0.8$&12.2 & $ 1.7\pm 0.8$&$ 2.7\pm 0.5$&$1.9 \pm 0.4$ & 2.0\\
ID7 & $ 369^{+383}_{-192} $  & > 3.63 &  42.5$\pm 0.1$ &  $ 16^{+17}_{-8} $  &  $ - $  & $ 1.8\pm 0.8$&12.1 & $ 4.1\pm 0.8$ & $ 0.7\pm 0.2$& $2.8 \pm 0.8$ & 1.2\\
ID8 &  $ 806^{+838}_{-420} $  & 3.89 &  43.2$\pm 0.1$ &  $ 83^{+86}_{-41} $  &  $ 1494^{+1688}_{-844} $  &  4.4$\pm 0.8$ & 20.1 & $ 1.6\pm 0.8$ & $ 1.8\pm 0.1$&$1.5 \pm 0.8$ & 2.0\\
\hline 
\end{tabular}}

%% file: Tables/Table_appendix_fnl.tex
\resizebox{0.85\paperwidth}{!}{\begin{tabular}{@{}lcccccccccccccccc@{}} 
\hline 
\hline 
(1) & (2 & (3) & (4) & (5) & (6) & (7)   & (8) & (9) & (10) & (11) & (12)   & (13) & (14) & (15)   & (16) & (17) \\
ID & H$\alpha$(nar) flux & H$\alpha$(bro) flux & NII flux & H$\alpha$(nar)  FWHM & H$\alpha$(bro)  FWHM & H$\alpha$(nar) centre & H$\alpha$(bro) centre & [O~{\sc iii}](nar) flux & [O~{\sc iii}](bro) flux & [O~{\sc iii}](nar)  FWHM & [O~{\sc iii}](bro)  FWHM & [O~{\sc iii}](nar) centre & [O~{\sc iii}](bro) centre & H$\beta$(nar) flux & H$\beta$(nar)  FWHM & H$\beta$(nar) centre \\
 & (ergs/s/cm$^2$) & (ergs/s/cm$^2$) & (ergs/s/cm$^2$) & ($\mathrm{km\,s^{-1}}$) & ($\mathrm{km\,s^{-1}}$) & (microns) & (microns) & (ergs/s/cm$^2$) & (ergs/s/cm$^2$) & ($\mathrm{km\,s^{-1}}$) & ($\mathrm{km\,s^{-1}}$) & (microns) & (microns) & (ergs/s/cm$^2$) & ($\mathrm{km\,s^{-1}}$) & (microns) \\
\hline
ID 1 & 4.62e-17 & 3.38e-16 & 7.64e-17 & $478 \pm 72$ & $5947 \pm 70$ & 1.5796 & 1.5775 & 4.88e-17 & 4.34e-17 & $339 \pm 50$ & $614 \pm 70$ & 1.2056 & 1.2046 & $<$3.2e-16 & 473 & 1.1700 \\
ID 2 & 7.40e-17 & 3.98e-16 & 6.43e-17 & $363 \pm 48$ & $4622 \pm 80$ & 1.7136 & 1.7131 & 2.51e-17 & 2.90e-17 & $226 \pm 60$ & $792 \pm 90$ & 1.3074 & 1.3057 & -        & -   & -      \\
ID 3 & 1.12e-17 & -        & 2.10e-18 & $483 \pm 63$ & -             & 2.0668 & 2.0645 & 3.56e-17 & -        & $329 \pm 50$ & -            & 1.5732 & -      & $<$8.12-18 & 488 & 1.5278 \\
ID 4 & 2.86e-17 & 2.58e-16 & 3.72e-17 & $399 \pm 48$ & $2291 \pm 90$ & 1.7162 & 1.7159 & 4.17e-17 & 2.08e-17 & $429 \pm 70$ & $747 \pm 70$ & 1.3087 & 1.3060 & -        & -   & -      \\
ID 5 & 1.41e-17 & 2.82e-17 & 1.60e-17 & $437 \pm 67$ & $897 \pm 70$  & 2.1166 & 2.1172 & 1.17e-17 & 1.45e-17 & $316 \pm 50$ & $951 \pm 90$ & 1.6148 & 1.6132 & 2.15e-18 & 258 & 1.5662 \\
ID 6 & 6.28e-16 & 2.50e-15 & 3.90e-16 & $640 \pm 138$& $5945 \pm 90$ & 1.7013 & 1.7062 & 1.00e-16 & 1.11e-16 & $383 \pm 50$ & $647 \pm 60$ & 1.2988 & 1.2977 & $<$4.2e-17 & 721 & 1.2593 \\
ID 7 & 8.22e-18 & -        & 1.29e-17 & $468 \pm 80$ & -             & 2.2663 & 2.2686 & 1.16e-17 & -        & $724 \pm 100$& -            & 1.7289 & -      & $<$1.0e-16 & 480 & 1.6760 \\
ID 8 & 2.83e-16 & -        & 1.27e-16 & $529 \pm 80$ & -             & 2.3281 & 2.3330 & 1.38e-18 & -        & $374 \pm 60$ & -            & 1.7776 & -      & 1.88e-17 & 439 & 1.7256 \\
\hline
\end{tabular}}

%% file: Main.bbl
\begin{thebibliography}{}
\makeatletter
\relax
\def\mn@urlcharsother{\let\do\@makeother \do\$\do\&\do\#\do\^\do\_\do\%\do\~}
\def\mn@doi{\begingroup\mn@urlcharsother \@ifnextchar [ {\mn@doi@}
  {\mn@doi@[]}}
\def\mn@doi@[#1]#2{\def\@tempa{#1}\ifx\@tempa\@empty \href
  {http://dx.doi.org/#2} {doi:#2}\else \href {http://dx.doi.org/#2} {#1}\fi
  \endgroup}
\def\mn@eprint#1#2{\mn@eprint@#1:#2::\@nil}
\def\mn@eprint@arXiv#1{\href {http://arxiv.org/abs/#1} {{\tt arXiv:#1}}}
\def\mn@eprint@dblp#1{\href {http://dblp.uni-trier.de/rec/bibtex/#1.xml}
  {dblp:#1}}
\def\mn@eprint@#1:#2:#3:#4\@nil{\def\@tempa {#1}\def\@tempb {#2}\def\@tempc
  {#3}\ifx \@tempc \@empty \let \@tempc \@tempb \let \@tempb \@tempa \fi \ifx
  \@tempb \@empty \def\@tempb {arXiv}\fi \@ifundefined
  {mn@eprint@\@tempb}{\@tempb:\@tempc}{\expandafter \expandafter \csname
  mn@eprint@\@tempb\endcsname \expandafter{\@tempc}}}

\bibitem[\protect\citeauthoryear{{Aird} et~al.,}{{Aird} et~al.}{2015}]{Aird15}
{Aird} J.,  et~al., 2015, \mn@doi [\apj] {10.1088/0004-637X/815/1/66}, \href
  {http://adsabs.harvard.edu/abs/2015ApJ...815...66A} {815, 66}

\bibitem[\protect\citeauthoryear{{Alatalo} et~al.,}{{Alatalo}
  et~al.}{2015}]{Alatalo15}
{Alatalo} K.,  et~al., 2015, \mn@doi [\apj] {10.1088/0004-637X/798/1/31}, \href
  {https://ui.adsabs.harvard.edu/abs/2015ApJ...798...31A} {798, 31}

\bibitem[\protect\citeauthoryear{{Alexander} \& {Hickox}}{{Alexander} \&
  {Hickox}}{2012}]{Alexander12}
{Alexander} D.~M.,  {Hickox} R.~C.,  2012, \mn@doi [\nar]
  {10.1016/j.newar.2011.11.003}, \href
  {http://adsabs.harvard.edu/abs/2012NewAR..56...93A} {56, 93}

\bibitem[\protect\citeauthoryear{{Alexander}, {Swinbank}, {Smail}, {McDermid}
  \& {Nesvadba}}{{Alexander} et~al.}{2010}]{Alexander10}
{Alexander} D.~M.,  {Swinbank} A.~M.,  {Smail} I.,  {McDermid} R.,   {Nesvadba}
  N.~P.~H.,  2010, \mn@doi [\mnras] {10.1111/j.1365-2966.2009.16046.x}, \href
  {https://ui.adsabs.harvard.edu/abs/2010MNRAS.402.2211A} {402, 2211}

\bibitem[\protect\citeauthoryear{{Asmus}, {Gandhi}, {Smette}, {H{\"o}nig}  \&
  {Duschl}}{{Asmus} et~al.}{2011}]{Asmus11}
{Asmus} D.,  {Gandhi} P.,  {Smette} A.,  {H{\"o}nig} S.~F.,   {Duschl} W.~J.,
  2011, \mn@doi [\aap] {10.1051/0004-6361/201116693}, \href
  {https://ui.adsabs.harvard.edu/abs/2011A&A...536A..36A} {536, A36}

\bibitem[\protect\citeauthoryear{{Balmaverde} \& {Capetti}}{{Balmaverde} \&
  {Capetti}}{2015}]{Balmaverde15}
{Balmaverde} B.,  {Capetti} A.,  2015, \mn@doi [\aap]
  {10.1051/0004-6361/201526496}, \href
  {http://adsabs.harvard.edu/abs/2015A%26A...581A..76B} {581, A76}

\bibitem[\protect\citeauthoryear{{Barro} et~al.,}{{Barro}
  et~al.}{2017}]{Barro17}
{Barro} G.,  et~al., 2017, \mn@doi [\apj] {10.3847/1538-4357/aa6b05}, \href
  {https://ui.adsabs.harvard.edu/abs/2017ApJ...840...47B} {840, 47}

\bibitem[\protect\citeauthoryear{{Beckmann} et~al.,}{{Beckmann}
  et~al.}{2017}]{Beckmann17}
{Beckmann} R.~S.,  et~al., 2017, preprint, \href
  {http://adsabs.harvard.edu/abs/2017arXiv170107838B} {} (\mn@eprint {arXiv}
  {1701.07838})

\bibitem[\protect\citeauthoryear{{Bonnet} et~al.,}{{Bonnet}
  et~al.}{2004}]{Bonnet04}
{Bonnet} H.,  et~al., 2004, in {Bonaccini Calia} D.,  {Ellerbroek} B.~L.,
  {Ragazzoni} R.,  eds,  Society of Photo-Optical Instrumentation Engineers
  (SPIE) Conference Series Vol. 5490, Advancements in Adaptive Optics. pp
  130--138, \mn@doi{10.1117/12.551187}

\bibitem[\protect\citeauthoryear{{Brusa} et~al.,}{{Brusa}
  et~al.}{2015}]{Brusa15}
{Brusa} M.,  et~al., 2015, \mn@doi [\mnras] {10.1093/mnras/stu2117}, \href
  {https://ui.adsabs.harvard.edu/abs/2015MNRAS.446.2394B} {446, 2394}

\bibitem[\protect\citeauthoryear{{Brusa} et~al.,}{{Brusa}
  et~al.}{2018}]{Brusa18}
{Brusa} M.,  et~al., 2018, \mn@doi [\aap] {10.1051/0004-6361/201731641}, \href
  {http://adsabs.harvard.edu/abs/2018A%26A...612A..29B} {612, A29}

\bibitem[\protect\citeauthoryear{{Bruzual} \& {Charlot}}{{Bruzual} \&
  {Charlot}}{2003}]{bruzual03}
{Bruzual} G.,  {Charlot} S.,  2003, \mn@doi [\mnras]
  {10.1046/j.1365-8711.2003.06897.x}, \href
  {http://adsabs.harvard.edu/abs/2003MNRAS.344.1000B} {344, 1000}

\bibitem[\protect\citeauthoryear{{Burgarella} et~al.,}{{Burgarella}
  et~al.}{2013}]{Burgarella13}
{Burgarella} D.,  et~al., 2013, \mn@doi [\aap] {10.1051/0004-6361/201321651},
  \href {https://ui.adsabs.harvard.edu/abs/2013A&A...554A..70B} {554, A70}

\bibitem[\protect\citeauthoryear{{Calzetti}}{{Calzetti}}{2013}]{Calzetti13}
{Calzetti} D.,  2013, {Star Formation Rate Indicators}.
p.~419

\bibitem[\protect\citeauthoryear{{Calzetti}, {Armus}, {Bohlin}, {Kinney},
  {Koornneef}  \& {Storchi-Bergmann}}{{Calzetti} et~al.}{2000}]{Calzetti00}
{Calzetti} D.,  {Armus} L.,  {Bohlin} R.~C.,  {Kinney} A.~L.,  {Koornneef} J.,
   {Storchi-Bergmann} T.,  2000, \mn@doi [\apj] {10.1086/308692}, \href
  {http://adsabs.harvard.edu/abs/2000ApJ...533..682C} {533, 682}

\bibitem[\protect\citeauthoryear{{Cano-D{\'{\i}}az}, {Maiolino}, {Marconi},
  {Netzer}, {Shemmer}  \& {Cresci}}{{Cano-D{\'{\i}}az}
  et~al.}{2012}]{Canodiaz12}
{Cano-D{\'{\i}}az} M.,  {Maiolino} R.,  {Marconi} A.,  {Netzer} H.,  {Shemmer}
  O.,   {Cresci} G.,  2012, \mn@doi [\aap] {10.1051/0004-6361/201118358}, \href
  {http://adsabs.harvard.edu/abs/2012A%26A...537L...8C} {537, L8}

\bibitem[\protect\citeauthoryear{{Cardamone} et~al.,}{{Cardamone}
  et~al.}{2010}]{Cardamone10}
{Cardamone} C.~N.,  et~al., 2010, \mn@doi [\apjs]
  {10.1088/0067-0049/189/2/270}, \href
  {http://adsabs.harvard.edu/abs/2010ApJS..189..270C} {189, 270}

\bibitem[\protect\citeauthoryear{{Carniani} et~al.,}{{Carniani}
  et~al.}{2015}]{Carniani15}
{Carniani} S.,  et~al., 2015, \mn@doi [\aap] {10.1051/0004-6361/201526557},
  \href {http://adsabs.harvard.edu/abs/2015A%26A...580A.102C} {580, A102}

\bibitem[\protect\citeauthoryear{{Carniani} et~al.,}{{Carniani}
  et~al.}{2016}]{Carniani16}
{Carniani} S.,  et~al., 2016, \mn@doi [\aap] {10.1051/0004-6361/201528037},
  \href {http://adsabs.harvard.edu/abs/2016A%26A...591A..28C} {591, A28}

\bibitem[\protect\citeauthoryear{{Casey}, {Narayanan}  \& {Cooray}}{{Casey}
  et~al.}{2014}]{Casey14}
{Casey} C.~M.,  {Narayanan} D.,   {Cooray} A.,  2014, \mn@doi [\physrep]
  {10.1016/j.physrep.2014.02.009}, \href
  {http://adsabs.harvard.edu/abs/2014PhR...541...45C} {541, 45}

\bibitem[\protect\citeauthoryear{{Chabrier}}{{Chabrier}}{2003}]{Chabrier03}
{Chabrier} G.,  2003, \mn@doi [\pasp] {10.1086/376392}, \href
  {http://adsabs.harvard.edu/abs/2003PASP..115..763C} {115, 763}

\bibitem[\protect\citeauthoryear{{Chen} et~al.,}{{Chen} et~al.}{2017}]{Chen17}
{Chen} C.-C.,  et~al., 2017, \mn@doi [\apj] {10.3847/1538-4357/aa863a}, \href
  {http://adsabs.harvard.edu/abs/2017ApJ...846..108C} {846, 108}

\bibitem[\protect\citeauthoryear{{Chen} et~al.,}{{Chen} et~al.}{2019}]{Chen19}
{Chen} C.-C.,  et~al., 2019, \mn@doi [\apj] {10.3847/1538-4357/aa863a}, \href
  {http://adsabs.harvard.edu/abs/2017ApJ...846..108C} {846, 108}

\bibitem[\protect\citeauthoryear{{Choi}, {Somerville}, {Ostriker}, {Naab}  \&
  {Hirschmann}}{{Choi} et~al.}{2018}]{Choi18}
{Choi} E.,  {Somerville} R.~S.,  {Ostriker} J.~P.,  {Naab} T.,   {Hirschmann}
  M.,  2018, \mn@doi [\apj] {10.3847/1538-4357/aae076}, \href
  {https://ui.adsabs.harvard.edu/abs/2018ApJ...866...91C} {866, 91}

\bibitem[\protect\citeauthoryear{{Cicone}, {Feruglio}, {Maiolino}, {Fiore},
  {Piconcelli}, {Menci}, {Aussel}  \& {Sturm}}{{Cicone}
  et~al.}{2012}]{Cicone12}
{Cicone} C.,  {Feruglio} C.,  {Maiolino} R.,  {Fiore} F.,  {Piconcelli} E.,
  {Menci} N.,  {Aussel} H.,   {Sturm} E.,  2012, \mn@doi [\aap]
  {10.1051/0004-6361/201218793}, \href
  {http://adsabs.harvard.edu/abs/2012A%26A...543A..99C} {543, A99}

\bibitem[\protect\citeauthoryear{{Cicone} et~al.,}{{Cicone}
  et~al.}{2014}]{Cicone14}
{Cicone} C.,  et~al., 2014, \mn@doi [\aap] {10.1051/0004-6361/201322464}, \href
  {http://adsabs.harvard.edu/abs/2014A%26A...562A..21C} {562, A21}

\bibitem[\protect\citeauthoryear{{Cicone}, {Brusa}, {Ramos Almeida}, {Cresci},
  {Husemann}  \& {Mainieri}}{{Cicone} et~al.}{2018}]{Cicone18}
{Cicone} C.,  {Brusa} M.,  {Ramos Almeida} C.,  {Cresci} G.,  {Husemann} B.,
  {Mainieri} V.,  2018, \mn@doi [Nature Astronomy] {10.1038/s41550-018-0406-3},
  \href {https://ui.adsabs.harvard.edu/abs/2018NatAs...2..176C} {2, 176}

\bibitem[\protect\citeauthoryear{{Circosta} et~al.,}{{Circosta}
  et~al.}{2018}]{Circosta18}
{Circosta} C.,  et~al., 2018, \mn@doi [\aap] {10.1051/0004-6361/201833520},
  \href {https://ui.adsabs.harvard.edu/abs/2018A&A...620A..82C} {620, A82}

\bibitem[\protect\citeauthoryear{{Condon}}{{Condon}}{1997}]{Condon97}
{Condon} J.~J.,  1997, \mn@doi [\pasp] {10.1086/133871}, \href
  {http://adsabs.harvard.edu/abs/1997PASP..109..166C} {109, 166}

\bibitem[\protect\citeauthoryear{{Crain} et~al.,}{{Crain}
  et~al.}{2015}]{Crain15}
{Crain} R.~A.,  et~al., 2015, \mn@doi [\mnras] {10.1093/mnras/stv725}, \href
  {http://adsabs.harvard.edu/abs/2015MNRAS.450.1937C} {450, 1937}

\bibitem[\protect\citeauthoryear{{Cresci} \& {Maiolino}}{{Cresci} \&
  {Maiolino}}{2018}]{Cresci18}
{Cresci} G.,  {Maiolino} R.,  2018, \mn@doi [Nature Astronomy]
  {10.1038/s41550-018-0404-5}, \href
  {https://ui.adsabs.harvard.edu/abs/2018NatAs...2..179C} {2, 179}

\bibitem[\protect\citeauthoryear{{Cresci} et~al.,}{{Cresci}
  et~al.}{2015a}]{Cresci15b}
{Cresci} G.,  et~al., 2015a, \mn@doi [\aap] {10.1051/0004-6361/201526581},
  \href {https://ui.adsabs.harvard.edu/abs/2015A&A...582A..63C} {582, A63}

\bibitem[\protect\citeauthoryear{{Cresci} et~al.,}{{Cresci}
  et~al.}{2015b}]{Cresci15}
{Cresci} G.,  et~al., 2015b, \mn@doi [\apj] {10.1088/0004-637X/799/1/82}, \href
  {http://adsabs.harvard.edu/abs/2015ApJ...799...82C} {799, 82}

\bibitem[\protect\citeauthoryear{{Croft} et~al.,}{{Croft}
  et~al.}{2006}]{Croft06}
{Croft} S.,  et~al., 2006, \mn@doi [\apj] {10.1086/505526}, \href
  {https://ui.adsabs.harvard.edu/abs/2006ApJ...647.1040C} {647, 1040}

\bibitem[\protect\citeauthoryear{{D'Agostino} et~al.,}{{D'Agostino}
  et~al.}{2019}]{DAgostino19}
{D'Agostino} J.~J.,  et~al., 2019, \mn@doi [\mnras] {10.1093/mnras/stz1611},
  \href {https://ui.adsabs.harvard.edu/abs/2019MNRAS.487.4153D} {487, 4153}

\bibitem[\protect\citeauthoryear{{Dale}, {Helou}, {Magdis}, {Rigopoulou},
  {5MUSES}  \& {HerMES}}{{Dale} et~al.}{2014}]{dale14}
{Dale} D.~A.,  {Helou} G.,  {Magdis} G.,  {Rigopoulou} D.,  {5MUSES}  {HerMES}
  2014, in American Astronomical Society Meeting Abstracts \#223. p. 453.01

\bibitem[\protect\citeauthoryear{{Davies} et~al.,}{{Davies}
  et~al.}{2013}]{Davies13}
{Davies} R.~I.,  et~al., 2013, \mn@doi [\aap] {10.1051/0004-6361/201322282},
  \href {https://ui.adsabs.harvard.edu/abs/2013A&A...558A..56D} {558, A56}

\bibitem[\protect\citeauthoryear{{Davies}, {Rich}, {Kewley}  \&
  {Dopita}}{{Davies} et~al.}{2014a}]{Davies14a}
{Davies} R.~L.,  {Rich} J.~A.,  {Kewley} L.~J.,   {Dopita} M.~A.,  2014a,
  \mn@doi [\mnras] {10.1093/mnras/stu234}, \href
  {https://ui.adsabs.harvard.edu/abs/2014MNRAS.439.3835D} {439, 3835}

\bibitem[\protect\citeauthoryear{{Davies}, {Kewley}, {Ho}  \&
  {Dopita}}{{Davies} et~al.}{2014b}]{Davies14b}
{Davies} R.~L.,  {Kewley} L.~J.,  {Ho} I.~T.,   {Dopita} M.~A.,  2014b, \mn@doi
  [\mnras] {10.1093/mnras/stu1740}, \href
  {https://ui.adsabs.harvard.edu/abs/2014MNRAS.444.3961D} {444, 3961}

\bibitem[\protect\citeauthoryear{{Di Matteo}, {Springel}  \& {Hernquist}}{{Di
  Matteo} et~al.}{2005}]{DiMatteo05}
{Di Matteo} T.,  {Springel} V.,   {Hernquist} L.,  2005, \mn@doi [\nat]
  {10.1038/nature03335}, \href
  {http://adsabs.harvard.edu/abs/2005Natur.433..604D} {433, 604}

\bibitem[\protect\citeauthoryear{{Dimitrijevi{\'c}}, {Popovi{\'c}}, {Kova{\v
  c}evi{\'c}}, {Da{\v c}i{\'c}}  \& {Ili{\'c}}}{{Dimitrijevi{\'c}}
  et~al.}{2007}]{Dimitrijevic07}
{Dimitrijevi{\'c}} M.~S.,  {Popovi{\'c}} L.~{\v C}.,  {Kova{\v c}evi{\'c}} J.,
  {Da{\v c}i{\'c}} M.,   {Ili{\'c}} D.,  2007, \mn@doi [\mnras]
  {10.1111/j.1365-2966.2006.11238.x}, \href
  {http://adsabs.harvard.edu/abs/2007MNRAS.374.1181D} {374, 1181}

\bibitem[\protect\citeauthoryear{{Dunlop} et~al.,}{{Dunlop}
  et~al.}{2017}]{Dunlop17}
{Dunlop} J.~S.,  et~al., 2017, \mn@doi [\mnras] {10.1093/mnras/stw3088}, \href
  {http://adsabs.harvard.edu/abs/2017MNRAS.466..861D} {466, 861}

\bibitem[\protect\citeauthoryear{{Eisenhauer} et~al.,}{{Eisenhauer}
  et~al.}{2003}]{Eisenhauer03}
{Eisenhauer} F.,  et~al., 2003, in {Iye} M.,  {Moorwood} A. F.~M.,  eds,
  Society of Photo-Optical Instrumentation Engineers (SPIE) Conference Series
  Vol. 4841, Instrument Design and Performance for Optical/Infrared
  Ground-based Telescopes. pp 1548--1561 (\mn@eprint {arXiv}
  {astro-ph/0306191}), \mn@doi{10.1117/12.459468}

\bibitem[\protect\citeauthoryear{{Elbaz} et~al.,}{{Elbaz}
  et~al.}{2018}]{Elbaz18}
{Elbaz} D.,  et~al., 2018, \mn@doi [\aap] {10.1051/0004-6361/201732370}, \href
  {http://adsabs.harvard.edu/abs/2018A%26A...616A.110E} {616, A110}

\bibitem[\protect\citeauthoryear{{Feruglio} et~al.,}{{Feruglio}
  et~al.}{2015}]{Feruglio15}
{Feruglio} C.,  et~al., 2015, \mn@doi [\aap] {10.1051/0004-6361/201526020},
  \href {https://ui.adsabs.harvard.edu/abs/2015A&A...583A..99F} {583, A99}

\bibitem[\protect\citeauthoryear{{Fischer} et~al.,}{{Fischer}
  et~al.}{2019}]{Fischer19}
{Fischer} T.~C.,  et~al., 2019, \mn@doi [\apj] {10.3847/1538-4357/ab11c3},
  \href {https://ui.adsabs.harvard.edu/abs/2019ApJ...875..102F} {875, 102}

\bibitem[\protect\citeauthoryear{{Fluetsch} et~al.,}{{Fluetsch}
  et~al.}{2019}]{Fluetsch19}
{Fluetsch} A.,  et~al., 2019, \mn@doi [\mnras] {10.1093/mnras/sty3449}, \href
  {https://ui.adsabs.harvard.edu/abs/2019MNRAS.483.4586F} {483, 4586}

\bibitem[\protect\citeauthoryear{{F{\"o}rster Schreiber} et~al.,}{{F{\"o}rster
  Schreiber} et~al.}{2009}]{FSchreiber09}
{F{\"o}rster Schreiber} N.~M.,  et~al., 2009, \mn@doi [\apj]
  {10.1088/0004-637X/706/2/1364}, \href
  {http://adsabs.harvard.edu/abs/2009ApJ...706.1364F} {706, 1364}

\bibitem[\protect\citeauthoryear{{F{\"o}rster Schreiber} et~al.,}{{F{\"o}rster
  Schreiber} et~al.}{2018a}]{ForsterSch18b}
{F{\"o}rster Schreiber} N.~M.,  et~al., 2018a, arXiv e-prints, \href
  {http://adsabs.harvard.edu/abs/2018arXiv180704738F} {}

\bibitem[\protect\citeauthoryear{{F{\"o}rster Schreiber} et~al.,}{{F{\"o}rster
  Schreiber} et~al.}{2018b}]{ForsterSch18a}
{F{\"o}rster Schreiber} N.~M.,  et~al., 2018b, \mn@doi [\apjs]
  {10.3847/1538-4365/aadd49}, \href
  {http://adsabs.harvard.edu/abs/2018ApJS..238...21F} {238, 21}

\bibitem[\protect\citeauthoryear{{Freudling}, {Romaniello}, {Bramich},
  {Ballester}, {Forchi}, {Garc{\'\i}a-Dabl{\'o}}, {Moehler}  \&
  {Neeser}}{{Freudling} et~al.}{2013}]{Freudling13}
{Freudling} W.,  {Romaniello} M.,  {Bramich} D.~M.,  {Ballester} P.,  {Forchi}
  V.,  {Garc{\'\i}a-Dabl{\'o}} C.~E.,  {Moehler} S.,   {Neeser} M.~J.,  2013,
  \mn@doi [\aap] {10.1051/0004-6361/201322494}, \href
  {https://ui.adsabs.harvard.edu/abs/2013A&A...559A..96F} {559, A96}

\bibitem[\protect\citeauthoryear{{Fujimoto} et~al.,}{{Fujimoto}
  et~al.}{2018}]{Fujimoto18}
{Fujimoto} S.,  et~al., 2018, \mn@doi [\apj] {10.3847/1538-4357/aac6c4}, \href
  {http://adsabs.harvard.edu/abs/2018ApJ...861....7F} {861, 7}

\bibitem[\protect\citeauthoryear{{Gabor} \& {Bournaud}}{{Gabor} \&
  {Bournaud}}{2014}]{Gabor14}
{Gabor} J.~M.,  {Bournaud} F.,  2014, \mn@doi [\mnras] {10.1093/mnras/stu677},
  \href {https://ui.adsabs.harvard.edu/abs/2014MNRAS.441.1615G} {441, 1615}

\bibitem[\protect\citeauthoryear{{Gallagher}, {Maiolino}, {Belfiore}, {Drory},
  {Riffel}  \& {Riffel}}{{Gallagher} et~al.}{2019}]{Gallagher19}
{Gallagher} R.,  {Maiolino} R.,  {Belfiore} F.,  {Drory} N.,  {Riffel} R.,
  {Riffel} R.~A.,  2019, \mn@doi [\mnras] {10.1093/mnras/stz564}, \href
  {https://ui.adsabs.harvard.edu/abs/2019MNRAS.485.3409G} {485, 3409}

\bibitem[\protect\citeauthoryear{{Ganguly} \& {Brotherton}}{{Ganguly} \&
  {Brotherton}}{2008}]{Ganguly08}
{Ganguly} R.,  {Brotherton} M.~S.,  2008, \mn@doi [\apj] {10.1086/524106},
  \href {http://adsabs.harvard.edu/abs/2008ApJ...672..102G} {672, 102}

\bibitem[\protect\citeauthoryear{{Genzel} et~al.,}{{Genzel}
  et~al.}{2014}]{Genzel14}
{Genzel} R.,  et~al., 2014, \mn@doi [\apj] {10.1088/0004-637X/796/1/7}, \href
  {http://adsabs.harvard.edu/abs/2014ApJ...796....7G} {796, 7}

\bibitem[\protect\citeauthoryear{{Guo} et~al.,}{{Guo} et~al.}{2013}]{Guo13}
{Guo} Y.,  et~al., 2013, \mn@doi [\apjs] {10.1088/0067-0049/207/2/24}, \href
  {http://adsabs.harvard.edu/abs/2013ApJS..207...24G} {207, 24}

\bibitem[\protect\citeauthoryear{{Hao}, {Kennicutt}, {Johnson}, {Calzetti},
  {Dale}  \& {Moustakas}}{{Hao} et~al.}{2011}]{Hao11}
{Hao} C.-N.,  {Kennicutt} R.~C.,  {Johnson} B.~D.,  {Calzetti} D.,  {Dale}
  D.~A.,   {Moustakas} J.,  2011, \mn@doi [\apj] {10.1088/0004-637X/741/2/124},
  \href {https://ui.adsabs.harvard.edu/abs/2011ApJ...741..124H} {741, 124}

\bibitem[\protect\citeauthoryear{{Harrison}}{{Harrison}}{2017}]{Harrison17}
{Harrison} C.~M.,  2017, \mn@doi [Nature Astronomy] {10.1038/s41550-017-0165},
  \href {http://adsabs.harvard.edu/abs/2017NatAs...1E.165H} {1, 0165}

\bibitem[\protect\citeauthoryear{{Harrison} et~al.,}{{Harrison}
  et~al.}{2012}]{Harrison12}
{Harrison} C.~M.,  et~al., 2012, \mn@doi [\apjl] {10.1088/2041-8205/760/1/L15},
  \href {http://adsabs.harvard.edu/abs/2012ApJ...760L..15H} {760, L15}

\bibitem[\protect\citeauthoryear{{Harrison} et~al.,}{{Harrison}
  et~al.}{2016a}]{Harrison16}
{Harrison} C.~M.,  et~al., 2016a, \mn@doi [\mnras] {10.1093/mnras/stv2727},
  \href {http://adsabs.harvard.edu/abs/2016MNRAS.456.1195H} {456, 1195}

\bibitem[\protect\citeauthoryear{{Harrison} et~al.,}{{Harrison}
  et~al.}{2016b}]{Harrison16Alm}
{Harrison} C.~M.,  et~al., 2016b, \mn@doi [\mnras] {10.1093/mnrasl/slw001},
  \href {http://adsabs.harvard.edu/abs/2016MNRAS.457L.122H} {457, L122}

\bibitem[\protect\citeauthoryear{{Harrison}, {Costa}, {Tadhunter},
  {Fl{\"u}tsch}, {Kakkad}, {Perna}  \& {Vietri}}{{Harrison}
  et~al.}{2018}]{Harrison18}
{Harrison} C.~M.,  {Costa} T.,  {Tadhunter} C.~N.,  {Fl{\"u}tsch} A.,  {Kakkad}
  D.,  {Perna} M.,   {Vietri} G.,  2018, \mn@doi [Nature Astronomy]
  {10.1038/s41550-018-0403-6}, \href
  {https://ui.adsabs.harvard.edu/abs/2018NatAs...2..198H} {2, 198}

\bibitem[\protect\citeauthoryear{{Hickox} \& {Alexander}}{{Hickox} \&
  {Alexander}}{2018}]{Hickox18}
{Hickox} R.~C.,  {Alexander} D.~M.,  2018, \mn@doi [\araa]
  {10.1146/annurev-astro-081817-051803}, \href
  {https://ui.adsabs.harvard.edu/abs/2018ARA&A..56..625H} {56, 625}

\bibitem[\protect\citeauthoryear{{Hirschmann}, {Dolag}, {Saro}, {Bachmann},
  {Borgani}  \& {Burkert}}{{Hirschmann} et~al.}{2014}]{Hirschmann14}
{Hirschmann} M.,  {Dolag} K.,  {Saro} A.,  {Bachmann} L.,  {Borgani} S.,
  {Burkert} A.,  2014, \mn@doi [\mnras] {10.1093/mnras/stu1023}, \href
  {https://ui.adsabs.harvard.edu/abs/2014MNRAS.442.2304H} {442, 2304}

\bibitem[\protect\citeauthoryear{{Hodge} et~al.,}{{Hodge}
  et~al.}{2013}]{Hodge13}
{Hodge} J.~A.,  et~al., 2013, \mn@doi [\apj] {10.1088/0004-637X/768/1/91},
  \href {http://adsabs.harvard.edu/abs/2013ApJ...768...91H} {768, 91}

\bibitem[\protect\citeauthoryear{{Hodge} et~al.,}{{Hodge}
  et~al.}{2016}]{Hodge16}
{Hodge} J.~A.,  et~al., 2016, \mn@doi [\apj] {10.3847/1538-4357/833/1/103},
  \href {http://adsabs.harvard.edu/abs/2016ApJ...833..103H} {833, 103}

\bibitem[\protect\citeauthoryear{{Hsu} et~al.,}{{Hsu} et~al.}{2014}]{Hsu14}
{Hsu} L.-T.,  et~al., 2014, \mn@doi [\apj] {10.1088/0004-637X/796/1/60}, \href
  {http://adsabs.harvard.edu/abs/2014ApJ...796...60H} {796, 60}

\bibitem[\protect\citeauthoryear{{Husemann}, {Scharw{\"a}chter}, {Bennert},
  {Mainieri}, {Woo}  \& {Kakkad}}{{Husemann} et~al.}{2016}]{Husemann16}
{Husemann} B.,  {Scharw{\"a}chter} J.,  {Bennert} V.~N.,  {Mainieri} V.,  {Woo}
  J.~H.,   {Kakkad} D.,  2016, \mn@doi [\aap] {10.1051/0004-6361/201527992},
  \href {https://ui.adsabs.harvard.edu/abs/2016A&A...594A..44H} {594, A44}

\bibitem[\protect\citeauthoryear{{Husemann} et~al.,}{{Husemann}
  et~al.}{2019}]{Husemann19}
{Husemann} B.,  et~al., 2019, \mn@doi [\aap] {10.1051/0004-6361/201935283},
  \href {https://ui.adsabs.harvard.edu/abs/2019A&A...627A..53H} {627, A53}

\bibitem[\protect\citeauthoryear{{Huynh}, {Hopkins}, {Lenc}, {Mao},
  {Middelberg}, {Norris}  \& {Randall}}{{Huynh} et~al.}{2012}]{Huynh12}
{Huynh} M.~T.,  {Hopkins} A.~M.,  {Lenc} E.,  {Mao} M.~Y.,  {Middelberg} E.,
  {Norris} R.~P.,   {Randall} K.~E.,  2012, \mn@doi [Monthly Notices of the
  Royal Astronomical Society] {10.1111/j.1365-2966.2012.21894.x}, \href
  {https://ui.adsabs.harvard.edu/abs/2012MNRAS.426.2342H} {426, 2342}

\bibitem[\protect\citeauthoryear{{Ikarashi} et~al.,}{{Ikarashi}
  et~al.}{2015}]{Ikarashi15}
{Ikarashi} S.,  et~al., 2015, \mn@doi [\apj] {10.1088/0004-637X/810/2/133},
  \href {http://adsabs.harvard.edu/abs/2015ApJ...810..133I} {810, 133}

\bibitem[\protect\citeauthoryear{{Jarvis} et~al.,}{{Jarvis}
  et~al.}{2019}]{Jarvis19}
{Jarvis} M.~E.,  et~al., 2019, \mn@doi [\mnras] {10.1093/mnras/stz556}, \href
  {https://ui.adsabs.harvard.edu/abs/2019MNRAS.485.2710J} {485, 2710}

\bibitem[\protect\citeauthoryear{{Jin} et~al.,}{{Jin} et~al.}{2018}]{Jin18}
{Jin} S.,  et~al., 2018, \mn@doi [\apj] {10.3847/1538-4357/aad4af}, \href
  {https://ui.adsabs.harvard.edu/abs/2018ApJ...864...56J} {864, 56}

\bibitem[\protect\citeauthoryear{{Kakkad} et~al.,}{{Kakkad}
  et~al.}{2016}]{Kakkad16}
{Kakkad} D.,  et~al., 2016, \mn@doi [\aap] {10.1051/0004-6361/201527968}, \href
  {https://ui.adsabs.harvard.edu/abs/2016A&A...592A.148K} {592, A148}

\bibitem[\protect\citeauthoryear{{Kang}, {Woo}  \& {Bae}}{{Kang}
  et~al.}{2017}]{Kang17}
{Kang} D.,  {Woo} J.-H.,   {Bae} H.-J.,  2017, \mn@doi [\apj]
  {10.3847/1538-4357/aa80e8}, \href
  {https://ui.adsabs.harvard.edu/abs/2017ApJ...845..131K} {845, 131}

\bibitem[\protect\citeauthoryear{{Karouzos}, {Woo}  \& {Bae}}{{Karouzos}
  et~al.}{2016}]{Karouzos16}
{Karouzos} M.,  {Woo} J.-H.,   {Bae} H.-J.,  2016, \mn@doi [\apj]
  {10.3847/0004-637X/819/2/148}, \href
  {https://ui.adsabs.harvard.edu/abs/2016ApJ...819..148K} {819, 148}

\bibitem[\protect\citeauthoryear{{Kashino} et~al.,}{{Kashino}
  et~al.}{2013}]{Kashino13}
{Kashino} D.,  et~al., 2013, \mn@doi [\apj] {10.1088/2041-8205/777/1/L8}, \href
  {https://ui.adsabs.harvard.edu/abs/2013ApJ...777L...8K} {777, L8}

\bibitem[\protect\citeauthoryear{{Kennicutt} \& {Evans}}{{Kennicutt} \&
  {Evans}}{2012}]{Kennicutt12}
{Kennicutt} R.~C.,  {Evans} N.~J.,  2012, \mn@doi [\araa]
  {10.1146/annurev-astro-081811-125610}, \href
  {https://ui.adsabs.harvard.edu/abs/2012ARA&A..50..531K} {50, 531}

\bibitem[\protect\citeauthoryear{{Kirkpatrick}, {Sharon}, {Keller}  \&
  {Pope}}{{Kirkpatrick} et~al.}{2019}]{Kirkpatrick19}
{Kirkpatrick} A.,  {Sharon} C.,  {Keller} E.,   {Pope} A.,  2019, \mn@doi
  [\apj] {10.3847/1538-4357/ab223a}, \href
  {https://ui.adsabs.harvard.edu/abs/2019ApJ...879...41K} {879, 41}

\bibitem[\protect\citeauthoryear{{Kocevski} et~al.,}{{Kocevski}
  et~al.}{2018}]{Kocevski18}
{Kocevski} D.~D.,  et~al., 2018, \mn@doi [\apjs] {10.3847/1538-4365/aab9b4},
  \href {http://adsabs.harvard.edu/abs/2018ApJS..236...48K} {236, 48}

\bibitem[\protect\citeauthoryear{{Koekemoer} et~al.,}{{Koekemoer}
  et~al.}{2007}]{Koekemoer07}
{Koekemoer} A.~M.,  et~al., 2007, \mn@doi [\apjs] {10.1086/520086}, \href
  {https://ui.adsabs.harvard.edu/abs/2007ApJS..172..196K} {172, 196}

\bibitem[\protect\citeauthoryear{{Kormendy} \& {Ho}}{{Kormendy} \&
  {Ho}}{2013}]{Kormendy13}
{Kormendy} J.,  {Ho} L.~C.,  2013, \mn@doi [\araa]
  {10.1146/annurev-astro-082708-101811}, \href
  {http://adsabs.harvard.edu/abs/2013ARA%26A..51..511K} {51, 511}

\bibitem[\protect\citeauthoryear{{Laigle} et~al.,}{{Laigle}
  et~al.}{2016}]{laigle16}
{Laigle} C.,  et~al., 2016, \mn@doi [\apjs] {10.3847/0067-0049/224/2/24}, \href
  {http://adsabs.harvard.edu/abs/2016ApJS..224...24L} {224, 24}

\bibitem[\protect\citeauthoryear{{Lang} et~al.,}{{Lang} et~al.}{2019}]{Lang19}
{Lang} P.,  et~al., 2019, \mn@doi [\apj] {10.3847/1538-4357/ab1f77}, \href
  {https://ui.adsabs.harvard.edu/abs/2019ApJ...879...54L} {879, 54}

\bibitem[\protect\citeauthoryear{{Lansbury}, {Jarvis}, {Harrison}, {Alexander},
  {Del Moro}, {Edge}, {Mullaney}  \& {Thomson}}{{Lansbury}
  et~al.}{2018}]{Lansbury18}
{Lansbury} G.~B.,  {Jarvis} M.~E.,  {Harrison} C.~M.,  {Alexander} D.~M.,  {Del
  Moro} A.,  {Edge} A.~C.,  {Mullaney} J.~R.,   {Thomson} A.~P.,  2018, \mn@doi
  [\apjl] {10.3847/2041-8213/aab357}, \href
  {https://ui.adsabs.harvard.edu/abs/2018ApJ...856L...1L} {856, L1}

\bibitem[\protect\citeauthoryear{{Lanzuisi} et~al.,}{{Lanzuisi}
  et~al.}{2017}]{Lanzuisi17}
{Lanzuisi} G.,  et~al., 2017, \mn@doi [\aap] {10.1051/0004-6361/201629955},
  \href {http://adsabs.harvard.edu/abs/2017A%26A...602A.123L} {602, A123}

\bibitem[\protect\citeauthoryear{{Leung} et~al.,}{{Leung}
  et~al.}{2017}]{Leung17}
{Leung} G.~C.~K.,  et~al., 2017, preprint, \href
  {http://adsabs.harvard.edu/abs/2017arXiv170310255L} {} (\mn@eprint {arXiv}
  {1703.10255})

\bibitem[\protect\citeauthoryear{{Liu}, {Zakamska}, {Greene}, {Nesvadba}  \&
  {Liu}}{{Liu} et~al.}{2013}]{Liu13b}
{Liu} G.,  {Zakamska} N.~L.,  {Greene} J.~E.,  {Nesvadba} N. P.~H.,   {Liu} X.,
   2013, \mn@doi [\mnras] {10.1093/mnras/stt1755}, \href
  {https://ui.adsabs.harvard.edu/abs/2013MNRAS.436.2576L} {436, 2576}

\bibitem[\protect\citeauthoryear{{Loiacono}, {Talia}, {Fraternali}, {Cimatti},
  {Di Teodoro}  \& {Caminha}}{{Loiacono} et~al.}{2019}]{Loiacono19}
{Loiacono} F.,  {Talia} M.,  {Fraternali} F.,  {Cimatti} A.,  {Di Teodoro}
  E.~M.,   {Caminha} G.~B.,  2019, \mn@doi [\mnras] {10.1093/mnras/stz2170},
  \href {https://ui.adsabs.harvard.edu/abs/2019MNRAS.tmp.2095L} {p.~2095}

\bibitem[\protect\citeauthoryear{{Lutz} et~al.,}{{Lutz} et~al.}{2011}]{lutz11}
{Lutz} D.,  et~al., 2011, \mn@doi [\aap] {10.1051/0004-6361/201117107}, \href
  {http://adsabs.harvard.edu/abs/2011A%26A...532A..90L} {532, A90}

\bibitem[\protect\citeauthoryear{{Madau} \& {Dickinson}}{{Madau} \&
  {Dickinson}}{2014}]{Madau14}
{Madau} P.,  {Dickinson} M.,  2014, \mn@doi [\araa]
  {10.1146/annurev-astro-081811-125615}, \href
  {http://adsabs.harvard.edu/abs/2014ARA%26A..52..415M} {52, 415}

\bibitem[\protect\citeauthoryear{{Madau}, {Ferguson}, {Dickinson},
  {Giavalisco}, {Steidel}  \& {Fruchter}}{{Madau} et~al.}{1996}]{Madau96}
{Madau} P.,  {Ferguson} H.~C.,  {Dickinson} M.~E.,  {Giavalisco} M.,  {Steidel}
  C.~C.,   {Fruchter} A.,  1996, \mn@doi [\mnras] {10.1093/mnras/283.4.1388},
  \href {http://adsabs.harvard.edu/abs/1996MNRAS.283.1388M} {283, 1388}

\bibitem[\protect\citeauthoryear{{Maiolino} et~al.,}{{Maiolino}
  et~al.}{2017}]{Maiolino17}
{Maiolino} R.,  et~al., 2017, \mn@doi [\nat] {10.1038/nature21677}, \href
  {https://ui.adsabs.harvard.edu/abs/2017Natur.544..202M} {544, 202}

\bibitem[\protect\citeauthoryear{{Marchesi} et~al.,}{{Marchesi}
  et~al.}{2016}]{Marchesi16}
{Marchesi} S.,  et~al., 2016, \mn@doi [\apj] {10.3847/0004-637X/817/1/34},
  \href {http://adsabs.harvard.edu/abs/2016ApJ...817...34M} {817, 34}

\bibitem[\protect\citeauthoryear{{McCarthy}, {Schaye}, {Bower}, {Ponman},
  {Booth}, {Dalla Vecchia}  \& {Springel}}{{McCarthy}
  et~al.}{2011}]{McCarthy11}
{McCarthy} I.~G.,  {Schaye} J.,  {Bower} R.~G.,  {Ponman} T.~J.,  {Booth}
  C.~M.,  {Dalla Vecchia} C.,   {Springel} V.,  2011, \mn@doi [\mnras]
  {10.1111/j.1365-2966.2010.18033.x}, \href
  {http://adsabs.harvard.edu/abs/2011MNRAS.412.1965M} {412, 1965}

\bibitem[\protect\citeauthoryear{{McElroy} et~al.,}{{McElroy}
  et~al.}{2016}]{McElroy16}
{McElroy} R.~E.,  et~al., 2016, \mn@doi [\aap] {10.1051/0004-6361/201629102},
  \href {https://ui.adsabs.harvard.edu/abs/2016A&A...593L...8M} {593, L8}

\bibitem[\protect\citeauthoryear{{Merloni}, {Rudnick}  \& {Di
  Matteo}}{{Merloni} et~al.}{2004}]{Merloni04}
{Merloni} A.,  {Rudnick} G.,   {Di Matteo} T.,  2004, \mn@doi [\mnras]
  {10.1111/j.1365-2966.2004.08382.x}, \href
  {http://adsabs.harvard.edu/abs/2004MNRAS.354L..37M} {354, L37}

\bibitem[\protect\citeauthoryear{{Miller}, {Fomalont}, {Kellermann},
  {Mainieri}, {Norman}, {Padovani}, {Rosati}  \& {Tozzi}}{{Miller}
  et~al.}{2008}]{Miller08}
{Miller} N.~A.,  {Fomalont} E.~B.,  {Kellermann} K.~I.,  {Mainieri} V.,
  {Norman} C.,  {Padovani} P.,  {Rosati} P.,   {Tozzi} P.,  2008, \mn@doi
  [\apjs] {10.1086/591054}, \href
  {http://adsabs.harvard.edu/abs/2008ApJS..179..114M} {179, 114}

\bibitem[\protect\citeauthoryear{{Miller} et~al.,}{{Miller}
  et~al.}{2013}]{Miller13}
{Miller} N.~A.,  et~al., 2013, \mn@doi [The Astrophysical Journal Supplement
  Series] {10.1088/0067-0049/205/2/13}, \href
  {https://ui.adsabs.harvard.edu/abs/2013ApJS..205...13M} {205, 13}

\bibitem[\protect\citeauthoryear{{Morganti}, {Tadhunter}  \&
  {Oosterloo}}{{Morganti} et~al.}{2005}]{Morganti05}
{Morganti} R.,  {Tadhunter} C.~N.,   {Oosterloo} T.~A.,  2005, \mn@doi [\aap]
  {10.1051/0004-6361:200500197}, \href
  {https://ui.adsabs.harvard.edu/abs/2005A&A...444L...9M} {444, L9}

\bibitem[\protect\citeauthoryear{{Mullaney}, {Alexander}, {Goulding}  \&
  {Hickox}}{{Mullaney} et~al.}{2011}]{Mullaney11}
{Mullaney} J.~R.,  {Alexander} D.~M.,  {Goulding} A.~D.,   {Hickox} R.~C.,
  2011, \mn@doi [\mnras] {10.1111/j.1365-2966.2011.18448.x}, \href
  {http://adsabs.harvard.edu/abs/2011MNRAS.414.1082M} {414, 1082}

\bibitem[\protect\citeauthoryear{{Mullaney}, {Alexander}, {Fine}, {Goulding},
  {Harrison}  \& {Hickox}}{{Mullaney} et~al.}{2013}]{Mullaney13}
{Mullaney} J.~R.,  {Alexander} D.~M.,  {Fine} S.,  {Goulding} A.~D.,
  {Harrison} C.~M.,   {Hickox} R.~C.,  2013, \mn@doi [\mnras]
  {10.1093/mnras/stt751}, \href
  {http://adsabs.harvard.edu/abs/2013MNRAS.433..622M} {433, 622}

\bibitem[\protect\citeauthoryear{{Mullaney} et~al.,}{{Mullaney}
  et~al.}{2015}]{Mullaney15}
{Mullaney} J.~R.,  et~al., 2015, \mn@doi [\mnras] {10.1093/mnrasl/slv110},
  \href {http://adsabs.harvard.edu/abs/2015MNRAS.453L..83M} {453, L83}

\bibitem[\protect\citeauthoryear{{Murphy}, {Chary}, {Dickinson}, {Pope},
  {Frayer}  \& {Lin}}{{Murphy} et~al.}{2011}]{Murphy11}
{Murphy} E.~J.,  {Chary} R.~R.,  {Dickinson} M.,  {Pope} A.,  {Frayer} D.~T.,
  {Lin} L.,  2011, \mn@doi [\apj] {10.1088/0004-637X/732/2/126}, \href
  {https://ui.adsabs.harvard.edu/abs/2011ApJ...732..126M} {732, 126}

\bibitem[\protect\citeauthoryear{{Nelson} et~al.,}{{Nelson}
  et~al.}{2012}]{Nelson12}
{Nelson} E.~J.,  et~al., 2012, \mn@doi [\apjl] {10.1088/2041-8205/747/2/L28},
  \href {https://ui.adsabs.harvard.edu/abs/2012ApJ...747L..28N} {747, L28}

\bibitem[\protect\citeauthoryear{{Nelson} et~al.,}{{Nelson}
  et~al.}{2019}]{Nelson19}
{Nelson} E.~J.,  et~al., 2019, \mn@doi [\apj] {10.3847/1538-4357/aaf38a}, \href
  {http://adsabs.harvard.edu/abs/2019ApJ...870..130N} {870, 130}

\bibitem[\protect\citeauthoryear{{Oliver} et~al.,}{{Oliver}
  et~al.}{2012}]{Oliver12}
{Oliver} S.~J.,  et~al., 2012, \mn@doi [\mnras]
  {10.1111/j.1365-2966.2012.20912.x}, \href
  {https://ui.adsabs.harvard.edu/abs/2012MNRAS.424.1614O} {424, 1614}

\bibitem[\protect\citeauthoryear{{Osterbrock} \& {Ferland}}{{Osterbrock} \&
  {Ferland}}{2006}]{Osterbrock06}
{Osterbrock} D.~E.,  {Ferland} G.~J.,  2006, {Astrophysics of gaseous nebulae
  and active galactic nuclei}

\bibitem[\protect\citeauthoryear{{Perna} et~al.,}{{Perna}
  et~al.}{2018}]{Perna18}
{Perna} M.,  et~al., 2018, \mn@doi [\aap] {10.1051/0004-6361/201833040}, \href
  {https://ui.adsabs.harvard.edu/abs/2018A&A...619A..90P} {619, A90}

\bibitem[\protect\citeauthoryear{{Planck Collaboration} et~al.,}{{Planck
  Collaboration} et~al.}{2014}]{Planck13}
{Planck Collaboration} et~al., 2014, \mn@doi [\aap]
  {10.1051/0004-6361/201321591}, \href
  {https://ui.adsabs.harvard.edu/abs/2014A%26A...571A..16P} {571, A16}

\bibitem[\protect\citeauthoryear{{Popping} et~al.,}{{Popping}
  et~al.}{2017}]{Popping17}
{Popping} G.,  et~al., 2017, \mn@doi [\aap] {10.1051/0004-6361/201730391},
  \href {http://adsabs.harvard.edu/abs/2017A%26A...602A..11P} {602, A11}

\bibitem[\protect\citeauthoryear{{Price} et~al.,}{{Price}
  et~al.}{2014}]{Price14}
{Price} S.~H.,  et~al., 2014, \mn@doi [\apj] {10.1088/0004-637X/788/1/86},
  \href {http://adsabs.harvard.edu/abs/2014ApJ...788...86P} {788, 86}

\bibitem[\protect\citeauthoryear{{Puglisi} et~al.,}{{Puglisi}
  et~al.}{2016}]{Puglisi16}
{Puglisi} A.,  et~al., 2016, \mn@doi [\aap] {10.1051/0004-6361/201526782},
  \href {https://ui.adsabs.harvard.edu/abs/2016A&A...586A..83P} {586, A83}

\bibitem[\protect\citeauthoryear{{Querejeta} et~al.,}{{Querejeta}
  et~al.}{2016}]{Querejeta16}
{Querejeta} M.,  et~al., 2016, \mn@doi [\aap] {10.1051/0004-6361/201628674},
  \href {https://ui.adsabs.harvard.edu/abs/2016A&A...593A.118Q} {593, A118}

\bibitem[\protect\citeauthoryear{{Ramos Almeida}, {Acosta-Pulido}, {Tadhunter},
  {Gonz{\'a}lez-Fern{\'a}ndez}, {Cicone}  \& {Fern{\'a}ndez-Torreiro}}{{Ramos
  Almeida} et~al.}{2019}]{RamosAlmeida19}
{Ramos Almeida} C.,  {Acosta-Pulido} J.~A.,  {Tadhunter} C.~N.,
  {Gonz{\'a}lez-Fern{\'a}ndez} C.,  {Cicone} C.,   {Fern{\'a}ndez-Torreiro} M.,
   2019, \mn@doi [\mnras] {10.1093/mnrasl/slz072}, \href
  {https://ui.adsabs.harvard.edu/abs/2019MNRAS.487L..18R} {487, L18}

\bibitem[\protect\citeauthoryear{{Reddy} et~al.,}{{Reddy}
  et~al.}{2015}]{Reddy15}
{Reddy} N.~A.,  et~al., 2015, \mn@doi [\apj] {10.1088/0004-637X/806/2/259},
  \href {http://adsabs.harvard.edu/abs/2015ApJ...806..259R} {806, 259}

\bibitem[\protect\citeauthoryear{{Revalski} et~al.,}{{Revalski}
  et~al.}{2018}]{Revalski18}
{Revalski} M.,  et~al., 2018, \mn@doi [\apj] {10.3847/1538-4357/aae3e6}, \href
  {https://ui.adsabs.harvard.edu/abs/2018ApJ...867...88R} {867, 88}

\bibitem[\protect\citeauthoryear{{Rohlfs} \& {Wilson}}{{Rohlfs} \&
  {Wilson}}{1996}]{Rohlfs96}
{Rohlfs} K.,  {Wilson} T.~L.,  1996, {Tools of Radio Astronomy}

\bibitem[\protect\citeauthoryear{{Rosario}}{{Rosario}}{2019}]{Fortes}
{Rosario} D.~J.,  2019, {FortesFit: Flexible spectral energy distribution
  modelling with a Bayesian backbone} (\mn@eprint {ascl} {1904.011})

\bibitem[\protect\citeauthoryear{{Rosario}, {Togi}, {Burtscher}, {Davies},
  {Shimizu}  \& {Lutz}}{{Rosario} et~al.}{2019}]{Rosario19}
{Rosario} D.~J.,  {Togi} A.,  {Burtscher} L.,  {Davies} R.~I.,  {Shimizu}
  T.~T.,   {Lutz} D.,  2019, \mn@doi [\apjl] {10.3847/2041-8213/ab1262}, \href
  {https://ui.adsabs.harvard.edu/abs/2019ApJ...875L...8R} {875, L8}

\bibitem[\protect\citeauthoryear{{Rose}, {Tadhunter}, {Ramos Almeida},
  {Rodr{\'\i}guez Zaur{\'\i}n}, {Santoro}  \& {Spence}}{{Rose}
  et~al.}{2018}]{Rose18}
{Rose} M.,  {Tadhunter} C.,  {Ramos Almeida} C.,  {Rodr{\'\i}guez Zaur{\'\i}n}
  J.,  {Santoro} F.,   {Spence} R.,  2018, \mn@doi [\mnras]
  {10.1093/mnras/stx2590}, \href
  {https://ui.adsabs.harvard.edu/abs/2018MNRAS.474..128R} {474, 128}

\bibitem[\protect\citeauthoryear{{Rupke}, {G{\"u}ltekin}  \&
  {Veilleux}}{{Rupke} et~al.}{2017}]{Rupke17}
{Rupke} D. S.~N.,  {G{\"u}ltekin} K.,   {Veilleux} S.,  2017, \mn@doi [\apj]
  {10.3847/1538-4357/aa94d1}, \href
  {https://ui.adsabs.harvard.edu/abs/2017ApJ...850...40R} {850, 40}

\bibitem[\protect\citeauthoryear{{Santini} et~al.,}{{Santini}
  et~al.}{2019}]{Santini19}
{Santini} P.,  et~al., 2019, \mn@doi [\mnras] {10.1093/mnras/stz801}, \href
  {https://ui.adsabs.harvard.edu/abs/2019MNRAS.486..560S} {486, 560}

\bibitem[\protect\citeauthoryear{{Schinnerer} et~al.,}{{Schinnerer}
  et~al.}{2010}]{Schinnerer10}
{Schinnerer} E.,  et~al., 2010, \mn@doi [The Astrophysical Journal Supplement
  Series] {10.1088/0067-0049/188/2/384}, \href
  {https://ui.adsabs.harvard.edu/abs/2010ApJS..188..384S} {188, 384}

\bibitem[\protect\citeauthoryear{{Scholtz} et~al.,}{{Scholtz}
  et~al.}{2018}]{Scholtz18}
{Scholtz} J.,  et~al., 2018, \mn@doi [\mnras] {10.1093/mnras/stx3177}, \href
  {http://adsabs.harvard.edu/abs/2018MNRAS.475.1288S} {475, 1288}

\bibitem[\protect\citeauthoryear{{Schreiber} et~al.,}{{Schreiber}
  et~al.}{2015}]{Schreiber15}
{Schreiber} C.,  et~al., 2015, \mn@doi [\aap] {10.1051/0004-6361/201425017},
  \href {http://adsabs.harvard.edu/abs/2015A%26A...575A..74S} {575, A74}

\bibitem[\protect\citeauthoryear{Schwarz}{Schwarz}{1978}]{Schwarz78}
Schwarz G.,  1978, Ann. Statist., 6, 461

\bibitem[\protect\citeauthoryear{{Segers}, {Schaye}, {Bower}, {Crain},
  {Schaller}  \& {Theuns}}{{Segers} et~al.}{2016}]{Segers16}
{Segers} M.~C.,  {Schaye} J.,  {Bower} R.~G.,  {Crain} R.~A.,  {Schaller} M.,
  {Theuns} T.,  2016, \mn@doi [\mnras] {10.1093/mnrasl/slw111}, \href
  {https://ui.adsabs.harvard.edu/abs/2016MNRAS.461L.102S} {461, L102}

\bibitem[\protect\citeauthoryear{{Sharples} et~al.,}{{Sharples}
  et~al.}{2004}]{Sharples04}
{Sharples} R.~M.,  et~al., 2004, in {Moorwood} A. F.~M.,  {Iye} M.,  eds,
  Society of Photo-Optical Instrumentation Engineers (SPIE) Conference Series
  Vol. 5492, Ground-based Instrumentation for Astronomy. pp 1179--1186,
  \mn@doi{10.1117/12.550495}

\bibitem[\protect\citeauthoryear{{Sharples} et~al.,}{{Sharples}
  et~al.}{2013}]{Sharples13}
{Sharples} R.,  et~al., 2013, The Messenger, \href
  {https://ui.adsabs.harvard.edu/abs/2013Msngr.151...21S} {151, 21}

\bibitem[\protect\citeauthoryear{{Shin}, {Woo}, {Chung}, {Baek}, {Cho}, {Kang}
  \& {Bae}}{{Shin} et~al.}{2019}]{Shin19}
{Shin} J.,  {Woo} J.-H.,  {Chung} A.,  {Baek} J.,  {Cho} K.,  {Kang} D.,
  {Bae} H.-J.,  2019, arXiv e-prints, \href
  {https://ui.adsabs.harvard.edu/abs/2019arXiv190700982S} {p. arXiv:1907.00982}

\bibitem[\protect\citeauthoryear{{Silk} \& {Rees}}{{Silk} \&
  {Rees}}{1998}]{Silk98}
{Silk} J.,  {Rees} M.~J.,  1998, \aap, \href
  {https://ui.adsabs.harvard.edu/abs/1998A&A...331L...1S} {331, L1}

\bibitem[\protect\citeauthoryear{{Simpson} et~al.,}{{Simpson}
  et~al.}{2006}]{Simpson06}
{Simpson} C.,  et~al., 2006, \mn@doi [Monthly Notices of the Royal Astronomical
  Society] {10.1111/j.1365-2966.2006.10907.x}, \href
  {https://ui.adsabs.harvard.edu/abs/2006MNRAS.372..741S} {372, 741}

\bibitem[\protect\citeauthoryear{{Simpson} et~al.,}{{Simpson}
  et~al.}{2015}]{Simpson15}
{Simpson} J.~M.,  et~al., 2015, \mn@doi [\apj] {10.1088/0004-637X/807/2/128},
  \href {http://adsabs.harvard.edu/abs/2015ApJ...807..128S} {807, 128}

\bibitem[\protect\citeauthoryear{{Slone} \& {Netzer}}{{Slone} \&
  {Netzer}}{2012}]{Slone12}
{Slone} O.,  {Netzer} H.,  2012, \mn@doi [\mnras]
  {10.1111/j.1365-2966.2012.21699.x}, \href
  {https://ui.adsabs.harvard.edu/abs/2012MNRAS.426..656S} {426, 656}

\bibitem[\protect\citeauthoryear{{Soltan}}{{Soltan}}{1982}]{Soltan82}
{Soltan} A.,  1982, \mn@doi [\mnras] {10.1093/mnras/200.1.115}, \href
  {https://ui.adsabs.harvard.edu/abs/1982MNRAS.200..115S} {200, 115}

\bibitem[\protect\citeauthoryear{{Spilker} et~al.,}{{Spilker}
  et~al.}{2016}]{Spilker16}
{Spilker} J.~S.,  et~al., 2016, \mn@doi [\apj] {10.3847/0004-637X/826/2/112},
  \href {http://adsabs.harvard.edu/abs/2016ApJ...826..112S} {826, 112}

\bibitem[\protect\citeauthoryear{{Stach} et~al.,}{{Stach}
  et~al.}{2019}]{Stach19}
{Stach} S.~M.,  et~al., 2019, arXiv e-prints, \href
  {http://adsabs.harvard.edu/abs/2019arXiv190302602S} {}

\bibitem[\protect\citeauthoryear{{Stanley}, {Harrison}, {Alexander},
  {Swinbank}, {Aird}, {Del Moro}, {Hickox}  \& {Mullaney}}{{Stanley}
  et~al.}{2015}]{Stanley15}
{Stanley} F.,  {Harrison} C.~M.,  {Alexander} D.~M.,  {Swinbank} A.~M.,  {Aird}
  J.~A.,  {Del Moro} A.,  {Hickox} R.~C.,   {Mullaney} J.~R.,  2015, \mn@doi
  [\mnras] {10.1093/mnras/stv1678}, \href
  {http://adsabs.harvard.edu/abs/2015MNRAS.453..591S} {453, 591}

\bibitem[\protect\citeauthoryear{{Stanley}, {Harrison}, {Alexander}, {Simpson},
  {Knudsen}, {Mullaney}, {Rosario}  \& {Scholtz}}{{Stanley}
  et~al.}{2018}]{Stanley18}
{Stanley} F.,  {Harrison} C.~M.,  {Alexander} D.~M.,  {Simpson} J.,  {Knudsen}
  K.~K.,  {Mullaney} J.~R.,  {Rosario} D.~J.,   {Scholtz} J.,  2018, \mn@doi
  [\mnras] {10.1093/mnras/sty1044}, \href
  {http://adsabs.harvard.edu/abs/2018MNRAS.478.3721S} {478, 3721}

\bibitem[\protect\citeauthoryear{{Storchi-Bergmann}, {Lopes}, {McGregor},
  {Riffel}, {Beck}  \& {Martini}}{{Storchi-Bergmann}
  et~al.}{2010}]{StorchiBergmann10}
{Storchi-Bergmann} T.,  {Lopes} R.~D.~S.,  {McGregor} P.~J.,  {Riffel} R.~A.,
  {Beck} T.,   {Martini} P.,  2010, \mn@doi [\mnras]
  {10.1111/j.1365-2966.2009.15962.x}, \href
  {https://ui.adsabs.harvard.edu/abs/2010MNRAS.402..819S} {402, 819}

\bibitem[\protect\citeauthoryear{{Stott} et~al.,}{{Stott}
  et~al.}{2016}]{Stott16}
{Stott} J.~P.,  et~al., 2016, \mn@doi [\mnras] {10.1093/mnras/stw129}, \href
  {http://adsabs.harvard.edu/abs/2016MNRAS.457.1888S} {457, 1888}

\bibitem[\protect\citeauthoryear{{Sturm} et~al.,}{{Sturm}
  et~al.}{2011}]{Sturm11}
{Sturm} E.,  et~al., 2011, \mn@doi [\apjl] {10.1088/2041-8205/733/1/L16}, \href
  {http://adsabs.harvard.edu/abs/2011ApJ...733L..16S} {733, L16}

\bibitem[\protect\citeauthoryear{{Tadaki} et~al.,}{{Tadaki}
  et~al.}{2017}]{Tadaki17}
{Tadaki} K.-i.,  et~al., 2017, \mn@doi [\apj] {10.3847/1538-4357/834/2/135},
  \href {http://adsabs.harvard.edu/abs/2017ApJ...834..135T} {834, 135}

\bibitem[\protect\citeauthoryear{{Talia} et~al.,}{{Talia}
  et~al.}{2018}]{Talia18}
{Talia} M.,  et~al., 2018, \mn@doi [\mnras] {10.1093/mnras/sty481}, \href
  {http://adsabs.harvard.edu/abs/2018MNRAS.476.3956T} {476, 3956}

\bibitem[\protect\citeauthoryear{{Tiley} et~al.,}{{Tiley}
  et~al.}{2019}]{Tiley19}
{Tiley} A.~L.,  et~al., 2019, \mn@doi [\mnras] {10.1093/mnras/stz428}, \href
  {http://adsabs.harvard.edu/abs/2019MNRAS.485..934T} {485, 934}

\bibitem[\protect\citeauthoryear{{Ueda} et~al.,}{{Ueda} et~al.}{2008}]{Ueda08}
{Ueda} Y.,  et~al., 2008, \mn@doi [\apjs] {10.1086/591083}, \href
  {https://ui.adsabs.harvard.edu/abs/2008ApJS..179..124U} {179, 124}

\bibitem[\protect\citeauthoryear{{Veilleux}, {Cecil}  \&
  {Bland-Hawthorn}}{{Veilleux} et~al.}{2005}]{Veilleux05}
{Veilleux} S.,  {Cecil} G.,   {Bland-Hawthorn} J.,  2005, \mn@doi [\araa]
  {10.1146/annurev.astro.43.072103.150610}, \href
  {http://adsabs.harvard.edu/abs/2005ARA%26A..43..769V} {43, 769}

\bibitem[\protect\citeauthoryear{{Veilleux} et~al.,}{{Veilleux}
  et~al.}{2013}]{Veilleux13}
{Veilleux} S.,  et~al., 2013, \mn@doi [\apj] {10.1088/0004-637X/776/1/27},
  \href {https://ui.adsabs.harvard.edu/abs/2013ApJ...776...27V} {776, 27}

\bibitem[\protect\citeauthoryear{{Venturi} et~al.,}{{Venturi}
  et~al.}{2018}]{Venturi18}
{Venturi} G.,  et~al., 2018, \mn@doi [\aap] {10.1051/0004-6361/201833668},
  \href {https://ui.adsabs.harvard.edu/abs/2018A&A...619A..74V} {619, A74}

\bibitem[\protect\citeauthoryear{{Villar-Mart{\'\i}n}, {Arribas}, {Emonts},
  {Humphrey}, {Tadhunter}, {Bessiere}, {Cabrera Lavers}  \& {Ramos
  Almeida}}{{Villar-Mart{\'\i}n} et~al.}{2016}]{VillarMartin16}
{Villar-Mart{\'\i}n} M.,  {Arribas} S.,  {Emonts} B.,  {Humphrey} A.,
  {Tadhunter} C.,  {Bessiere} P.,  {Cabrera Lavers} A.,   {Ramos Almeida} C.,
  2016, \mn@doi [\mnras] {10.1093/mnras/stw901}, \href
  {https://ui.adsabs.harvard.edu/abs/2016MNRAS.460..130V} {460, 130}

\bibitem[\protect\citeauthoryear{{Vogelsberger} et~al.,}{{Vogelsberger}
  et~al.}{2014}]{Vogelsberger14}
{Vogelsberger} M.,  et~al., 2014, \mn@doi [\mnras] {10.1093/mnras/stu1536},
  \href {http://adsabs.harvard.edu/abs/2014MNRAS.444.1518V} {444, 1518}

\bibitem[\protect\citeauthoryear{{Whitaker}, {van Dokkum}, {Brammer}  \&
  {Franx}}{{Whitaker} et~al.}{2012}]{Whitaker12}
{Whitaker} K.~E.,  {van Dokkum} P.~G.,  {Brammer} G.,   {Franx} M.,  2012,
  \mn@doi [\apjl] {10.1088/2041-8205/754/2/L29}, \href
  {http://adsabs.harvard.edu/abs/2012ApJ...754L..29W} {754, L29}

\bibitem[\protect\citeauthoryear{{Whitaker} et~al.,}{{Whitaker}
  et~al.}{2014}]{Whitaker14}
{Whitaker} K.~E.,  et~al., 2014, \mn@doi [\apj] {10.1088/0004-637X/795/2/104},
  \href {http://adsabs.harvard.edu/abs/2014ApJ...795..104W} {795, 104}

\bibitem[\protect\citeauthoryear{{Wild}, {Charlot}, {Brinchmann}, {Heckman},
  {Vince}, {Pacifici}  \& {Chevallard}}{{Wild} et~al.}{2011}]{Wild11}
{Wild} V.,  {Charlot} S.,  {Brinchmann} J.,  {Heckman} T.,  {Vince} O.,
  {Pacifici} C.,   {Chevallard} J.,  2011, \mn@doi [\mnras]
  {10.1111/j.1365-2966.2011.19367.x}, \href
  {https://ui.adsabs.harvard.edu/abs/2011MNRAS.417.1760W} {417, 1760}

\bibitem[\protect\citeauthoryear{{Wisnioski} et~al.,}{{Wisnioski}
  et~al.}{2018}]{Wisnioski17}
{Wisnioski} E.,  et~al., 2018, \mn@doi [\apj] {10.3847/1538-4357/aab097}, \href
  {http://adsabs.harvard.edu/abs/2018ApJ...855...97W} {855, 97}

\bibitem[\protect\citeauthoryear{{Woo}, {Bae}, {Son}  \& {Karouzos}}{{Woo}
  et~al.}{2016}]{Woo16}
{Woo} J.-H.,  {Bae} H.-J.,  {Son} D.,   {Karouzos} M.,  2016, \mn@doi [\apj]
  {10.3847/0004-637X/817/2/108}, \href
  {https://ui.adsabs.harvard.edu/abs/2016ApJ...817..108W} {817, 108}

\bibitem[\protect\citeauthoryear{{Wylezalek} \& {Zakamska}}{{Wylezalek} \&
  {Zakamska}}{2016}]{Wylezalek16}
{Wylezalek} D.,  {Zakamska} N.~L.,  2016, \mn@doi [\mnras]
  {10.1093/mnras/stw1557}, \href
  {https://ui.adsabs.harvard.edu/abs/2016MNRAS.461.3724W} {461, 3724}

\bibitem[\protect\citeauthoryear{{Wylezalek}, {Veilleux}, {Zakamska},
  {Barrera-Ballesteros}, {Luetzgendorf}, {Nesvadba}, {Rupke}  \&
  {Sun}}{{Wylezalek} et~al.}{2017}]{Wylezalek17JWST}
{Wylezalek} D.,  {Veilleux} S.,  {Zakamska} N.,  {Barrera-Ballesteros} J.,
  {Luetzgendorf} N.,  {Nesvadba} N.,  {Rupke} D.,   {Sun} A.,  2017, {Q-3D:
  Imaging Spectroscopy of Quasar Hosts with JWST Analyzed with a Powerful New
  PSF Decomposition and Spectral Analysis Package}, JWST Proposal ID 1335.
  Cycle 0 Early Release Scienc

\bibitem[\protect\citeauthoryear{{Xue}, {Luo}, {Brandt}, {Alexander}, {Bauer},
  {Lehmer}  \& {Yang}}{{Xue} et~al.}{2016}]{Xue16}
{Xue} Y.~Q.,  {Luo} B.,  {Brandt} W.~N.,  {Alexander} D.~M.,  {Bauer} F.~E.,
  {Lehmer} B.~D.,   {Yang} G.,  2016, \mn@doi [\apjs]
  {10.3847/0067-0049/224/2/15}, \href
  {https://ui.adsabs.harvard.edu/abs/2016ApJS..224...15X} {224, 15}

\bibitem[\protect\citeauthoryear{{Zinn}, {Middelberg}, {Norris}, {Hales}, {Mao}
   \& {Randall}}{{Zinn} et~al.}{2012}]{Zinn12}
{Zinn} P.~C.,  {Middelberg} E.,  {Norris} R.~P.,  {Hales} C.~A.,  {Mao} M.~Y.,
   {Randall} K.~E.,  2012, \mn@doi [Astronomy and Astrophysics]
  {10.1051/0004-6361/201219349}, \href
  {https://ui.adsabs.harvard.edu/abs/2012A&A...544A..38Z} {544, A38}

\makeatother
\end{thebibliography}
